\documentclass[English,11pt]{iopart}

\usepackage{iopams}  
\usepackage{setstack}
\usepackage[utf8]{inputenc}
\usepackage{enumitem}
\usepackage[normalem]{ulem}
\usepackage{esint}
\usepackage{graphicx}
\usepackage{appendix}
\usepackage[colorlinks = true,
            linkcolor = blue,
            urlcolor  = blue,
            citecolor = blue,
            anchorcolor = blue]{hyperref}
\usepackage[square,sort&compress,sectionbib]{natbib}		
			
\newcommand\dfdx[2]{\frac{{\mathrm{d}} #1}{{\mathrm{d}} #2}}

\def\d{\mathrm{d}}
\newcommand{\Matrix}[9]{\left( \matrix{ #1 & #2 & #3 \cr 
#4 & #5 & #6 \cr 
#7 & #8 & #9 \cr} \right)}
\newcommand{\Matrixd}[4]{\left( \matrix{ #1 & #2 \cr 
#3 & #4 \cr } \right)}

\newcommand{\N}{\mathbb{N}}    
\newcommand{\Z}{\mathbb{Z}}    
\newcommand{\R}{\mathbb{R}}    
\renewcommand{\P}{\mathbb{P}}  
\newcommand{\dd}{{\mathrm{d}}}

\begin{document}

\title{Quasi-isometric embedding of Kerr poloidal sub-manifolds}

\author{Lo\"\i c Chantry$^1$, Frédéric Dauvergne$^{1,2}$, Youssef Temmam$^3$, Véronique Cayatte$^1$}

\address{$^1$ LUTH, Observatoire de Paris, Université PSL, UMR 8102 CNRS/INSU, Université de Paris, 5 place Jules Janssen, 92190 Meudon, France}
\address{$^2$ IMCCE, Observatoire de Paris, PSL Research University, UMR 8028 CNRS, Sorbonne Université, Université de Lille, 77 avenue  Denfert-Rochereau, 75014 Paris, France}
\address{$^3$ Observatoire de Paris, Université PSL, 61 avenue de l'Observatoire, 75014 Paris, France}

\ead{loic.chantry@obspm.fr}

\vspace{10pt}

\begin{abstract} 
We propose two approaches to obtain an isometric embedding of the poloidal Kerr sub-manifold. The first one relies on the convex integration process using the corrugation from a primitive embedding. This allows us to obtain one parameter family of embeddings reaching the limits of an isometric embedding. The second one consists in consecutive numerical resolutions of the Gauss-Codazzi-Mainardi and frame equations. This method requires geometric assumptions near the equatorial axis of the poloidal sub-manifold to get initial and boundary conditions. The second approach allows to understand some physical properties in the vicinity of a Kerr black hole, in particular the fast increasing ergoregion extent with angular momentum.
\end{abstract}

%
%
%
%
%

\section{Introduction}

Since the introduction of the general relativity, isometric embedding has helped to picture the geometry of the curved space-time or parts of it. Essentially, two approaches were used for those embeddings : the embedding of two-dimensional sub-manifold and the embedding of the whole space-time into a higher dimension space as in the work of \cite{doi:10.1063/1.4891157} or \cite{2016arXiv161205256R}.

This paper follows the first approach. Many isometric sub-manifold embeddings have already been produced. \cite{PhysRevD.7.289} shows the existence of an isometric embedding of  Kerr black hole horizon section into the ordinary $3$-dimensional Euclidean space for a spin $a$ lower than $\sqrt{3}/2$. We remind hereby the work of \cite{2009PhRvD..80d4014G} who used $3$-dimensional hyperbolic space instead of the ordinary $3$-dimensional Euclidean space; they achieved the isometric embedding of Kerr black hole horizon for values of $a$ higher than $\sqrt{3}/2$. The equatorial sub-manifold, usually called the ``equatorial plane'', has been isometrically embedded; one can see for example  \cite{2016arXiv160706535H} for the Kerr metric and a deformed Kerr metric. In \cite{2001ragt.meet...25H} and \cite{2008mgm..conf.2299H}, the isometric embedding of the equatorial sub-manifold is presented for a Schwarzschild black hole with or without addition of uniform matter density, and for different space-time manifolds (Reisner-Nordström, Ernst and Kerr-Newmann) with and without cosmological constant.

Nevertheless, in each of those cases the considered sub-manifold possesses an \emph{internal symmetry} (axisymmetry) that enables to detail the calculation of its embedding. The embedding of distorted black hole horizons (without axisymmetry) has also inspired a lot of works that describe methods for the isometric sub-manifold embedding with spherical topology as in \cite{2002CQGra..19..375B}, \cite{2012CQGra..29o5010J}, and \cite{Tichy_2014}. \\

The history of isometric embedding problem started naturally after the introduction and formalization of differential geometry by Riemann at the end of the {\uppercase\expandafter{\romannumeral 19}} century. In particular, the question came from Poincaré's half-plane study. \cite{Hilbert1933} showed the non-existence of $\mathcal{C}^2$ global isometric embedding of this manifold.

In 1873, Schlaefli conjectured that the dimension of {the} Euclidean space for an isometrically $n$-dimensional embedded manifold must be larger than $s_n=\frac{n(n+1)}{2}$. \cite{cartan1927possibilite} proved that the system of isometric embedding equations for $n=2$ has a local solution if the dimension of the Euclidean space is at least $s_2=3$ and if the considered Riemannian manifold is \emph{analytic} (i.e., the metric coefficients are analytic). Indeed the proof requires the use of Cauchy-Kowalesky theorem. \cite{Janet_Maurice_Sur_1927} extended the proof for any $n\in \N$ and using a $s_n$ dimensional Euclidean space to embed the manifold. This result requires that the considered manifold metric must be analytic, and \cite{Pogorelov1971AnEO} found a counter-example for a $2$-dimensional manifold. Another counter-example was given by \cite{Nadirashvili2002CounterexamplesFL} whose metric is $\mathcal{C}^\infty$ but \emph{not analytic} and whose isometric embedding does not exist. For \emph{non-analytic metrics}, \cite{lin1985} proved that a local embedding exists around points of non-zero curvature and also around points of zero curvature if this latter is negative nearby. He will go further in \cite{doi:10.1002/cpa.3160390607} where he proved this local existence around points of zero curvature, but whose gradient of curvature does not vanish. You can find all these results and most of the demonstrations in \cite{han2006isometric}.

Concerning global embedding problem, the famous result known as Withney theorem (\cite{key1503303m}) ensures the existence of \emph{non-isometric} embedding in $\R^{2n}$ for $n$-dimensional manifold. Using a completely new approach, \cite{10.2307/1969840} and \cite{zbMATH03114836} proved the existence of a $\mathcal{C}^1$ isometric embedding, if a non-isometric embedding exists. Using Withney theorem, it implies the existence of isometric embedding in Euclidean spaces of dimension greater than or equal to $2n$. The proof introduced by Nash is a prelude to the formalization of convex integration formulated by \cite{Gromov_1973}. Nevertheless the low smoothness of the embedding makes its use counter-intuitive. \cite{10.2307/1969989} proved the existence of global $\mathcal{C}^3$ isometric embeddings with the need of an Euclidean space of dimension $\frac{n(3n+12)}{2}$. This result will be improved by \cite{zbMATH03341024} who showed that the dimension can be reduced to $n^2+10n+3$ and $\frac{(n+2)(n+3)}{2}$ if the metric is at least $\mathcal{C}^4$. Later, the \href{http://hevea-project.fr/ENIndexHevea.html}{Hevea project} (\cite{borrelli:hal-00945774}) included the tools developed by Gromov to build an impressive image of the flat torus.\\


For magnetized flow model in the vicinity of a Kerr black hole, it is very common to use the axisymmetry and stationarity assumptions (\cite{1977MNRAS.179..433B}, \cite{1991PhRvD..44.2295N}, \cite{1993PhyU...36..529B}, \cite{2013PhRvD..88h4046G} or \cite{2018A&A...612A..63C}) for magneto-hydrodynamic equations; it allows to build solutions for modelling incoming flows, accretion disks and even relativistic jets. The induced field solutions have to be drawn only in the poloidal sub-manifold. The use of Cartesian pseudo-coordinates is sufficient for drawing fields at some distances of the black hole. Whereas in the case of ideal magnetized flows the field lines must penetrate the horizon orthogonally, the use of these coordinates does not allow to visualize this property. 

In a previous study (\cite{2020CQGra..37j5003C}), we have introduced different ways for obtaining a conformal representation, i.e. producing {an accurate} representation of the angles between curves of the same manifold. The purpose of this paper is to pursue this study in the same perspective by presenting isometric embeddings of this manifold. This leads to some {accurate} representations in terms of distances and angles for the fields defined on this sub-manifold. After section 2, one could read independently the sections 3 or 4 and 5.

In section 2, we start by reminding the basic properties of the $2$-dimensional poloidal sub-manifold. We will take advantage of this opportunity to introduce a coordinate system adapted to the geometric study of this sub-manifold.

The method of corrugation from a primitive embedding is described in section 3. It will enable us to obtain embeddings as close as one wishes from the isometric embedding.

In section 4, with the aid of geometric assumptions at the equatorial axis for initial conditions, we developed the Gauss-Coddazzi-Mainardi equations (GCM) to obtain the second fundamental form that respect these geometric requirements. We also introduced the frame equations which allow to get an embedding from second fundamental form.

In section 5, we numerically integrate the GCM and frame equations in order to obtain an approximate isometric embedding (the isometric default of induced metrics is less than $10^{-3}$). The geometric shape and induced numerical error are analysed and commented.
 
\section{The poloidal sub-manifold of Kerr metric} \label{sec2}
    \subsection{Definition and properties}
We shall start by recalling some useful basics about Kerr manifolds. The Kerr manifolds are the set of space-times whose metric is circular, stationary and axisymmetric solution of the vacuum Einstein's equation (see \cite{2010arXiv1003.5015G} for a detailed review of the circular space-time).

This set of space-times $\mathcal{M}_{m,a}$ is a two-parameter family, $m,a\in\R\times[0,1[$ (mass and spin of the black-hole; we do not consider the case where $a=1$), of 4-dimensional Pseudo-Riemmannian manifolds $(\mathcal{M}_{m,a},\bf{g}_{m,a})$. We will focus on the portion of Kerr space-time outside the event horizon. This region can be described by a singular map, that we call dimensionless Boyer-Lindquist coordinates. Then, there is a smooth diffeomorphism,
\begin{equation}
\begin{array}{ccccc}
\chi & : & \mathcal{M}_{m,a} & \longrightarrow & \mathcal{U}_a \\
 & & M & \mapsto & (u,r,\theta,\phi)
\end{array},
\end{equation}
where $\mathcal{U}_a=\R\times]r_H(a),\infty[\times]0,\pi[\times[0,2\pi[$ is a subset of $\R^4$, with $r_H(a)=1+\sqrt{1-a^2}$. The dimensionless Boyer-Lindquist coordinates $(u,r,\theta,\phi)$ are linked to the usual ones $(t_{BL},r_{BL},\theta,\phi)$ by,
\begin{equation}
\eqalign{
u=\frac{ct_{BL}}{r_g} \; ,\qquad\qquad r=\frac{r_{BL}}{r_g}\\
}\; ,
\end{equation}
with $r_g=\frac{\mathcal{G}m}{c^2}$. In the dimensionless map, the metric ${\bf g}_{m,a}$ could be written using coordinate forms, $\d\boldsymbol{u},\d\boldsymbol{r},\d\boldsymbol{\theta}$ and $\d\boldsymbol{\phi}$,
\begin{equation}
\begin{array}{ccccc}
\fl{\bf g}_{m,a} = \displaystyle r_g^2\left[-\left(1-\frac{2}{r\Omega^2(r,\theta)}\right) {\d}\boldsymbol{u}\otimes{\d}\boldsymbol{u}+ \left( r^{2} + a^{2} + \frac{2 a^{2}}{r\Omega^2(r,\theta)}\sin^{2}\theta \right) \sin^{2}\theta {\d}\boldsymbol{\phi}\otimes{\d}\boldsymbol{\phi}\right.\\
\displaystyle \left.- \frac{4 a \sin^{2} \theta}{r\Omega^2(r,\theta)} \dd \boldsymbol{u}\otimes\d\boldsymbol{\phi} + \Omega^2(r,\theta)\left(\frac{r^2}{\Delta(r)}{\d}\boldsymbol{r}\otimes{\d}\boldsymbol{r}+r^2{\d}\boldsymbol{\theta}\otimes{\d}\boldsymbol{\theta}\right)\right]
\end{array},
\label{eq-Kerr4}
\end{equation}
where we have,
\begin{equation}
\Omega^2(r,\theta)=1+\frac{a^2}{r^2}\cos^2\theta \quad {\rm and} \quad
\Delta(r)=r^2+a^2-2r\; .
\end{equation}
Now we will shed light on $(\mathcal{P}_{a,m,u_0,\phi_0},\boldsymbol{\mu}_{m,a,u_0,\phi_0})$, the poloidal sub-manifolds of Kerr manifold, which can be obtained fixing the value of $u=u_0$ and $\phi=\phi_0$ in Boyer-Lindquist map. The induced metric is obviously,
\begin{equation}\label{Eq-TargetMetric}
\eqalign{
\boldsymbol{\mu}_{m,a}=r_g^2\Omega^2(r,\theta)\left(\frac{r^2}{\Delta(r)}{\d}\boldsymbol{r}\otimes{\d}\boldsymbol{r}+r^2{\d}\boldsymbol{\theta}\otimes{\d}\boldsymbol{\theta}\right),
}
\end{equation}
which is independent of the values $u_0$ and $\phi_0$, so that we directly write $\boldsymbol{\mu}_{m,a,u_0,\phi_0}=\boldsymbol{\mu}_{m,a}$. 

In the following, we will note $(\mathcal{P},\boldsymbol{\mu})=(\mathcal{P}_{a,m},\boldsymbol{\mu}_{m,a})$ for writing convenience. We will remove also the factor $r_g^2$ which appears in (\ref{eq-Kerr4}). Indeed, this constant factor can be retrieved by using a simple homothetic transformation with the factor $r_g$. This two-dimensionnal manifold is also represented by a singular map defined on $\mathcal{V}=]r_H,\infty[\times]0,\pi[$.

It is easy to show that this sub-manifold has the same geometry as a poloidal restriction of the simultaneity sub-manifold of the ZAMO-observers (cf. \cite{2007gr.qc.....3035G}). In what follows, we will refer to the sub-manifold $(\mathcal{P},\boldsymbol{\mu})$ as the Kerr poloidal manifold, and will try to find an associated isometric embedding in $\R^3$.

\subsection{Coordinate system}
    
We will use more adapted coordinates to the geometrical study of the poloidal sub-manifold. Nevertheless, we will provide to the reader with the embeddings according to these coordinates or to the usual Boyer-Lindquist coordinates.

First of all, we will use latitude $\lambda=\frac{\pi}{2}-\theta$ instead of colatitude and introduce the radial coordinate $s : r \mapsto  s(r)$ which is defined by, 
\begin{equation}\label{Eq-sr-1}
    s(r)=\int_{r_H}^r\frac{x}{\sqrt{\Delta(x)}}\dd x\quad\Rightarrow\quad s(r)=\sqrt{\Delta(r)}+\ln\left(\frac{\sqrt{\Delta(r)}+r-1}{\sqrt{1-a^2}}\right)\,.
\end{equation}
The $s$ coordinate function is strictly increasing and then invertible. We call $r=s^{-1}$ the inverse function of $r \mapsto s(r)$. Then the diffeomorphism that represents the coordinate transformation is,
\begin{equation}\label{Eq-sr-2}
\begin{array}{ccccc}
g & : & \mathcal{U} & \longrightarrow & \mathcal{V} \\
 & & (s,\lambda) & \mapsto & (r(s),\frac{\pi}{2}-\lambda) 
\end{array}\,,
\end{equation}
where $\mathcal{U}=]0,+\infty[\times]-\pi/2,\pi/2[$ is an open set. We will use $\boldsymbol{e}_s,\boldsymbol{e}_\lambda$ to note the orthonormal basis of $\mathcal{U}\subset\R^2$ and $\dd \boldsymbol{s},\dd \boldsymbol{\lambda}$ the associated coordinate form. 

In this coordinate system, the metric of poloidal manifold could be written as,
\begin{equation}
\boldsymbol{\mu}=\Omega^2(s,\lambda)\left(\dd \boldsymbol{s}\otimes \dd \boldsymbol{s}+r^2(s){\d}\boldsymbol{\lambda}\otimes{\d}\boldsymbol{\lambda}\right)\; ,
\end{equation}
where
\begin{equation}
    \Omega^2(s,\lambda)=1+\frac{a^2}{r^2(s)}\sin^2\lambda \; .
\end{equation}
The $s$ coordinate corresponds to the length of the equatorial line, $\lambda=0$, for a given radius in Boyer-Lindquist coordinates.
    
    \subsection{Gaussian curvature and Christoffel's coefficients} \label{ss-voc&oq}
    
We used SageMath (see \cite{CCIRM_2018__6_1_A1_0}) to get the Christoffel's coefficients of the poloidal manifold,
\begin{equation}\label{Eq-Chritoffels}
\eqalign{\Gamma^s_{ss}&=-\displaystyle\frac{a^2\sin^2\lambda \sqrt{\Delta\left(r(s)\right)}}{r^2(s)\left(r^2(s)+a^2\sin^2\lambda\right)}\,,\\
    \Gamma^s_{s\lambda}&=\displaystyle\frac{a^2\cos\lambda\sin\lambda}{r^2(s)+a^2\sin^2\lambda}\,,\\   
    \Gamma^s_{\lambda\lambda}&=-\displaystyle\frac{r^2(s)\sqrt{\Delta\left(r(s)\right)}}{r^2(s)+a^2\sin^2\lambda}\,,}\qquad
\eqalign{
\Gamma^\lambda_{ss}&=-\displaystyle\frac{a^2\cos\lambda\sin\lambda}{r^2(s)\left(r^2(s)+a^2\sin^2\lambda\right)}\,,\\
    \Gamma^\lambda_{s\lambda}&=\displaystyle\frac{\sqrt{\Delta\left(r(s)\right)}}{r^2(s)+a^2\sin^2\lambda}\,,\\   
    \Gamma^\lambda_{\lambda\lambda}&=\displaystyle\frac{a^2\cos\lambda\sin\lambda}{r^2(s)+a^2\sin^2\lambda}\, ,}
\end{equation}
For the Gaussian curvature we get,
\begin{equation}\label{Eq-GaussianCurvature}
K=-\frac{r(s)\left(r^2(s)-3a^2\sin^2\lambda\right)}{(r^2(s)+a^2\sin^2\lambda)^3}\,.
\end{equation}
Note that for $a\geq{\sqrt{3}/2}$ the Gaussian curvature could change sign in some parts of the manifold. According to \cite{audoly:tel-00002515}, the degree of freedom for infinitesimal isometric deformation is equal to the number of unconstrained boundaries minus the number of principal asymptotic lines (i.e.: $K=0$ and $\nabla K\neq 0$). 
Thus, if there is a $\mathcal{C}^\infty$ isometric embedding, it looses two degrees of freedom for "isometric infinitesimal deformation", when $a$ passes through ${\sqrt{3}/2}$. 

This aspect will not bring any difficulty concerning the use of the convex integration method, but this will not be the case for the resolution of Gauss-Coddazzi-Mainardi equations, with the initial conditions we will consider.

\section{Quasi-isometric embedding by convex integration} \label{sec3}
    \subsection{Primitive embedding}
        \subsubsection{Definition}\leavevmode\par
        
We introduce the ordinary three dimensional Euclidean space $E_3=\left(\R^3,<\cdot,\cdot>\right)$ and we define the \emph{isometric default} of an embedding $f:\mathcal{U}\longrightarrow\R^3$ as,
\begin{equation}
    \boldsymbol{\delta}={\boldsymbol{\mu}}-f^\star<\cdot,\cdot>_{\R^3},
\end{equation}           
where $f^\star<\cdot,\cdot>_{\R^3}$ is a positive bilinear form corresponding to the induced metric calculated from Euclidean scalar product on $\R^3$, and ${\boldsymbol{\mu}}$ is the target metric. If $\boldsymbol{\delta}$ is a positive bilinear form, then the embedding $f$ is \emph{strictly short}.

The two-dimensional convex integration process, as explained in Chapter 2 of \cite{borrelli:hal-00945774}, can be simplified in the case where we get a \emph{primitive embedding} $f:\mathcal{U}\longrightarrow\R^3$. A primitive embedding has an isometric default which is a positive degenerate bilinear form,
\begin{equation}
    \boldsymbol{\delta}=\alpha\boldsymbol{\ell}\otimes\boldsymbol{\ell} \quad \Rightarrow \quad {\boldsymbol{\mu}}=f^\star<\cdot,\cdot>_{\R^3}+\alpha\boldsymbol{\ell}\otimes\boldsymbol{\ell}\,,
\end{equation}
where $\alpha : \mathcal{U}\longrightarrow \R_+$ is a positive function and $\boldsymbol{\ell}$ is  some non zero linear form on $\R^2$. Thus, $f$ is said a \emph{strictly short primitive embedding along $\boldsymbol{\ell}$}. At this point, we are looking for such an embedding in {this way},
\begin{equation}\label{Eq-immersion}
\begin{array}{ccccc}
f & : & \mathcal{U} & \longrightarrow & \R^3 \\
 & & (s,\lambda) & \mapsto & \left|\begin{array}{c}
 x(s)\cos\lambda\\
 y(s)\sin\lambda\\
 Z(s)
 \end{array}\right. \,.
\end{array}
\end{equation}
Thus, the induced metric is equal to,
\begin{equation}\label{Eq-EmbMetric-1}\left\{
\begin{array}{ccc}
    <\boldsymbol{\partial_s},\boldsymbol{\partial_s}>&=&\dot{x}^2\cos^2\lambda+\dot{y}^2\sin^2\lambda+\dot{Z}^2\\
    <\boldsymbol{\partial_s},\boldsymbol{\partial_\lambda}>&=&-\left[\dfdx{\,}{s}\left(x^2-y^2\right)\right]\cos\lambda\sin\lambda\\
    <\boldsymbol{\partial_\lambda},\boldsymbol{\partial_\lambda}>&=&x^2\sin^2\lambda+y^2\cos^2\lambda\\
\end{array}\right.    \,,
\end{equation}
where $\boldsymbol{\partial_s}=\partial_s f$ and $\boldsymbol{\partial_\lambda}=\partial_\lambda f$. By choosing $x(s)=\sqrt{R^2(s)+a^2}$ and $y(s)=R(s)$, we ensure that the induced metric is orthogonal,
\begin{equation}\label{Eq-EmbMetric-2}\left\{
\begin{array}{ccc}
    <\boldsymbol{\partial_s},\boldsymbol{\partial_s}>&=&\left(\frac{\dot{R}^2R^2}{R^2+a^2}+\dot{Z}^2\right)+\frac{a^2 \dot{R}^2}{R^2+a^2}\sin^2\lambda\\
    <\boldsymbol{\partial_s},\boldsymbol{\partial_\lambda}>&=&0\\
    <\boldsymbol{\partial_\lambda},\boldsymbol{\partial_\lambda}>&=&R^2+a^2\sin^2\lambda\\
\end{array}\right.    \,.
\end{equation}
In what follows, we will choose two directions for convex integration process, in order to find the embedding which differs from $\boldsymbol{\mu}$ by a primitive metric. 

We will choose $R$ and $Z$, so that we get firstly a strictly short primitive embedding $\rho$ along $\boldsymbol{\partial}_s$ and secondly a strictly short primitive embedding $\Lambda$ along $\boldsymbol{\partial}_\lambda$.



    \subsubsection{Strictly short primitive embedding along \texorpdfstring{$\boldsymbol{\partial}_s$}{ds}}\leavevmode\par

From equations (\ref{Eq-EmbMetric-2}) and the expression of target metric in Eq.(\ref{Eq-TargetMetric}), it is obvious that the choice of $R(s)=r(s)$ insures that $<\boldsymbol{\partial_\lambda},\boldsymbol{\partial_\lambda}>=\boldsymbol{\mu}(\boldsymbol{e}_\lambda,\boldsymbol{e}_\lambda)$. It allows us to find an embedding which gives an isometric default proportional to $\dd \boldsymbol{s}\otimes\dd \boldsymbol{s}$. We also choose $\dot{Z}^2$, so that we have,
\begin{equation}\label{eq-zs-prim}
    \left(\frac{\dot{R}^2R^2}{R^2+a^2}+\dot{Z}^2\right)=1\quad\Rightarrow\quad Z(s)=\int_0^s\sqrt{\frac{2r(s)}{r^2(s)+a^2}}\dd s \,.
\end{equation}
One could check that this integral is not defined for $a=1$. The strictly short primitive embedding $\rho$ is then,
\begin{equation}\label{Eq-Immersion-r}
\begin{array}{ccccc}
\rho & : & \mathcal{U} & \longrightarrow & \R^3 \\
 & & (s,\lambda) & \mapsto & \left|\begin{array}{c}
 \sqrt{r^2(s)+a^2}\cos\lambda\\
 r(s)\sin\lambda\\
 Z(s)
 \end{array}\right. \,.
\end{array}
\end{equation}
In this case the induced metric is,
\begin{equation}
    \rho^\star<\cdot,\cdot>_{\R^3}=\tilde{\Omega}(s,\lambda){\d}\boldsymbol{s}\otimes{\d}\boldsymbol{s}+r^2\Omega^2(s,\lambda){\d}\boldsymbol{\lambda}\otimes{\d}\boldsymbol{\lambda}\,,
\end{equation}
with,
\begin{equation}
    \tilde{\Omega}(s,\lambda)=\left(1+\frac{a^2\Delta}{\left(r^2+a^2\right)r^2}\sin^2\lambda\right).
\end{equation}
The isometric default can be written as,
\begin{equation}
    \boldsymbol{\delta}=\frac{2a^2 \sin^2\lambda}{r\left(r^2+a^2\right)}\dd \boldsymbol{s}\otimes\dd \boldsymbol{s}\,, 
\end{equation}
and the function $\alpha : (s,\lambda)\mapsto \frac{2a^2 \sin^2\lambda}{r\left(r^2+a^2\right)} $ verifies $0\leq\alpha(s,\lambda)\leq \frac{a^2}{r_H^2}\leq 1$. 

        \subsubsection{Strictly short primitive embedding along \texorpdfstring{$\boldsymbol{\partial}_\lambda$}{dl}}\leavevmode\par
        
From Eq.(\ref{Eq-EmbMetric-2}), to get a primitive embedding along $ \boldsymbol{\partial}_\lambda$ we need to verify,
\begin{equation}
\left(\frac{\dot{R}^2R^2}{R^2+a^2}+\dot{Z}^2\right)+\frac{a^2 \dot{R}^2}{R^2+a^2}\sin^2\lambda=1+\frac{a^2}{r^2}\sin^2\lambda \, ,
\end{equation}
which is equivalent to solve the ordinary differential system,
\begin{equation}\label{Eq-SystePrimitEmbTheta}\left\{
\begin{array}{c}
\frac{\dot{R}^2R^2}{R^2+a^2}+\dot{Z}^2 = 1\\
\frac{\dot{R}^2}{R^2+a^2}=\frac{1}{r^2}
\end{array}\right.    \,.
\end{equation}
The second equation immediately yields to,
\begin{equation}
    \frac{dR}{\sqrt{R^2+a^2}}=\frac{dr}{\sqrt{\Delta}}\,.
\end{equation}
Then the integration can be performed,
\begin{equation}
    \frac{R+\sqrt{R^2+a^2}}{R_H+\sqrt{R_H^2+a^2}}=\frac{r-1+\sqrt{\Delta}}{\sqrt{1-a^2}} \, , 
\end{equation}
where $R_H=R(r_H)$.
In order to have $R \sim r$ when $r\longrightarrow \infty$, we put,
\begin{equation}
    R_H+\sqrt{R_H^2+a^2}=\sqrt{1-a^2} \qquad \Rightarrow \quad R_H=\frac{1-2a^2}{2\sqrt{1-a^2}}\,,
\end{equation}
so as to avoid self-intersection, we impose that $R_H>0$ which is equivalent to $a<1/\sqrt{2}$. Then the primitive embedding along $\boldsymbol{\partial}_\lambda$ will not exist for $a\geq 1/\sqrt{2}$. 

We could express directly $R$ in function of $r$, 
\begin{equation}\label{Eq-Rprimit-theta}
    R(r)=r-1-\frac{1}{2\left[r-1+\sqrt{\Delta(r)}\right]}\,,
\end{equation}
which is a strictly increasing function verifying $R(r)\leq r$ for all $r$. The value of $Z$ will be easily deduced from the $R$ function, 
\begin{equation}\label{Eq-Zprimit-theta}
    Z(r)=\int_{r_H}^r\sqrt{\frac{r^2-R^2(r)}{\Delta}}\dd r\,.
\end{equation}
As for the primitive embedding along $\boldsymbol{\partial}_r$, we use the coordinate change from Eq.(\ref{Eq-sr-1}) and Eq.(\ref{Eq-sr-2}). The strictly short primitive embedding $\Lambda$ is then,
\begin{equation}\label{Eq-Immersion-theta}
\begin{array}{ccccc}   
\Lambda & : & \mathcal{U} & \longrightarrow & \R^3 \\
 & & (s,\lambda) & \mapsto & \left|\begin{array}{c}
 \sqrt{R(r(s))^2+a^2}\cos\lambda\\
 R(r(s))\sin\lambda\\
 Z(r(s))
 \end{array}\right. \,,
\end{array}
\end{equation}
where $R$ and $Z$ are defined by Eq.(\ref{Eq-Rprimit-theta}) and Eq.(\ref{Eq-Zprimit-theta}). The induced metric verifies,
\begin{equation}\label{eq-29}
   \boldsymbol{\mu}= \Lambda^\star<\cdot,\cdot>_{\R^3}+\left(r^2-R^2\right)\dd \boldsymbol{\lambda}\otimes\dd \boldsymbol{\lambda}\, .
\end{equation}
The inequality $r^2-R^2>0$ implies that $\Lambda$ is strictly short and Eq.(\ref{eq-29}) also proves the primitive character of this embedding.

    \subsection{Convex integration} 
        \subsubsection{Corrugation}\leavevmode\par \label{321}
    
For a regular curve $\gamma : I\longrightarrow \R^3$, the 1D integration process has to construct another curve, $\Gamma : I\longrightarrow \R^3$, which is longer and remains close to the first curve $\gamma$ ($\mathcal{C}^0$-close to $\gamma$) with an explicit control of \emph{speed} $||\dot{\boldsymbol{\Gamma}}(x)||$. For a given scalar function $\kappa : I \longrightarrow \R_+$, such that for each $x$ in $I$, we have a strictly short curve if $\kappa(x)\geq ||\dot{\boldsymbol{\gamma}}(x)||$ (see chapter 2 in \cite{borrelli:hal-00945774}). The convex integration allows us to construct $\Gamma : I\longrightarrow \R^3$ such that for each $x$ in $I$ we have $||\dot{\boldsymbol{\Gamma}}(x)||=\kappa(x)$. We completely control the speed magnitude and then the local length of the curve. The control of the speed magnitude comes from the explicit construction of the curves using what we call \emph{the loop function}. The corrugated curve $\Gamma$ is $\mathcal{C}^0$-close to $\gamma$ and it is due to the $\mathcal{C}^0$-Lemma.

To explain the 2D convex integration for isometric embedding, we shall follow \cite{borrelli:hal-00945774}. We consider here the previous strictly short primitive embedding $\rho :\mathcal{U}\longrightarrow \R^3$. For simplification, we will take $f=\rho$ in the following, but the demonstration could be easily adapted with $f=\Lambda$ by swapping the variables. The basic idea is to construct, from this embedding, a 1-parameter family $\left(F_\nu\right)_{\nu\in\R_+^\star}$ of embeddings, such that $||\boldsymbol{\mu}-F_\nu^\star<\cdot,\cdot>_{\R^3}||=O(1/\nu)$. The limit of this family when $\nu \rightarrow +\infty$ is $f$, a non-isometric embedding, but {the induced metrics} $F_\nu^\star<\cdot,\cdot>$ {are} {as close as we need to the target metric}. Furthermore, our new embedding will verify a $\mathcal{C}^0$-close to $f$ condition. We remind that $f$ verifies,
\begin{equation}\label{eq-alpha}
    {\boldsymbol{\mu}}=f^\star<\cdot,\cdot>_{\R^3}+\alpha(s,\lambda)\dd\boldsymbol{s}\otimes\dd\boldsymbol{s}\,,
\end{equation}
where $\alpha(s,\lambda)=\frac{2a^2 \sin^2\lambda}{r\left(r^2+a^2\right)}$.

To construct the corrugated embedding $F_\nu$, we span our set of departure $\mathcal{U}$ with a 1-parameter family of lines $\phi_\lambda$. This family must be chosen carefully (see p 20-24 of \cite{borrelli:hal-00945774}). In our case, because our target metric verifies $\boldsymbol{\mu}(\boldsymbol{e}_s,\boldsymbol{e}_\lambda)=0$ for each $\lambda$, we could choose $\phi_\lambda : s\mapsto (s,\lambda)$.

\begin{figure}[ht]
    \centering
   	\includegraphics[scale=0.5]{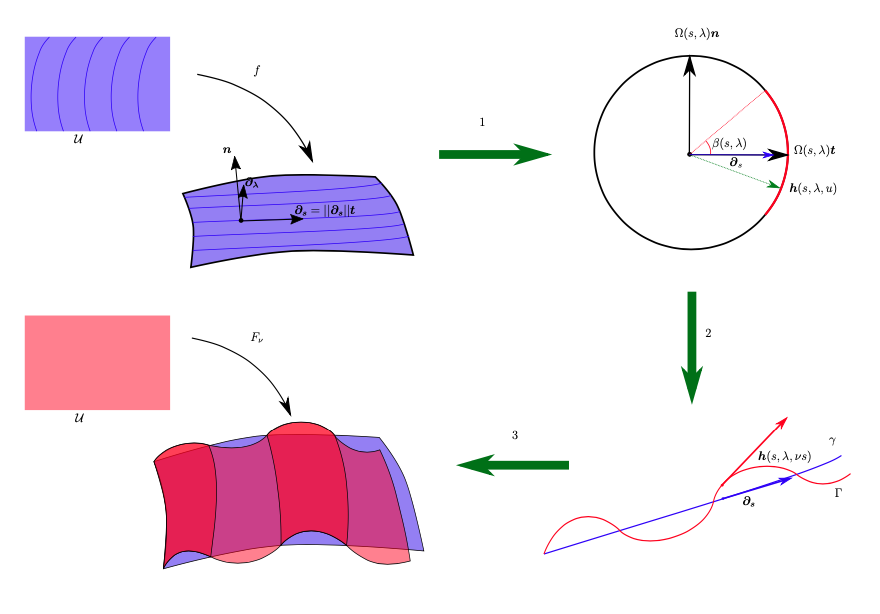}
    \caption{Corrugation schematic view. Top-left, a strictly short primitive embedding is shown. Blue lines represent a "well-adapted" one-parameter family of curves $(\phi_\lambda)_{\lambda\in]-\pi/2;\pi/2[}$ in the whole domain. This construction implies the existence of vectorial functions $\boldsymbol{\partial_s},\boldsymbol{\partial_\lambda},\boldsymbol{n}$ defined on $\mathcal{U}$. Top-right, these functions  are used to construct the loop function, which oscillates with the corrugated frequency in the red part of the circle. Bottom-right, the integration of  the loop function for one blue line gives the corrugated line in red. Bottom-left, the corrugated lines are gathered together to give the corrugated surface.}
    \label{fig-explicationcorrugationprocess}
\end{figure}

We apply a 1D corrugation method on each curve $\phi_\lambda$ "transported" by the primitive embedding $f\circ\phi_\lambda$. Because 1D corrugation method can only increase the local length, it is important to work with strictly short primitive embedding. We summarize the 2D-corrugation process in Fig.(\ref{fig-explicationcorrugationprocess}). 

Let us introduce the \emph{loop function},
\begin{equation}\label{Eq-loopsdef1}
\begin{array}{ccccc}
\boldsymbol{h} & : & \mathcal{U}\times \R/\Z & \longrightarrow & \R^3 \\
 & & (s,\lambda,u) & \mapsto & r(s,\lambda)\boldsymbol{u}(s,\lambda,u)
\end{array}\, ,
\end{equation}
with, 
\begin{equation}\label{Eq-loopsdef2}
    \boldsymbol{u}(s,\lambda,u)=\cos\left(\beta(s,\lambda)\cos 2\pi u\right)\boldsymbol{t}(s,\lambda)+\sin\left(\beta(s,\lambda)\cos 2\pi u\right)\boldsymbol{n}(s,\lambda)\, ,
\end{equation}
and,
\begin{equation}
\fl{r(s,\lambda)=\sqrt{\boldsymbol{\mu}(\boldsymbol{e}_s,\boldsymbol{e}_s)}=\Omega(s,\lambda)\quad \boldsymbol{t}(s,\lambda)=\frac{\boldsymbol{\partial}_s f}{||\boldsymbol{\partial}_s f||}\quad \boldsymbol{n}(s,\lambda)=\frac{\boldsymbol{\partial}_s f\times \boldsymbol{\partial}_\lambda f}{||\boldsymbol{\partial}_s f\times \boldsymbol{\partial}_\lambda f||}}\, .
\end{equation}

The function $\boldsymbol{h}$ is called loop function because it constructs a vector of magnitude $r(s,\lambda)$, which oscillates around $\boldsymbol{\partial}_s f$ in function of $u$,
\begin{equation}\label{eq-average-loop}
    \int_0^1 \boldsymbol{h}(s,\lambda,u) \dd u = r(s,\lambda)J_0\left(\beta(s,\lambda)\right)\boldsymbol{t}(s,\lambda)=\boldsymbol{\partial}_s f \, ,
\end{equation}
where $J_0$ is the one order Bessel function. The last equality is verified if we choose the amplitude of the angular oscillation as,
\begin{equation}\label{Eq-amposcillation}
    \beta(s,\lambda)=\pm J_0^{-1}\left(\frac{||\boldsymbol{\partial}_s f||}{r(s,\lambda)}\right)=\pm J_0^{-1}\left(\frac{\tilde{\Omega}(s,\lambda)}{\Omega(s,\lambda)}\right)\,.
\end{equation}

\begin{figure}[ht]
    \centering
    \includegraphics[scale=0.5]{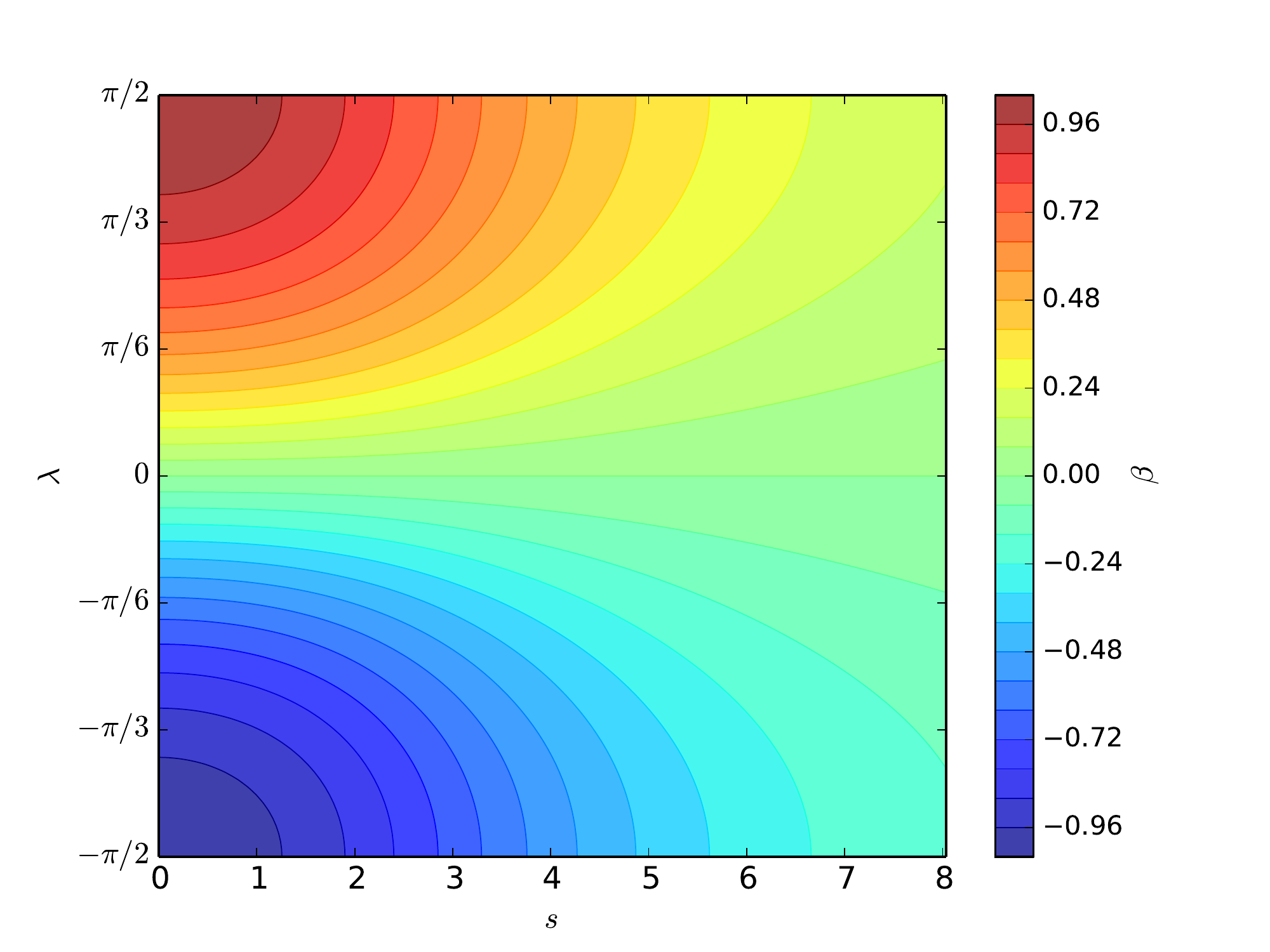}
    \caption{ Colormap of the amplitude of the angular oscillations $\beta$ for $a=0.99$.}
    \label{fig-beta-a=0.99}
\end{figure}

When $\lambda=0$ we have $\frac{\tilde{\Omega}}{\Omega}=1$. Because $J_0^{-1}$ is not differentiable in $1$, we loose $\mathcal{C}^1$ regularity on the line $\lambda=0$. Since we have $J_0^{-1}(1-\epsilon)\underset{\epsilon \rightarrow 0^+}{=}2\sqrt{\epsilon}\left(1+o(\epsilon)\right)$, {we obtain the following result,}
\begin{eqnarray*}\fl{
    J_0^{-1}\left(\frac{||\boldsymbol{\partial}_s f||}{r(s,\lambda)}\right)=J_0^{-1}\left(\sqrt{1-\frac{2a^2\sin^2\lambda}{r(r^2+a^2)\Omega^2}}\right)\underset{\lambda \rightarrow 0}{\sim} J_0^{-1}\left(1-\frac{a^2\sin^2\lambda}{r(r^2+a^2)\Omega^2}\right)\underset{\lambda\rightarrow 0}{\sim}2\frac{a|\sin\lambda|}{\sqrt{r(r^2+a^2)}\Omega}}\, . 
\end{eqnarray*}
That confirms the loss of $\mathcal{C}^1$ regularity of angular oscillations amplitude $\beta$ for $\lambda=0$. In order to get back {a} $\mathcal{C}^2$ regularity (and probably a $\mathcal{C}^\infty$ one), we choose the sign of $\beta$ so that we can remove the absolute value that appears in the previous equations. Then,
\begin{equation}\label{Eq-DefBeta}
\begin{array}{ccccc}
\beta & : & \mathcal{U} & \longrightarrow & \R^3 \\
 & & (s,\lambda\geq 0) & \mapsto &+ J_0^{-1}\left(\frac{\tilde{\Omega}(s,\lambda)}{\Omega(s,\lambda)}\right)\\
 & & (s,\lambda\leq 0) & \mapsto & -J_0^{-1}\left(\frac{\tilde{\Omega}(s,\lambda)}{\Omega(s,\lambda)}\right)\\
\end{array} \, ,
\end{equation}
which has $\mathcal{C}^2$ regularity. The amplitude $\beta$ of angular oscillations is represented  by a colormap in Fig.(\ref{fig-beta-a=0.99}). This function is $\lambda$-odd, and reaches its maximum for $(s,\lambda) = (0,\pi/2)$. We also have $\beta \rightarrow 0$ when $s\rightarrow +\infty$. \\

Let us introduce the corrugated embedding,
\begin{equation}\label{eq-defcorrugatedembedding}
    F_\nu(s,\lambda)=f(\lambda,0)+\int_0^s \boldsymbol{h}(u,\lambda,\nu u)\dd u  \, .
\end{equation}
This embedding verifies,
\begin{equation}
    \boldsymbol{\partial}_s F_\nu(s,\lambda)=\boldsymbol{h}(s,\lambda,\nu s) \, ,
\end{equation}
and has the expected length,
\begin{equation}
    ||\boldsymbol{\partial}_s F_\nu(s,\lambda)||^2=F_\nu^\star<\boldsymbol{e}_s,\boldsymbol{e}_s>_{\R^3}=r(s,\lambda)=\boldsymbol{\mu}(\boldsymbol{e}_s,\boldsymbol{e}_s) \, .
\end{equation}
In the following subsection, we will discuss the application of the $\mathcal{C}^0$-density Lemma. In our case, this provides,
\begin{equation}\label{Eq-C0Lemma}
\begin{array}{c}
    F_\nu(s,\lambda)\underset{\nu\rightarrow\infty}{=}f(s,\lambda)+O(1/\nu) \, ,\\
    \boldsymbol{\partial}_\lambda F_\nu(s,\lambda) \underset{\nu\rightarrow\infty}{=}\boldsymbol{\partial}_\lambda f(s,\lambda)+O(1/\nu) \, ,
    \end{array}    
\end{equation}
which implies that $F_\nu$ verifies both required properties, $\mathcal{C}^0$-close to $f$ and as close as we want to the isometric embedding. Thus we have,
\begin{equation}\label{Eq-C0Lemma-1}
\begin{array}{c}
    ||\boldsymbol{\mu}-F_\nu^\star<\cdot,\cdot>_{\R^3}||\underset{\nu\rightarrow\infty}{=}O(1/\nu)
    \end{array}.    
\end{equation}

Before discussing the relations in Eq.(\ref{Eq-C0Lemma}), we justify why we obtain these properties. When $\nu$ becomes large, the local average of the curve $s \mapsto \boldsymbol{h}(s,\lambda,\nu s)$ needs to be ``close to'' $\boldsymbol{\partial}_s f$ (see also Eq.\ref{eq-average-loop}) which explains why $s\mapsto F_\nu(s,\lambda)$ is getting closer to $s\mapsto f(s,\lambda)$ as $\nu$ is increasing.

\subsubsection{\texorpdfstring{$\mathcal{C}^0$}{C0}-density Lemma extension}\leavevmode\par  

In our case, we have two main differences in comparison to the $\mathcal{C}^0$-density Lemma proof presented in \cite{borrelli:hal-00945774}. 

The first one concerns the embedding definition domain $\mathcal{U}$. $\mathcal{U}$ is not compact, unlike $\R/\Z\times\R/\Z$. This difference is important because in different parts of $\mathcal{C}^0$-density Lemma proof, the authors use compacity in the majorization process. Nevertheless, if we look carefully at all the functions introduced in this section, they can be extended smoothly in $\lambda=\pi/2$, $\lambda=-\pi/2$ and $s=0$. Furthermore, we limit our study to a subset of $\mathcal{U}$ by setting an upper limit $s_{\rm max}$ for $s$ coordinate and this limitation will happen in every numerical computations. So we will replace $\mathcal{U}$ by a compact set and we will define it $\mathcal{U}_{s_{\rm max}}=[0,s_{\rm max}]\times[-\pi/2,\pi/2]$. 

The second one concerns the value of $\alpha(\lambda,s)$. Indeed, even if $\alpha$ is positive everywhere in our case, this function reaches $0$ for $\lambda=0$. Nevertheless, by examining carefully the demonstration made in \cite{borrelli:hal-00945774}, it can be easily extended to prove Eq.(\ref{Eq-C0Lemma}) and Eq.(\ref{Eq-C0Lemma-1}), if and only if the functions $\boldsymbol{h}$, $\boldsymbol{\partial}_s\boldsymbol{h}$,  $\boldsymbol{\partial}_\lambda\boldsymbol{h}$ and $\boldsymbol{\partial}^2_{\lambda,s}\boldsymbol{h}$ are $\mathcal{C}^0$ and have a maximum on $\mathcal{U}_{s_{\rm max}}\times \R/\Z$. 

In \cite{borrelli:hal-00945774} these functions are majorized, because the departure set is compact and the functions and their derivatives are continuous. The regularity issue that could arise from the definition of $\beta$ was overcome by the judicious choice of the sign inversion for $\lambda=0$ in Eq.(\ref{Eq-DefBeta}). Thus the functions $\boldsymbol{h}$, $\boldsymbol{\partial}_s\boldsymbol{h}$,  $\boldsymbol{\partial}_\lambda\boldsymbol{h}$ and $\boldsymbol{\partial}^2_{\lambda,s}\boldsymbol{h}$ are continuous and 
the $\mathcal{C}^0$-density Lemma is valid on every compact subset of $\mathcal{U}\times \R/\Z$.

In order to get a reasonable computational time, we are working on the compact subset $\mathcal{U}_{s_{\rm max}}$ of poloidal manifold with finite radius $s<s_{\rm max}$. The convergence of the corrugation process is obtained thanks to the extension of the $\mathcal{C}^0$-density Lemma (Eq.(\ref{Eq-C0Lemma}) and Eq.(\ref{Eq-C0Lemma-1})). 

The maximisation of $\boldsymbol{h}$, $\boldsymbol{\partial}_s\boldsymbol{h}$,  $\boldsymbol{\partial}_\lambda\boldsymbol{h}$ and $\boldsymbol{\partial}^2_{\lambda,s}\boldsymbol{h}$ on the entire $\mathcal{U}\times \R/\Z$ set may be proved. This proof would come from the extremely smooth behaviour of the primitive embedding and all the different functions in the vicinity of infinity. We will not further discuss this point because it is not necessary for our present work.

In our coordinates $\boldsymbol{\mu}$ is diagonal and rest metric is proportional to $\dd \boldsymbol{s}\otimes\dd \boldsymbol{s}$, then there is no need to determine the family of curves by integration as in Eq.(2.6) of \cite{borrelli:hal-00945774}. In fact, this aspect allows us to use the simple construction presented in paragraph 2.3.1 of \cite{borrelli:hal-00945774}.


        \subsection{Results}
    
 The corrugated embedding $F_\nu$ is built from the strictly short primitive embedding $\rho$. Table (\ref{tab-CorrugResult}) gives the main characteristic of the computations for $a=0.99$ in order to present results where the corrugation is the most visible. We choose $s_{\rm max}=s(r_{\rm max})$ with $r_{\rm max}=5$ and we estimate $F_\nu$ on a $N_s\times N_\lambda$ regular grid. We use $N_\lambda=21$ and adapted values for $N_s$.

 \begin{table}[ht]
 \begin{center}
 \begin{tabular}{ c c c}
 \hline\hline
   $\nu$  & $N_s$ & \quad$||\boldsymbol{\mu}-F_\nu^\star<\cdot,\cdot>_{\R^3}||_\infty$\\ \hline\hline
    $1$ & $80$   & $<0,5\;\;\;$ \\
    $3$  & $241$ & $<0,135$\\
    $5$  & $401$ & $<0,081$ \\
    $8$  & $642$ & $<0,065$\\
 \end{tabular}    
 \end{center}
 \caption{Number of grid points $N_s$ and residual $||\boldsymbol{\mu}-F_\nu^\star<\cdot,\cdot>_{\R^3}||_\infty$ versus $\nu$ for the corrugated surface}
 \label{tab-CorrugResult}
 \end{table}
 
The computational time increases with frequency because in Eq.(\ref{eq-defcorrugatedembedding}), the function in the integral is oscillating. In order to obtain the embedding, the number of points in $s$ must increase with the frequency (we choose $N_s=\left\lfloor10 \, \nu s_{\rm max} \right\rfloor$). As expected we get a decrease in $1/\nu$ of the isometric default
 
\hspace{-2cm}\begin{figure}[ht]
    \centering
    \begin{minipage}[c]{.46\linewidth}
      \includegraphics[scale=0.55, trim=2cm 2cm 4cm 1cm]{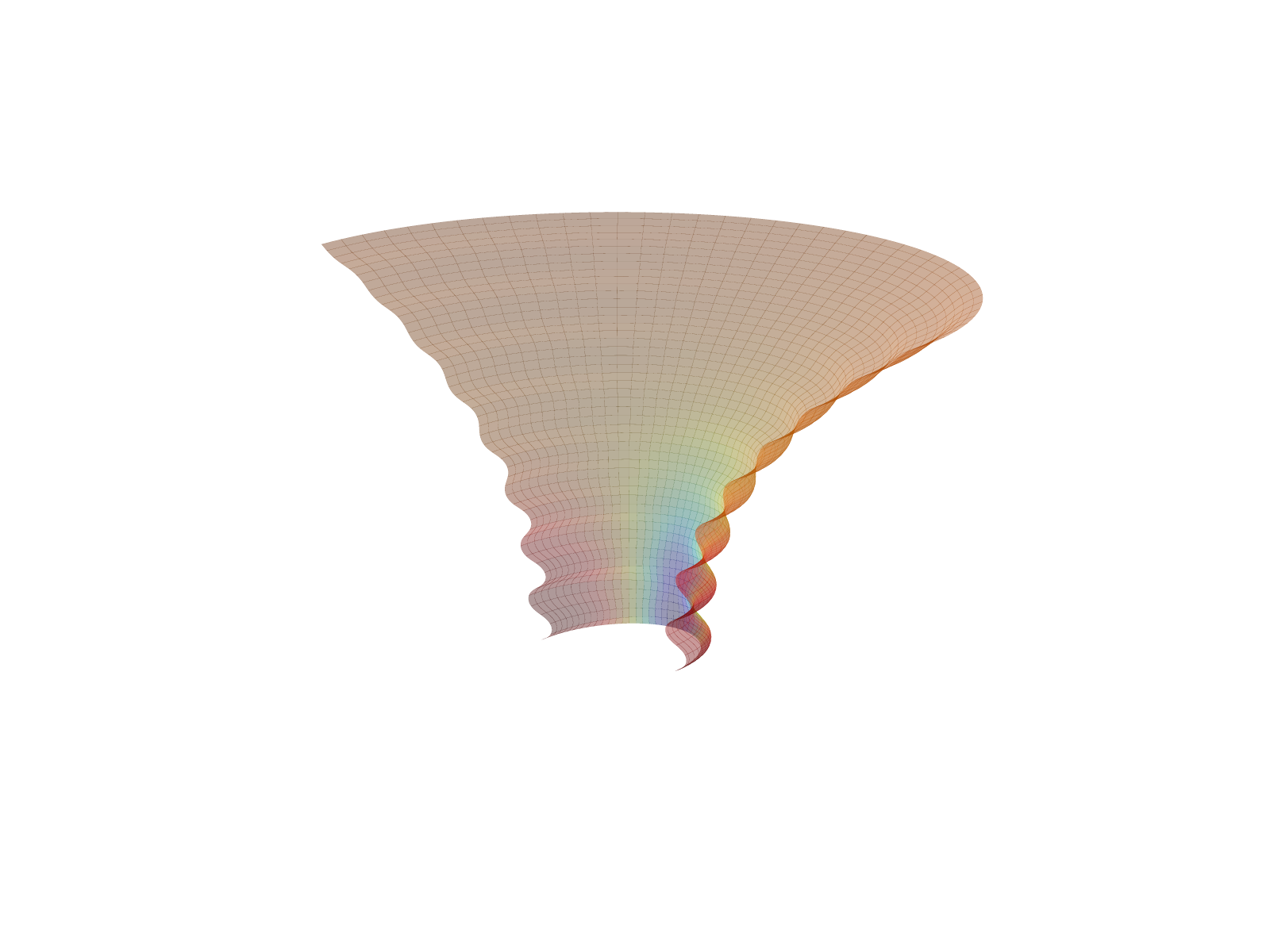}\\
      \includegraphics[scale=0.55, trim=2cm 2cm 4cm 1cm]{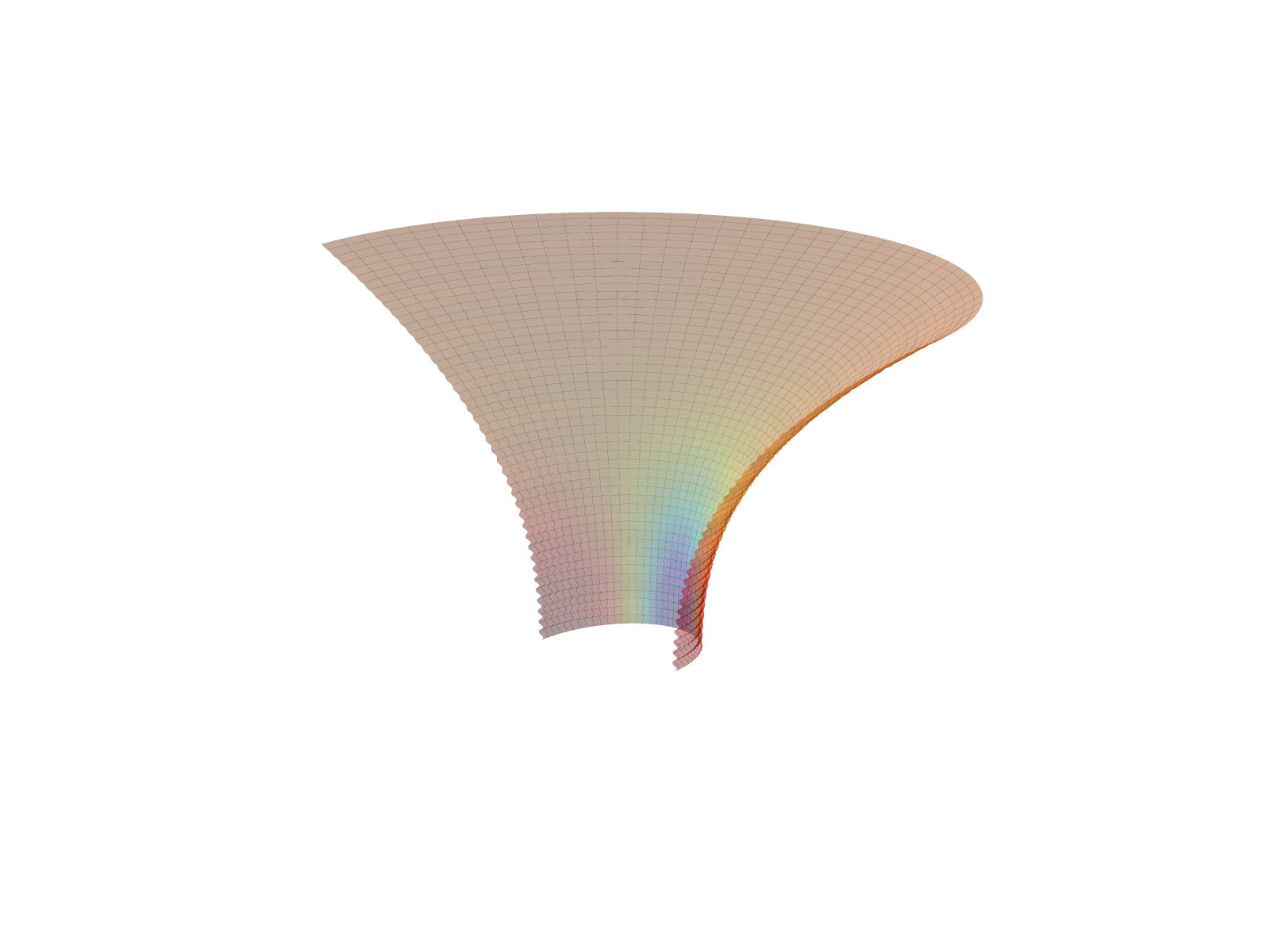}
   \end{minipage} \hfill
   \begin{minipage}[c]{.46\linewidth}
      \includegraphics[scale=0.55, trim=2cm 2cm 2cm 1cm]{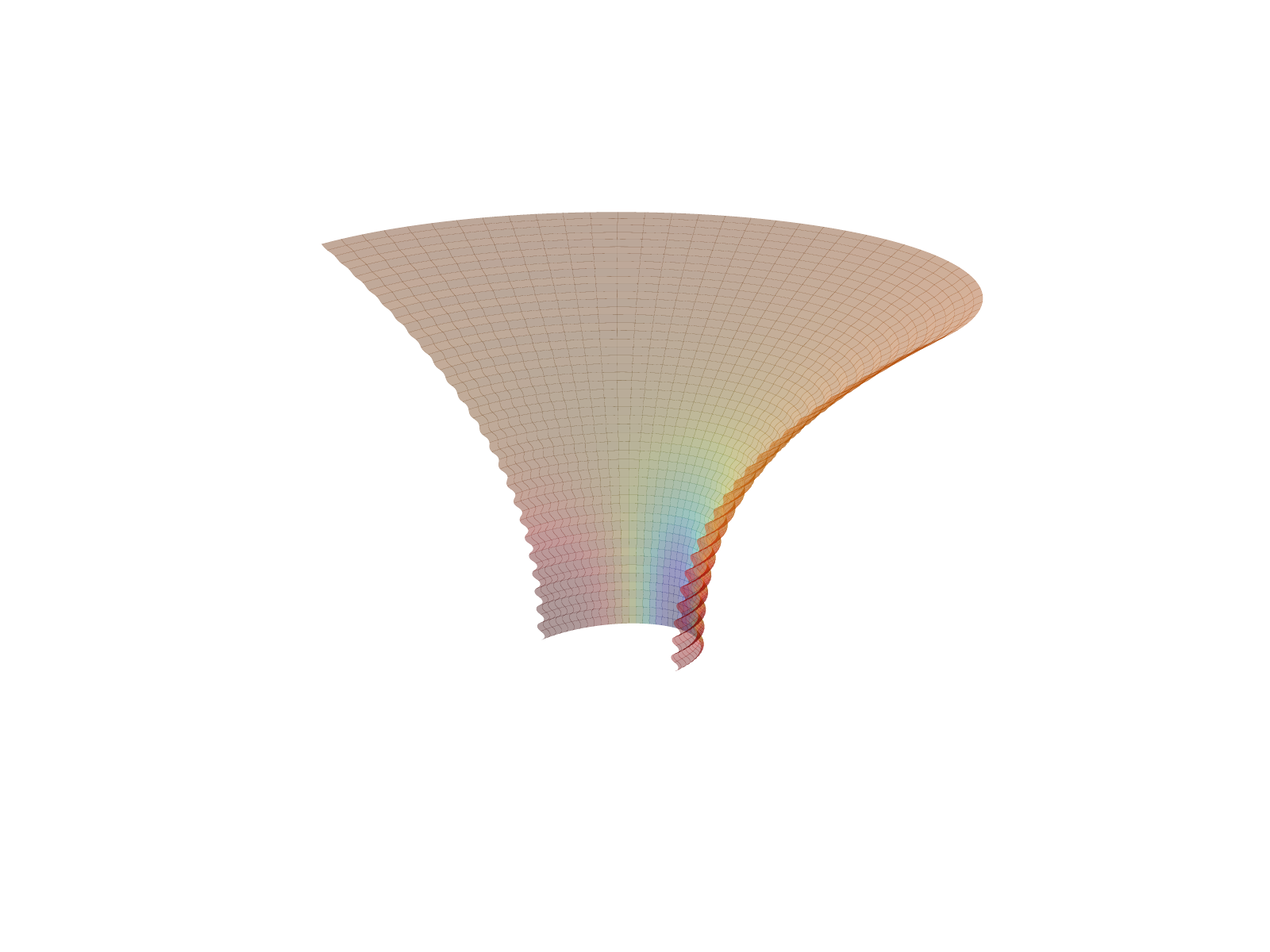}\\
      \includegraphics[scale=0.55, trim=2cm 2cm 2cm 1cm]{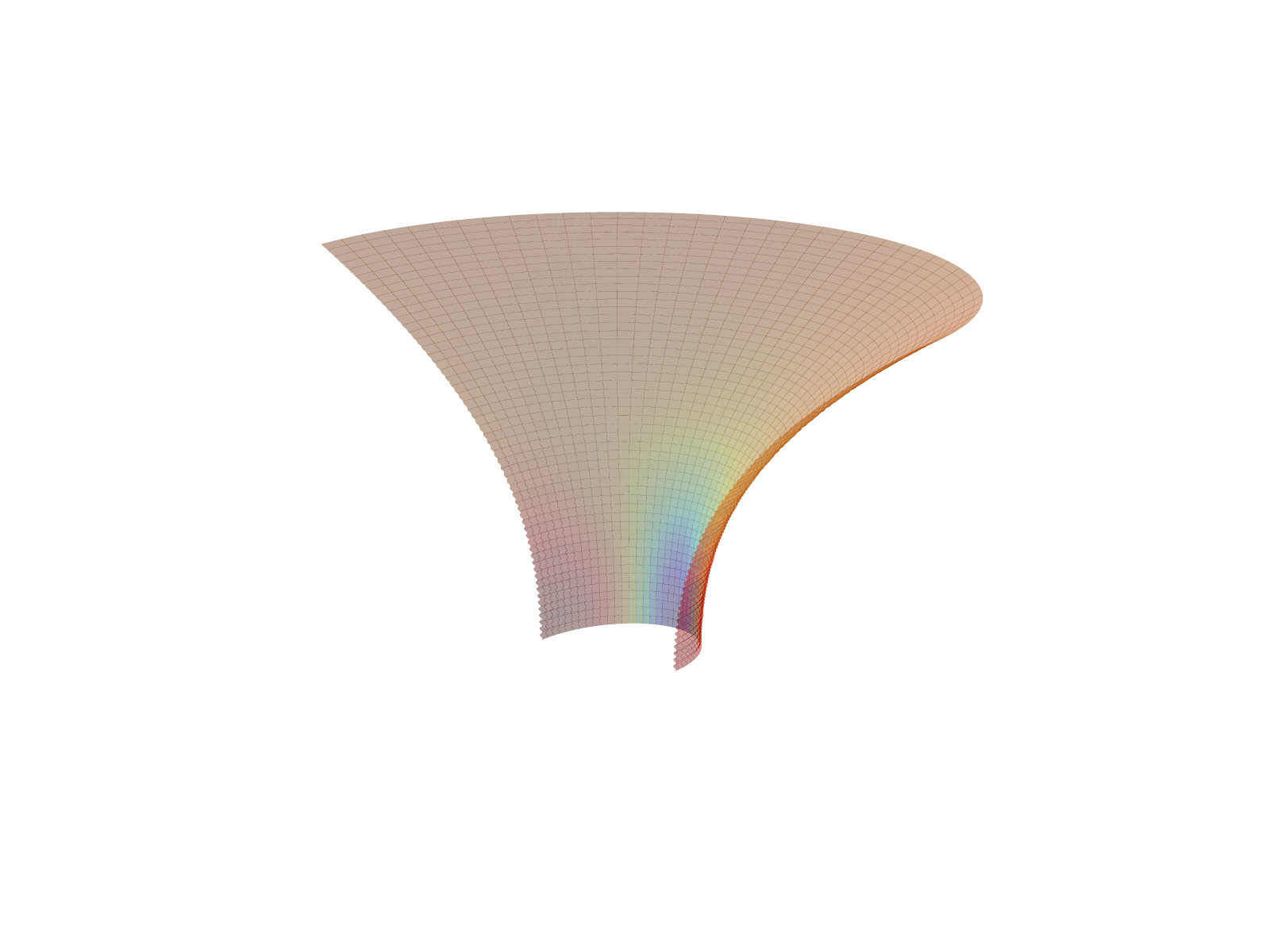}
   \end{minipage}
    \caption{Evolution of the corrugated embedding in respect of the corrugation frequency. Top-left, $\nu=1$. Top-right, $\nu=3$. Bottom-left, $\nu=5$. Bottom-right, $\nu=8$. The colors indicate the value of Gaussian curvature $K$ of the target metric. The red color is for the larger value of $K$. }
    \label{fig-CorrugatedEmbeddings}
\end{figure}

We draw in Fig.(\ref{fig-CorrugatedEmbeddings}) the corrugated embedding for different frequencies. One should notice that the odd property of $\lambda \mapsto \beta(s,\lambda)$ implies a phase opposition between the northern and the southern part of the corrugated embedding. If we note $\boldsymbol{\mathcal{S}}$ the symmetry with respect to the plane $O_{x,z}$, corresponding to $\lambda\leftrightarrow -\lambda$ on the primitive embedding, then the phase opposition of the corrugation is visible on the vector
$\boldsymbol{\partial}_s F_\nu$,
\begin{eqnarray*}
    \boldsymbol{\partial}_s F_\nu (s,-\lambda)=\boldsymbol{h}(s,-\lambda,\nu s)
    =\boldsymbol{\mathcal{S}}\boldsymbol{h}(s,\lambda,\nu s+1) \, .
\end{eqnarray*}

We observe the effect of $\mathcal{C}^0$-density Lemma : the corrugated surface is getting closer to the primitive embedding Eq.(\ref{Eq-C0Lemma}). Nevertheless the high frequency and the small amplitude of the oscillations make this embedding difficult to use for physicists who aim to get "smooth" isometric representations.

\begin{figure}[ht]
    \centering
    \hspace{0cm}\begin{minipage}[c]{.46\linewidth}
      \includegraphics[scale=0.4, trim=0cm 0cm 0cm 0cm]{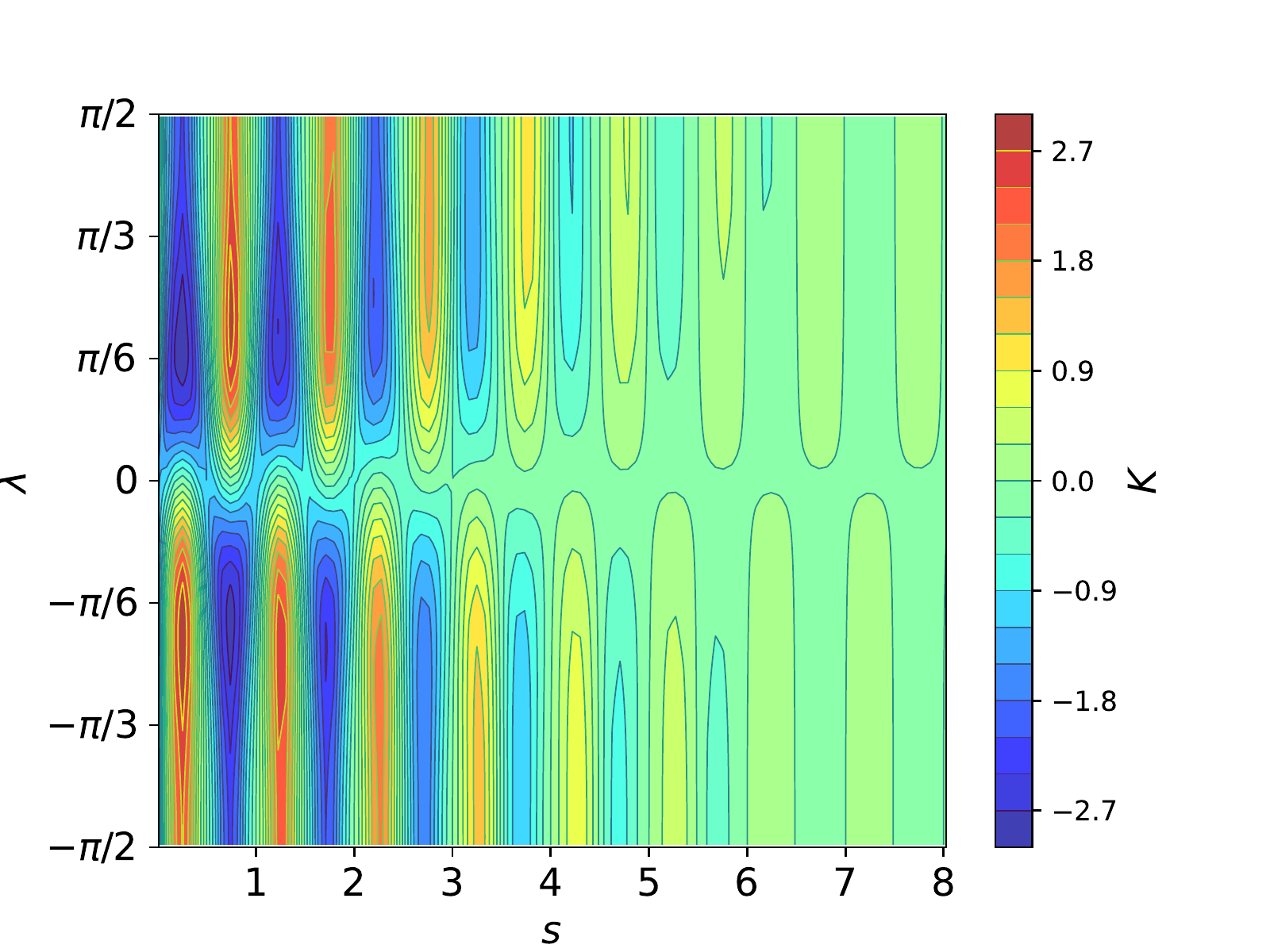}\\
      \includegraphics[scale=0.4, trim=0cm 0cm 0cm 0cm]{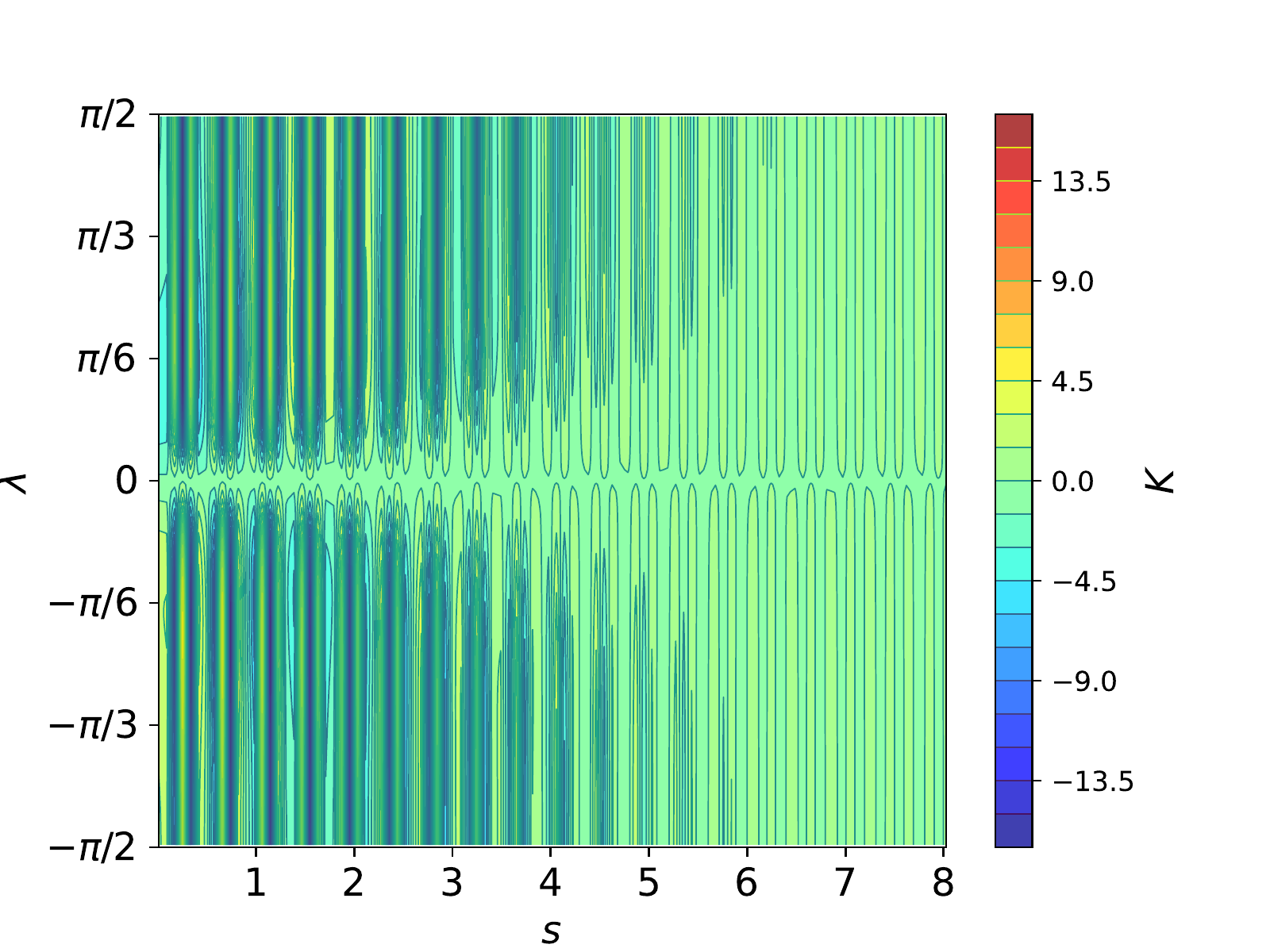}
   \end{minipage}\hfill\begin{minipage}[c]{.46\linewidth}
      \includegraphics[scale=0.4, trim=0cm 0cm 0cm 0cm]{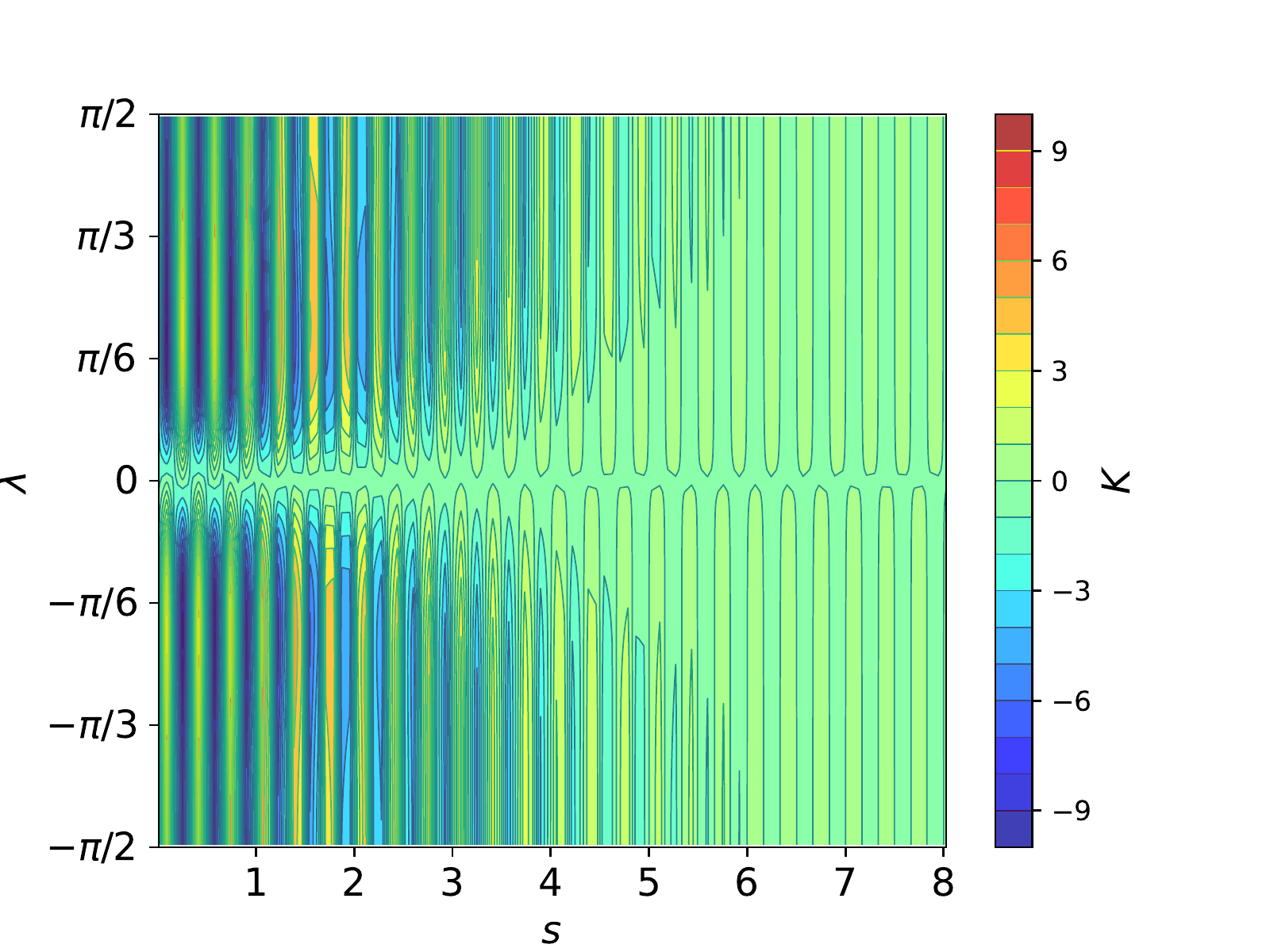}\\
      \includegraphics[scale=0.4, trim=0cm 0cm 0cm 0cm]{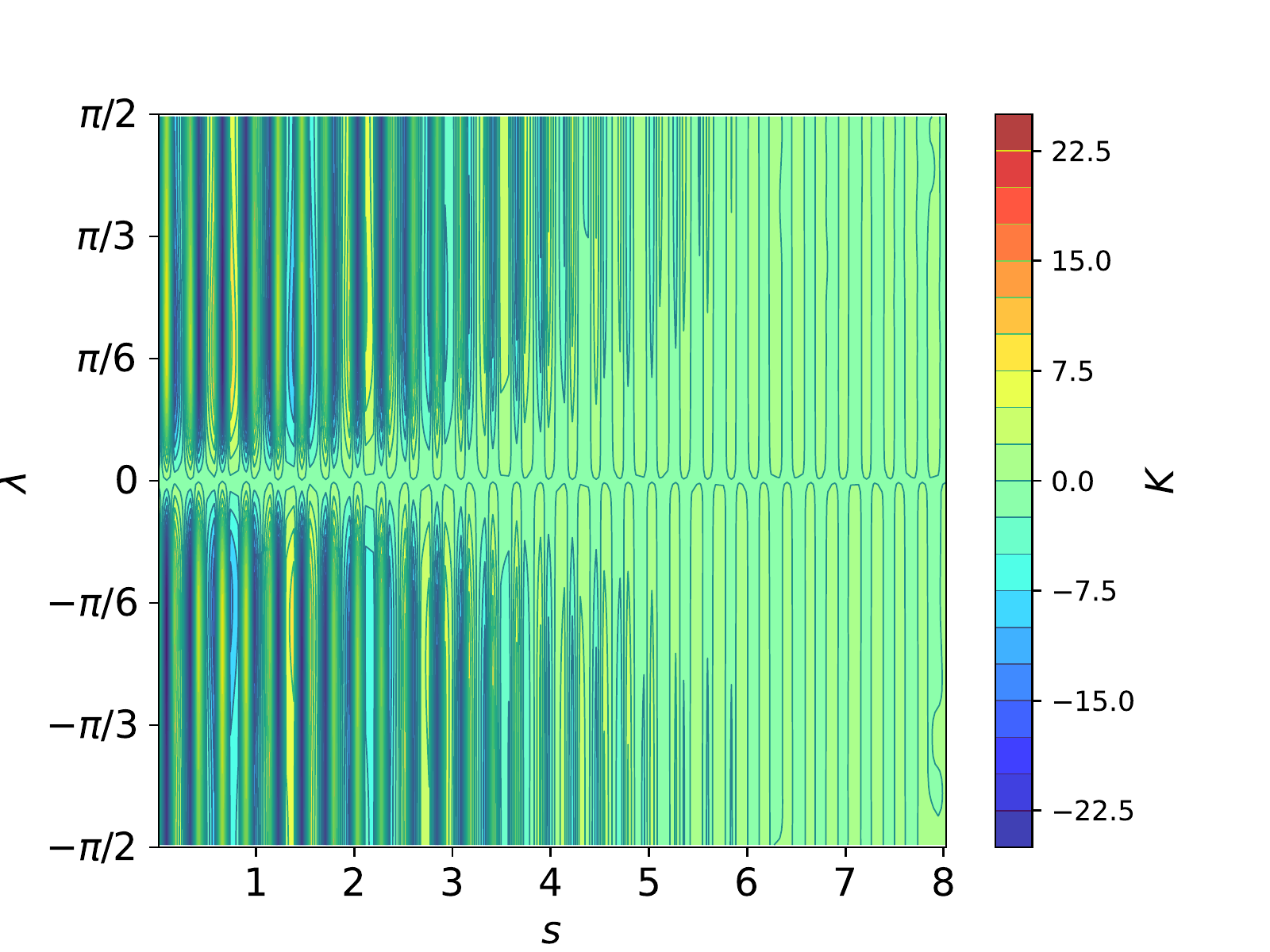}
   \end{minipage}
    \caption{Evolution of Gaussian curvature $K_\nu$ of corrugated embedding with the corrugation frequency. Top-left, $\nu=1$. Top-right, $\nu=3$. Bottom-left, $\nu=5$. Bottom-right, $\nu=8$.  }
    \label{fig-curvature}
\end{figure}

 The Fig.(\ref{fig-curvature}) shows the evolution of the Gaussian curvature of the corrugated surface for $a=0.99$ and for different frequencies $\nu$. The Gaussian curvature of the corrugated embedding $K_\nu$ does not converge to the one of the poloidal submanifold $K$, and takes higher values as we increase the frequency ($||K_\nu||_\infty\propto \nu$). This is one of the most important properties about corrugation process. Indeed, even if we have the results Eq.(\ref{Eq-C0Lemma}) and Eq.(\ref{Eq-C0Lemma-1}), this does not imply the convergence of the Gaussian curvature $K_\nu$ to the one of the target metric $K$.
 
In general relativity, the Gaussian curvature needs to be well defined, which is problematic with the use of corrugation embedding. Mathematicians use this method to construct a $\mathcal{C}^\alpha$ isometric embedding with $\alpha \in ]\,1,2\,[$, i.e. this embedding (for the flat torus see \cite{borrelli:hal-00945774}) has a tangent plane at any point but not necessarily a second fundamental form and, a fortiori, a Gauss curvature.

 
 Using corrugation, we can obtain an embedding defined on the whole $\mathcal{U}$ as close as we want to the isometric one. Another possibility is to apply corrugation from $\Lambda$ Eq.(\ref{Eq-Immersion-theta}) but we decided to use an alternative method with greater regularity. This is the object of the next section.
 
 





\section{Gauss-Coddazzi-Mainardi system and frame equations}\label{sec4}

To bypass the corrugation process from primitive embedding, an alternative is to solve both Gauss-Coddazzi-Mainardi (GCM) and frame equations. Indeed, the local existence of an $\mathcal{C}^3$-isometric embedding is ensured by the existence of a local solution of GCM equations (\cite{han2006isometric}), as well as the frame equations (see \cite{palais} and \cite{spivak1975comprehensive}).
In the following, we will present a numerical resolution of GCM equations which can be transformed in a quasi-linear partial derivative system. We subsequently solve the frame equations (Frobenius partial derivative type).
Boundary conditions for the resolution of the GCM equations are deduced from symmetric considerations. The latter are also used to deduce the initial conditions of the frame equations.

    \subsection{Frame application, first and second fundamental forms}
  
Let us consider a $\mathcal{C}^3$ isometric embbeding $f:\mathcal{U}\longrightarrow \Sigma\subset\R^3$. The frame equations determine the variations of $\boldsymbol{\partial}_s f$ and $  \boldsymbol{\partial}_\lambda f$ in function of first and second fundamental forms. We shall introduce the frame application,
\begin{equation}\label{Eq-Frame}
\begin{array}{ccccc}
\boldsymbol{\mathcal{R}} & : & \mathcal{U} & \longrightarrow & \mathfrak{M}_3\left(\R\right) \\
 & & (s,\lambda) & \mapsto & \left(\boldsymbol{\partial}_s f, \boldsymbol{\partial}_\lambda f, \boldsymbol{n}\right) \,,
\end{array}
\end{equation}
where $\boldsymbol{n}$ is the usual Gauss application,
\begin{equation}\label{Eq-GaussAppli}
\begin{array}{ccccc}
\boldsymbol{n} & : & \mathcal{U} & \longrightarrow & \R^3 \\
 & & (s,\lambda) & \mapsto & \displaystyle\frac{\boldsymbol{\partial}_s f\times\boldsymbol{\partial}_\lambda f}{||\boldsymbol{\partial}_s f\times\boldsymbol{\partial}_\lambda f||} \,.
\end{array}
\end{equation}
We also introduce the usual functions for the first fundamental form,
\begin{equation}\label{Eq-defpremform-2}\left\{
\begin{array}{cc}
    E(s,\lambda)=&<\boldsymbol{\partial_s}f,\boldsymbol{\partial_s}f>\\
    F(s,\lambda)=&<\boldsymbol{\partial_s}f,\boldsymbol{\partial_\lambda}f>\\
    G(s,\lambda)=&<\boldsymbol{\partial_\lambda}f,\boldsymbol{\partial_\lambda}f>\\
\end{array}\right.    \,,
\end{equation} 
and $\boldsymbol{\mathcal{G}}$ the Gram matrix associated to $\boldsymbol{\mathcal{R}}$,
\begin{equation}
    \boldsymbol{\mathcal{G}}(s,\lambda)=^t\boldsymbol{\mathcal{R}}(s,\lambda) \boldsymbol{\mathcal{R}}(s,\lambda)=
    \Matrix{E(s,\lambda)}{F(s,\lambda)}{0}{F(s,\lambda)}{G(s,\lambda)}{0}{0}{0}{1} \,.
\end{equation}
The second fundamental form $\boldsymbol{h}$ is written as,
\begin{equation}\label{Eq-defsecondform-1}
    \boldsymbol{h}=L(s,\lambda)\dd\boldsymbol{s}\otimes\dd\boldsymbol{s}+M(s,\lambda)\left[\dd\boldsymbol{s}\otimes\dd\boldsymbol{\lambda}+\dd\boldsymbol{\lambda}\otimes\dd\boldsymbol{s}\right]+N(s,\lambda)\dd\boldsymbol{\lambda}\otimes\dd\boldsymbol{\lambda} 
\end{equation}
with 
\begin{equation}\label{Eq-defsecondform-2}\hspace{-0.8cm}\left\{
\begin{array}{cccc}
    L(s,\lambda)&=<\boldsymbol{n},\boldsymbol{\partial^2_{ss}}f>&=-<{\partial_{s}}\boldsymbol{n},\boldsymbol{\partial_{s}}f>\phantom{=-<{\partial_{s}}\boldsymbol{n},\boldsymbol{\partial_{\lambda}}f>}\\
    M(s,\lambda) &=<\boldsymbol{n},\boldsymbol{\partial^2_{s\lambda}}f>&=-<{\partial_{\lambda}}\boldsymbol{n},\boldsymbol{\partial_{s}}f>=-<{\partial_{s}}\boldsymbol{n},\boldsymbol{\partial_{\lambda}}f>\\
    N(s,\lambda)&=<\boldsymbol{n},\boldsymbol{\partial^2_{\lambda\lambda}}f>&=-<{\partial_{\lambda}}\boldsymbol{n},\boldsymbol{\partial_{\lambda}}f>\phantom{=-<{\partial_{s}}\boldsymbol{n},\boldsymbol{\partial_{\lambda}}f>}\\
\end{array}\right.    \,.
\end{equation}

    \subsection{General frame and GCM equations}

For each $p\in\mathcal{U}$, $\det\boldsymbol{\mathcal{R}}=||\boldsymbol{\partial}_s f\times\boldsymbol{\partial}_\lambda f||\neq 0$ and $(\boldsymbol{\partial}_s f,\boldsymbol{\partial}_\lambda f ,\boldsymbol{n})$ is a base of $\R^3$. Every vector can be decomposed with the help of frame application and usual calculations gives,
\begin{equation}\label{Eq-FrameEquation-1}\left\{
\begin{array}{cccc}
   \partial_s\boldsymbol{\mathcal{R}}=\boldsymbol{\mathcal{R}}\boldsymbol{\mathcal{K}}_s=\boldsymbol{\mathcal{R}}\boldsymbol{\mathcal{G}}^{-1}\boldsymbol{\mathcal{F}}_s\\
    \partial_\lambda\boldsymbol{\mathcal{R}}=\boldsymbol{\mathcal{R}}\boldsymbol{\mathcal{K}}_\lambda=\boldsymbol{\mathcal{R}}\boldsymbol{\mathcal{G}}^{-1}\boldsymbol{\mathcal{F}}_\lambda\\
\end{array}\right.    \,,
\end{equation}
with,
\begin{equation}\label{Eq-FrameEquation-2}
   \boldsymbol{\mathcal{K}}_s=\Matrix{\Gamma_{ss}^s}{\Gamma_{s\lambda}^s}{a_s^s}{\Gamma_{ss}^\lambda}{\Gamma_{s\lambda}^\lambda}{a^\lambda_s}{L}{M}{0} \,,\qquad \boldsymbol{\mathcal{K}}_\lambda=\Matrix{\Gamma_{s\lambda}^s}{\Gamma_{\lambda\lambda}^s}{a_\lambda^s}{\Gamma_{s\lambda}^\lambda }{\Gamma_{\lambda\lambda}^\lambda}{a^\lambda_\lambda}{M}{N}{0} \,,
\end{equation}where we used the notation $\partial_i\boldsymbol{n}=a^j_i \boldsymbol{\partial}_j$ and the Christoffel symbols $\Gamma_{ij}^k$. The computation of the  $a^j_i$  coefficients requires the use of the inverse matrix $\boldsymbol{\mathcal{G}}^{-1}$. The matrices  $\boldsymbol{\mathcal{F}}_s$ and $\boldsymbol{\mathcal{F}}_\lambda$ are written from the Christoffel symbols and the second fundamental form as,
\begin{equation}\label{Eq-FrameEquation-3}\fl{
   \boldsymbol{\mathcal{F}}_s=\Matrix{\frac{1}{2}\partial_s E}{\frac{1}{2}\partial_\lambda E}{-L}{\partial_s F-\frac{1}{2}\partial_\lambda E}{\frac{1}{2}\partial_s G}{-M}{L}{M}{0} \, , \qquad \boldsymbol{\mathcal{F}}_\lambda=\Matrix{\frac{1}{2}\partial_\lambda E}{\partial_\lambda F-\frac{1}{2}\partial_s G}{-M}{\frac{1}{2}\partial_s G }{\frac{1}{2}\partial_\lambda G}{-N}{M}{N}{0}} \, . 
\end{equation}
The system of equations Eqs.(\ref{Eq-FrameEquation-1}) is know as \emph{frame equations}. We suppose here that $f$ is $\mathcal{C}^3$, then $\boldsymbol{\mathcal{R}}$ is $\mathcal{C}^2$, and the Schwarz's theorem conditions hold and lead straightforwardly to,
\begin{equation}\label{Eq-GCM-1}
    \partial_s \boldsymbol{\mathcal{K}}_\lambda-\partial_\lambda \boldsymbol{\mathcal{K}}_s +\left[ \boldsymbol{\mathcal{K}}_s; \boldsymbol{\mathcal{K}}_\lambda\right] = \boldsymbol{0} \, . 
\end{equation}
For the first line and the second column in Eq.(\ref{Eq-GCM-1}) we get the \emph{Theorema Egregium} that is an expression of the Gaussian curvature in function of the first fundamental form coefficients and their derivatives. For the last line and the last column we obtain two equations usually called \emph{Mainardi-Coddazzi} equations,
\begin{equation}\label{Eq-CM-1}\fl{\left\{
\begin{array}{cccc}
    LN-M^2&=&\tilde{K}\quad\hat{=}\quad K \left(EG-F^2\right)\qquad&{\rm{Gauss}}\\
    \partial_\lambda L-\partial_s M &=&\Gamma_{s\lambda}^s L+\left(\Gamma_{s\lambda}^\lambda-\Gamma_{ss}^s\right)M-\Gamma_{ss}^\lambda N &{\rm{Coddazzi-Mainardi-1}}  \\
    \partial_\lambda M-\partial_s N &=&\Gamma_{\lambda\lambda}^s L+\left(\Gamma_{\lambda\lambda}^\lambda-\Gamma_{\lambda s}^s\right)M-\Gamma_{\lambda s}^\lambda N &{\rm{Coddazzi-Mainardi-2}}  
\end{array}\right.  }  \,.
\end{equation}
We refer to the system (\ref{Eq-CM-1}) as the GCM system.

In the context of Kerr poloidal manifold described by $(s,\lambda)$ coordinates, frame equations matrices are,
\begin{equation}\label{Eq-FrameEquation-Poloidale-1}
   \boldsymbol{\mathcal{K}}_s(s,\lambda)= \Matrix{-\frac{a^2\sqrt{\Delta}\sin^2\lambda}{r^2(r^2+a^2\sin^2\lambda)}}{\frac{a^2\cos\lambda\sin\lambda}{r^2+a^2\sin^2\lambda}}{-\frac{Lr^2}{r^2+a^2\sin^2\lambda}}{-\frac{a^2\cos\lambda\sin\lambda}{r^2(r^2+a^2\sin^2\lambda)}}{\frac{\sqrt{\Delta}}{r^2+a^2\sin^2\lambda}}{-\frac{M}{r^2+a^2\sin^2\lambda}}{L}{M}{0} \, , 
\end{equation}
\begin{equation}\label{Eq-FrameEquation-Poloidale-2}
\boldsymbol{\mathcal{K}}_\lambda(s,\lambda)=\Matrix{\frac{a^2\cos\lambda\sin\lambda}{r^2+a^2\sin^2\lambda}}{-\frac{r^2\sqrt{\Delta}}{r^2+a^2\sin^2\lambda}}{-\frac{r^2 \, M}{r^2+a^2\sin^2\lambda}}{\frac{\sqrt{\Delta}}{r^2+a^2\sin^2\lambda} }{\frac{a^2\cos\lambda\sin\lambda}{r^2+a^2\sin^2\lambda}}{-\frac{N}{r^2+a^2\sin^2\lambda}}{M}{N}{0} \,.
\end{equation}

We remark that if $M(s,\lambda=0)=0$ we have
\mbox{$(\boldsymbol{\partial_s}(s,0),\boldsymbol{n}(s,0))\in\R\boldsymbol{\partial_s}(0,0)\oplus\R\boldsymbol{n}(0,0)$}. Indeed for $\lambda=0$ we get,
\begin{equation}\label{Eq-frame-l=0}\fl{\left\{
\begin{array}{cc}
    \partial_s \boldsymbol{\partial}_s=&L\boldsymbol{n}\\
    \partial_s \boldsymbol{n}=&-L\boldsymbol{\partial}_s\\
    \partial_s \boldsymbol{\partial}_\lambda=&\frac{\sqrt{\Delta}}{r^2} \boldsymbol{\partial}_\lambda\\
\end{array}\right.    \, \quad\Rightarrow\quad  \left\{
\begin{array}{cc}
    \boldsymbol{\partial}_s(s,0)=&\cos \psi(s)\boldsymbol{\partial}_s(0,0)+\sin \psi(s) \boldsymbol{n}(0,0) \\
    \boldsymbol{n}(s,0)=&-\sin \psi(s)\boldsymbol{\partial}_s(0,0)+\cos \psi(s) \boldsymbol{n}(0,0)\\
\end{array}\right. } \, , 
\end{equation}
with,
\begin{equation}
\psi(s)=\int_0^s L(s,\lambda=0){\rm d}s  \, .
\end{equation}
This proves that $\boldsymbol{\partial_s}(s,0),\boldsymbol{n}(s,0)\in\R\boldsymbol{\partial_s}(0,0)\oplus\R\boldsymbol{n}(0,0)$, meaning that the $s\mapsto f(s,0)$ curve is contained in the $f(0,0)+\R\boldsymbol{\partial_s}(0,0)\oplus\R\boldsymbol{n}(0,0)$ plane.

To solve the frame equations we need the second fundamental form. This could be achieved by a numerical integration of the GCM equations. 

    \subsection{Quasi-linear form of GCM equations}

Under some assumptions (see \ref{ap-derivGCMform}), Gauss-Codazzi-Mainardi equations Eq.(\ref{Eq-GCM-1}) could be written as,
\begin{equation}\label{Eq-uvGCM-SystemQuasiLinear-General2}
    \partial_\lambda\boldsymbol{U}+\boldsymbol{A}\left(s,\lambda,\boldsymbol{U}\right)\partial_s \boldsymbol{U}=\boldsymbol{S}\left(s,\lambda,\boldsymbol{U}\right)\,, 
\end{equation}
where $\boldsymbol{U}$ is linked to the second fundamental form with, 
\begin{equation}\label{Eq-uvGCM-BF3}\hspace{-1.5cm}
\begin{array}{ccc}
    L(s,\lambda)=\frac{\tilde{K}(s,\lambda)+u^2(s,\lambda)}{\sqrt{2r(s)}v(s,\lambda)},& M(s,\lambda)=u(s,\lambda),& N(s,\lambda)=\sqrt{2r(s)}v(s,\lambda)
\end{array}\,.
\end{equation}
The matrix $\boldsymbol{A}\left(s,\lambda,\boldsymbol{U}\right)\in\mathfrak{M}_2\left(\R\right)$ and the source term $\boldsymbol{S}\left(s,\lambda,\boldsymbol{U}\right)$ is a column of $\R^2$. Their explicit expressions are given in the Appendix C (Eq.(\ref{eq-matrix-GCM-reduced}) and Eq.(\ref{eq-source-GCM-reduced})). The Eq.(\ref{Eq-uvGCM-SystemQuasiLinear-General2}) is a quasi-linear partial derivative system of equations, which is adapted to be numerically solved.

    \subsection{Boundary conditions for GCM system}
    
We are seeking for a $\mathcal{C}^3$ isometric embedding $(s,\lambda)\mapsto f \in\R^3$ which respects a particular symmetry. Indeed, the transformation $\lambda\leftrightarrow -\lambda$ corresponds to an orthogonal symmetry with respect to a plane. The coordinate system of $\R^3$ will be fixed so that the  symmetry plane is $0_{xz}$.  Then the matrix of this symmetry is,
\begin{equation}
    \boldsymbol{\mathcal{S}}=\Matrix{1}{0}{0}{0}{-1}{0}{0}{0}{1}\,.
\end{equation}
Let us introduce the components of $f$,
\begin{equation}\label{Eq-GCM-Immersion}
\begin{array}{ccccc}
f & : & \mathcal{V} & \longrightarrow & \R^3 \\
 & & (s,\lambda) & \mapsto & \left(\begin{array}{c}
X(s,\lambda)\\
Y(s,\lambda)\\
Z(s,\lambda)
 \end{array}\right) \,.
\end{array}
\end{equation}
Then we have,
\begin{equation}
    f(s,-\lambda)=\boldsymbol{\mathcal{S}}f(s,\lambda)\,,
\end{equation}
which implies that $X$ and $Z$ are $\lambda$-even function and $Y$ is $\lambda$-odd function. Eqs.(\ref{Eq-defsecondform-2}) and Gauss application definition imply that $M$ is $\lambda$-odd and $L$ and $N$ are $\lambda$-even. Consequently it implies,
\begin{equation}\label{Eq-BoundaryCondition-1}\left\{
    \begin{array}{cc}
        M(s,\lambda=0)&=0  \\
        \partial_\lambda L(s,\lambda=0)&=0 \\
        \partial_\lambda N(s,\lambda=0)&=0 
    \end{array}\right.\,.
\end{equation}
Nevertheless, the second and third equations of the previous system derive immediately from the first equation combined with the GCM Eqs.(\ref{Eq-CM-1}) where the Christoffel symbols and curvatures are given by Eq.(\ref{Eq-Chritoffels},\ref{Eq-GaussianCurvature},\ref{Eq-GaussianCurvature-2}). Therefore it is necessary to add a second boundary condition that determines the value of $L$ or $N$ on the line $\lambda=0$ in order to use Cauchy-Kowalevsky's theorem or a numerical integration.

For geometrical reasons, the second boundary condition is $\partial_\lambda M (\lambda=0)=0$. Combined with the last equation of Eqs.(\ref{Eq-CM-1}), it implies that,
\begin{equation}\label{Eq-BoundaryCondition-1p}\hspace{-1.5cm}
    \frac{\partial}{\partial r}\left.\left(\frac{N^2-2r}{r^2}\right)\right|_{\lambda=0}=0\quad\Longrightarrow\quad\left\{ \eqalign{N(s,\lambda=0)=\pm  \left( - \sqrt{2r(s)+Ar^2(s)} \right)\cr
L(s,\lambda=0)=\pm\frac{1}{\sqrt{2r^3(s)+Ar^4(s)}}\cr}\right.\,,
\end{equation}
where $A$ is an arbitrary constant. In the previous equation, the choice of the sign $\pm$ fixes the "global orientation" of the surface. To clarify, the GCM equations Eqs.(\ref{Eq-CM-1}) are such that if $L,M,N$ are solutions, then $-L,-M,-N$ are also solutions. Note that $L_{\pm,A},M_{\pm,A},N_{\pm,A}$ are the solutions corresponding to these initial conditions. Thus Cauchy-Kowalvsky's theorem and the preceding property easily implies that,
\begin{equation}
L_{\pm,A}=\pm L_{+,A},\quad M_{\pm,A}=\pm M_{+,A},\quad N_{\pm,A}=\pm N_{+,A} \, .
\end{equation}
Furthermore, if we note $\mathcal{S}_z={\rm diag} (1,1,-1)$ and $\mathcal{R}^+$ a solution of frame equations Eq.(\ref{Eq-FrameEquation-1}) for the values $L_{+,A},M_{+,A},N_{+,A}$, $\mathcal{S}_z\mathcal{R}^+\mathcal{S}_z$ is a frame equations solution for $L_{-,A},M_{-,A},N_{-,A}$. Respectively, if  $\mathcal{R}^-$ is a solution of frame equations Eq.(\ref{Eq-FrameEquation-1}) for the values $L_{-,A},M_{-,A},N_{-,A}$, $\mathcal{S}_z\mathcal{R}^-\mathcal{S}_z$ is a frame equations solution for $L_{+,A},M_{+,A},N_{+,A}$. Then, Frobenius theorem ensures the uniqueness of the solution and the corresponding surface $f^+$ and $f^{-}$ are deduced from each other by a translation plus a $\mathcal{S}_z$ symmetry.

To choose the sign $\pm$ and $A$, we evaluate the second fundamental form from the immersion given by Eq.(\ref{Eq-Immersion-r}). This immersion is the primitive embedding used for corrugation process. On the line ($\lambda=0$), we get,
\begin{equation}\label{Eq-BoundaryCondition-2}
    L(s,\lambda=0)=-\frac{1}{\sqrt{2r^3}}\frac{\left(r^2(s)-a^2\right)}{\left(r^2(s)+a^2\right)}\,,
\end{equation}
which requires $A=0$. And the $\pm$ sign becomes minus, in order to obtain, for a large radius, a similar behaviour of the primitive embedding given in  Eq.(\ref{Eq-Immersion-r}). Then the initial conditions become,
\begin{equation}\label{Eq-BoundaryCondition-3}
\left\{ \eqalign{M(s,\lambda=0)=0 \cr N(s,\lambda=0)=+\sqrt{2r(s)}\cr
L(s,\lambda=0)=-\frac{1}{\sqrt{2r^3(s)}}\cr}\right. \, . 
\end{equation}
The boundary conditions explain also why we chose the form gives in Eqs.(\ref{Eq-uvGCM-BF3}). 

    \subsection{Pseudo-boundary conditions for frame equations}
        
\begin{figure}[ht]
    \centering
    \hspace{0cm}\begin{minipage}[c]{.46\linewidth}
      \includegraphics[scale=0.4, trim=0cm 0cm 0cm 0cm]{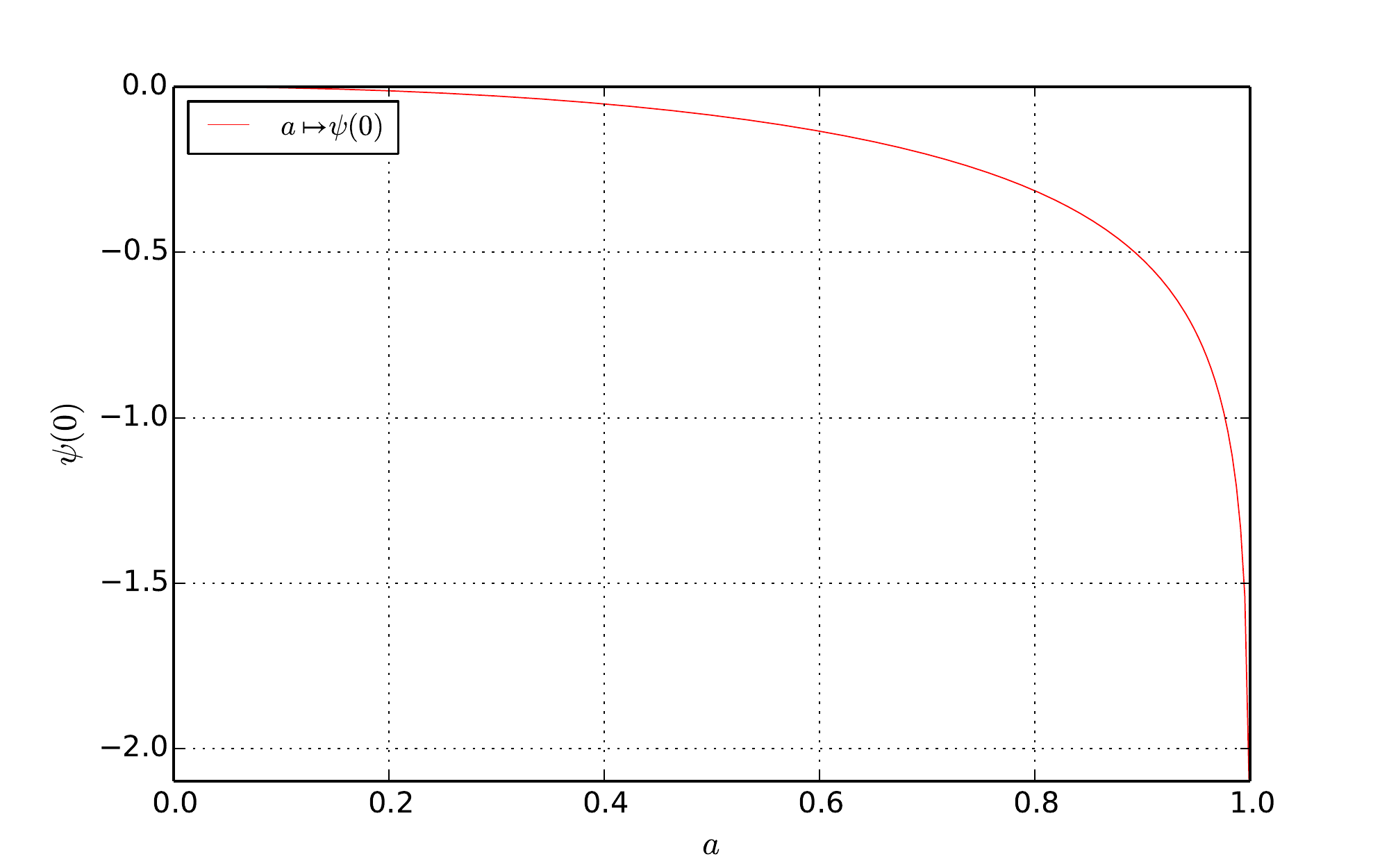}
   \end{minipage}\hfill\begin{minipage}[c]{.46\linewidth}
      \includegraphics[scale=0.4, trim=0cm 0cm 0cm 0cm]{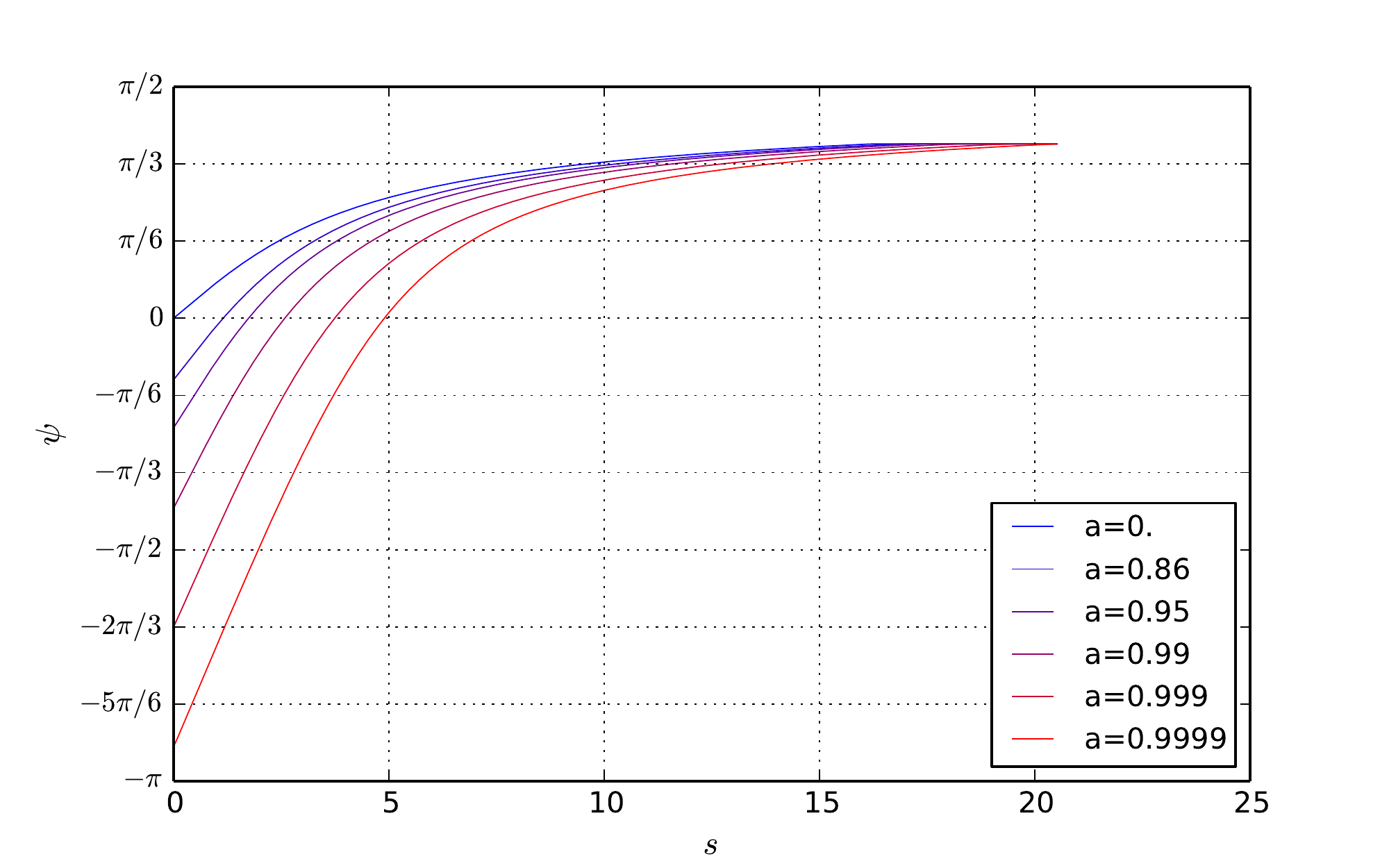}
   \end{minipage}
    \caption{On the left, the evolution of $\psi$ along the equatorial curve in $s=0$ with $O_z$ as a function of $a$: $a\mapsto\psi(0)$. On the right, the evolution of the angle $s\mapsto \psi(s)$ along the equatorial curve with $O_z$ for different values of $a$.} 
    \label{fig-evolutionangleprimitif}
\end{figure}

The boundary conditions and the symmetry of the problem lead to solve analytically frame equations for $\lambda=0$. Indeed, for $\lambda=0$, the symmetry $\lambda\leftrightarrow -\lambda$ implies that the curve $s \mapsto f(s,0)$ is contained in the $0_{xz}$ plane. This implies that for $\lambda=0$, $\boldsymbol{\partial}_s$ belongs to this plane and $\boldsymbol{\partial}_\lambda$ is orthogonal to this plane too. Since $\boldsymbol{\partial}_s$ is an unitary vector ($\lambda=0$), there is a function $s\mapsto \psi(s)$ such that,  
\begin{equation}
\left\{ \eqalign{\boldsymbol{\partial}_\lambda(s,\lambda=0)=r(s)\boldsymbol{e}_y \cr \boldsymbol{\partial}_s(s,\lambda=0)=\boldsymbol{u}(\psi(s))\cr
}\right. \, ,
\end{equation}
where $\boldsymbol{u} : \psi \mapsto \cos(\psi)\boldsymbol{e}_z+\sin(\psi)\boldsymbol{e}_x$ is a rotating vector function of the $0_{xz}$ plane. We call these equations, pseudo boundary conditions for the frame equations. The term ``pseudo'' is used because the frame equations are Frobenius-like. Then they just need initial conditions (at $s=\lambda=0$) to determine the solution. The frame equations became for $\lambda=0$,
\begin{equation}
    \dfdx{\boldsymbol{\partial}_s}{s}=-L(s,0)\boldsymbol{e}_y\times\boldsymbol{\partial}_s\quad\Rightarrow\quad \dfdx{\psi}{s}=-L(s,0)\, ,
\end{equation}
and by integrating from $0$ to $s$,
\begin{equation}
    \psi(s)=\psi_0-\int_0^s L(\tilde{s},0){\rm d}\tilde{s} \, .
\end{equation}
We set $\psi_0$ such that $\psi\underset{s\rightarrow +\infty}{\rightarrow}\pi/2$ (see right side of Fig.(\ref{fig-evolutionangleprimitif})) and this leads to,
\begin{equation}\label{eq-angleds-pseudobound}
    \psi(s)=\frac{\pi}{2}+\int_s^\infty L(\tilde{s},0){\rm d}\tilde{s}\,, 
\end{equation}
which gives for the choice of $L(s,0)$ (see Eq.(\ref{Eq-BoundaryCondition-3})),
\begin{equation}\label{eq-angleds-pseudoboundexplicite}
    \psi(s)=\frac{\pi}{2}-\sqrt{\frac{2}{r_+}}F\left(\sqrt{\frac{r_+}{r(s)}};\sqrt{\frac{r_-}{r_+}}\right)\,,
    \label{eq-anglepsi}
\end{equation}
where $x,k\mapsto F(x;k)$ is the usual elliptic integral of the first kind. This implies, using arithmetic-geometric mean, that $\psi(0)\leq 0$ with $\psi(0)=0$ for $a=0$. We also get $\psi(0)\rightarrow -\infty$ when $a\rightarrow 1$. The trend of $\psi(0)$ in function of $a$ is plotted on the left side of Fig.(\ref{fig-evolutionangleprimitif}). This leads to calculate the pseudo boundary conditions for the surface,
\begin{equation}
    \boldsymbol{f}(s,0)=\int_0^s \boldsymbol{u}(\psi(s)){\rm d}s \, .
\end{equation}

The Eq.(\ref{eq-angleds-pseudobound}) insures that $\boldsymbol{e}_x$ is the asymptotic direction of $s\mapsto \boldsymbol{f}(s,0)$ when $s$ reaches infinity (see right side of Fig.\ref{fig-evolutionangleprimitif}).

    \subsection{Numerical resolution}

Since an analytical solution is difficult to provide, we decided to solve  GCM and frame equations with numerical methods. On each step, the numerical errors is estimate to reach the convergence.
The dedicated developed program called \textit{Isopol}\footnote{\url{https://perso.imcce.fr/frederic-dauvergne/isopol/index.html}} is provided, with documentation, as an open source license.



        \subsubsection{GCM equations resolution}\leavevmode\par

Firstly, the program solves the Eqs.(\ref{Eq-uvGCM-SystemQuasiLinear-General2}) with the source term and the matrix explicitly given in Eq.(\ref{eq-matrix-GCM-reduced}) and Eq.(\ref{eq-source-GCM-reduced}). The integration along $\lambda$ is numerically done using a second order finite difference method. We use a regular grid of $n_s$ points for $s\in[0,s_{\rm max}]$ and $n_\lambda$ points for $\lambda\in[0,\pi/2]$. The value of $n_\lambda$ is evaluated from a reference parameter $(\Delta\lambda/\Delta s)_{ref}$. We ensure that $n_\lambda$ is an integer by setting $n_\lambda=\displaystyle\left\lfloor\frac{\pi}{2}\left(\frac{\Delta \lambda}{\Delta s}\right)_{ref}\frac{s_{\rm max}}{n_s}\right\rfloor$. 

For $a>\sqrt{3}/2$, we restrict to a sub-domain of the negative gaussian curvature region, defined by \mbox{$\mathcal{U}_{s_{\rm max}}^{a,f}=\left\{(s,\lambda)\in\mathcal{U}_{s_{\rm max}}\,|\, r^2(s)-3fa^2\sin^2\lambda>0\right\}$} with $f=1.27$.

Then to evaluate the error order, we numerically estimate the left side of the equation Eq.(\ref{Eq-GCM-1}). We call $\tilde{\boldsymbol{\mathcal{K}}}_s^{n_s,n_\lambda},\tilde{\boldsymbol{\mathcal{K}}}_\lambda^{n_s,n_\lambda}$ the values of the matrices ${\boldsymbol{\mathcal{K}}}_s,{\boldsymbol{\mathcal{K}}}_\lambda$ obtained with the linear interpolation of GCM system numerically integrated. Thus, we evaluate,
\begin{equation}
    \partial_s \tilde{\boldsymbol{\mathcal{K}}}_\lambda^{n_s,n_\lambda}-\partial_\lambda \tilde{\boldsymbol{\mathcal{K}}}_s^{n_s,n_\lambda} +\left[ \tilde{\boldsymbol{\mathcal{K}}}_s^{n_s,n_\lambda}; \tilde{\boldsymbol{\mathcal{K}}}_\lambda^{n_s,n_\lambda}\right]=\boldsymbol{\epsilon}^{n_s,n_\lambda}(s,\lambda) \, .
\end{equation}
The error $\delta \boldsymbol{\mathcal{R}}$ on the $\boldsymbol{\mathcal{R}}^{n_s,n_\lambda}$ matrix, provided by the GCM integration is written as,
\begin{equation}
     \delta \boldsymbol{\mathcal{R}} = \int_{\gamma_1} \boldsymbol{\omega}-\int_{\gamma_2} \boldsymbol{\omega}= \oint_{\gamma_1\cup -\gamma_2} \boldsymbol{\omega} \, , 
\end{equation}

with $\boldsymbol{\omega}=\boldsymbol{\mathcal{R}}\left(\tilde{\boldsymbol{\mathcal{K}}}_\lambda^{n_s,n_\lambda} {\rm d}\lambda+\tilde{\boldsymbol{\mathcal{K}}}_s^{n_s,n_\lambda} {\rm d}s \right)$, $\gamma_1$ the path from $(0,0)$ to $(s,\lambda)$ passing through $(s,0)$ and $\gamma_2$ the path from $(0,0)$ to $(s,\lambda)$ passing trough $(0,\lambda)$.

Using the Stockes theorem,

\begin{equation}\label{eq-erreur-R}
    \oint_{\gamma_1\cup -\gamma_2} \boldsymbol{\omega}=\iint_{[0,s]\times[0,\lambda]}\boldsymbol{\mathcal{R}}(\tilde{s},\tilde{\lambda})\boldsymbol{\epsilon}^{n_s,n_\lambda}(\tilde{s},\tilde{\lambda}){\rm d}\tilde{s}{\rm d}\tilde{\lambda} \, , 
\end{equation}
the error can be majorized by,
\begin{equation}
    ||\delta \boldsymbol{\mathcal{R}}|| \leq     \frac{\pi s_{\rm max}}{2}||\boldsymbol{\epsilon}^{n_s,n_\lambda}||_{\infty}\sqrt{r_{\rm max}^2+a^2} \, .
\end{equation}
    
        \subsubsection{Frame equations resolution}\leavevmode\par
    
Once solved the GCM system Eq.(\ref{Eq-CM-1}), we used the pseudo boundary conditions to get the frame $\boldsymbol{\mathcal{R}}$ and the surface $f$. 

Then the Frobenius-type equations are transformed into an ordinary differential equation by setting the value of $s$. Indeed, the frame equations could be written in this form,
\begin{equation}
    \frac{\rm d}{{\rm d}\lambda}\left|\begin{array}{c}
\boldsymbol{f}\\
\boldsymbol{\partial}_s\\
\boldsymbol{\partial}_\lambda
\end{array}\right. =\left|\begin{array}{c}
\boldsymbol{\partial}_\lambda\\
\Gamma_{s\lambda}^s \boldsymbol{\partial}_s+\Gamma_{s\lambda}^\lambda \boldsymbol{\partial}_\lambda +M \frac{\boldsymbol{\partial}_s\times\boldsymbol{\partial}_\lambda}{||\boldsymbol{\partial}_s\times\boldsymbol{\partial}_\lambda||} \\
\Gamma_{\lambda\lambda}^s \boldsymbol{\partial}_s+\Gamma_{\lambda\lambda}^\lambda \boldsymbol{\partial}_\lambda +N \frac{\boldsymbol{\partial}_s\times\boldsymbol{\partial}_\lambda}{||\boldsymbol{\partial}_s\times\boldsymbol{\partial}_\lambda||}
\end{array}\right.
\,.
\end{equation}
For each abscissa $s$, on the grid, the integration will be done along the $\lambda$ direction. The resolution of the differential equations is solved by a 4th order Runge-Kutta with initial conditions deduced from the pseudo-boundary conditions,
\begin{equation}\label{Eq-InitialConditionFrame}
\forall s\geq0, \quad \left\{ \eqalign{\boldsymbol{f}(s,0)&=\int_0^s \boldsymbol{u}(\psi(s)){\rm d}s \cr \boldsymbol{\partial}_s(s,\lambda=0)&=\boldsymbol{u}(\psi(s))\cr \boldsymbol{\partial}_\lambda(s,\lambda=0)&=r(s)\boldsymbol{e}_y \cr 
}\right. \, .
\end{equation}

Once this integration is done, the isometric default of the induced metric could be estimated by a discrete derivative on ${\bf f}$ in order to get the value of $(\boldsymbol{\partial}_s,\boldsymbol{\partial}_\lambda)$, and then we get induce metrics and the corresponding isometric default.
    
\section{Isometric embedding from GCM and frame equations} \label{sec5}

    \subsection{Global analysis and numerical errors}

The main input parameters are,
\begin{itemize}
\item the $n_s$ number,
\item $r_{\rm max}$ value (associated with a value of $s_{\rm max}$),
\item the spin of the black hole $a$,
\item a reference value $(\Delta \lambda/\Delta s)_{\rm ref}$ for the ratio between $\Delta \lambda$ and $\Delta s$ in the grid. 
\end{itemize}
In the solution, we choose $n_s=3000$, $s_{\rm max}=20$ and $(\Delta \lambda/\Delta s)_{\rm ref} = 0.01$. This choice is a good compromise between numerical stability and computational time. 

\begin{table}[ht]
 \begin{center}
\begin{tabular}{ c c c c c}
 \hline\hline
   \qquad $a$ \qquad&\qquad $s_{\rm max}$ \qquad&\qquad $n_\lambda$ \qquad&\qquad $||\boldsymbol{\epsilon}^{n_s,n_\lambda}||_{\infty}$ \qquad&\qquad $||\boldsymbol{\mu}-f^\star<\cdot,\cdot>_{\R^3}||$ \\ \hline\hline
    $0.00$ & $22.6$ & $20\,841$ & $<10^{-5}$ & $<7,4 \times 10^{-4}$ \\
    $0.50$ & $22.8$ & $20\,703$ & $<10^{-5}$ & $<6.8 \times 10^{-4}$ \\
    $0.86$ & $23.3$ & $20\,021$ & $<10^{-5}$ & $<7.65 \times 10^{-4}$ \\
    $0.99$ & $24.6 $ & $19\,159$ & $<1.72\times10^{-3}$ & $<8.16 \times 10^{-4}$ \\
    $0.999$ & $25.7$ & $18\,304$ & $<2.45\times10^{-3}$ & $<8.16 \times 10^{-4}$ \\
    $0.9999$ & $26.9$ & $17\,520$ & $<5.50\times10^{-3}$ & $<8.16 \times 10^{-4}$ \\
    $0.99999$ & $28.0$ & $16\,801$ & $<4.06\times10^{-4}$ & $<8.16 \times 10^{-4}$ \\
 \end{tabular}    
 \end{center}
 \caption{Characteristics of the solutions, all of them are calculated with $n_s=3000$ and $(\Delta \lambda/\Delta s)_{ref} = 0.01$.}
 \label{tab-GCM+FrameResult}
 \end{table}
 
\begin{figure}[ht]
    \centering
    \hspace{0cm}\begin{minipage}[c]{.46\linewidth}
      \includegraphics[scale=0.45, trim=0cm 0cm 0cm 0cm]{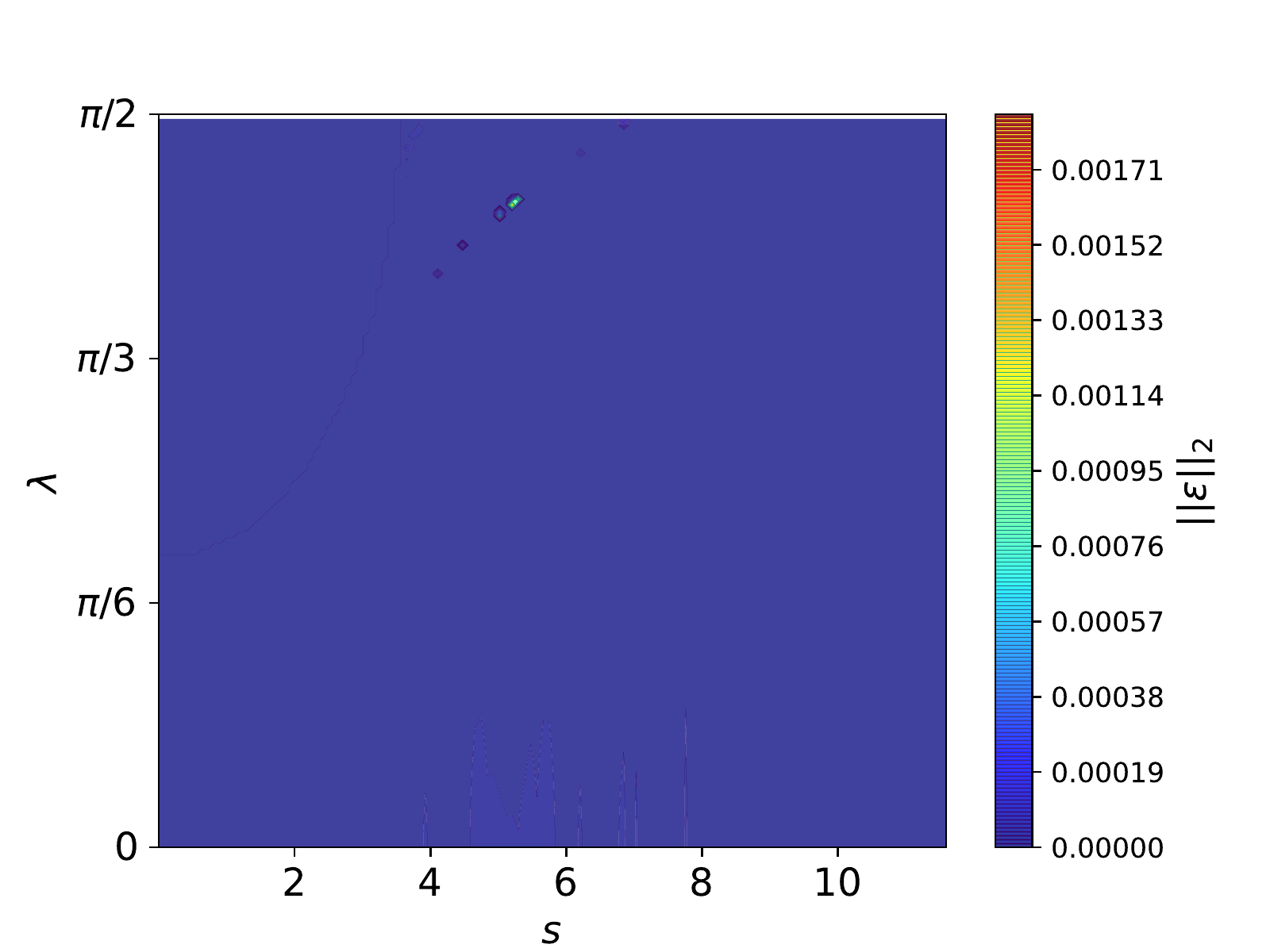}
   \end{minipage}\hfill\begin{minipage}[c]{.46\linewidth}
      \includegraphics[scale=0.45, trim=0cm 0cm 0cm 0cm]{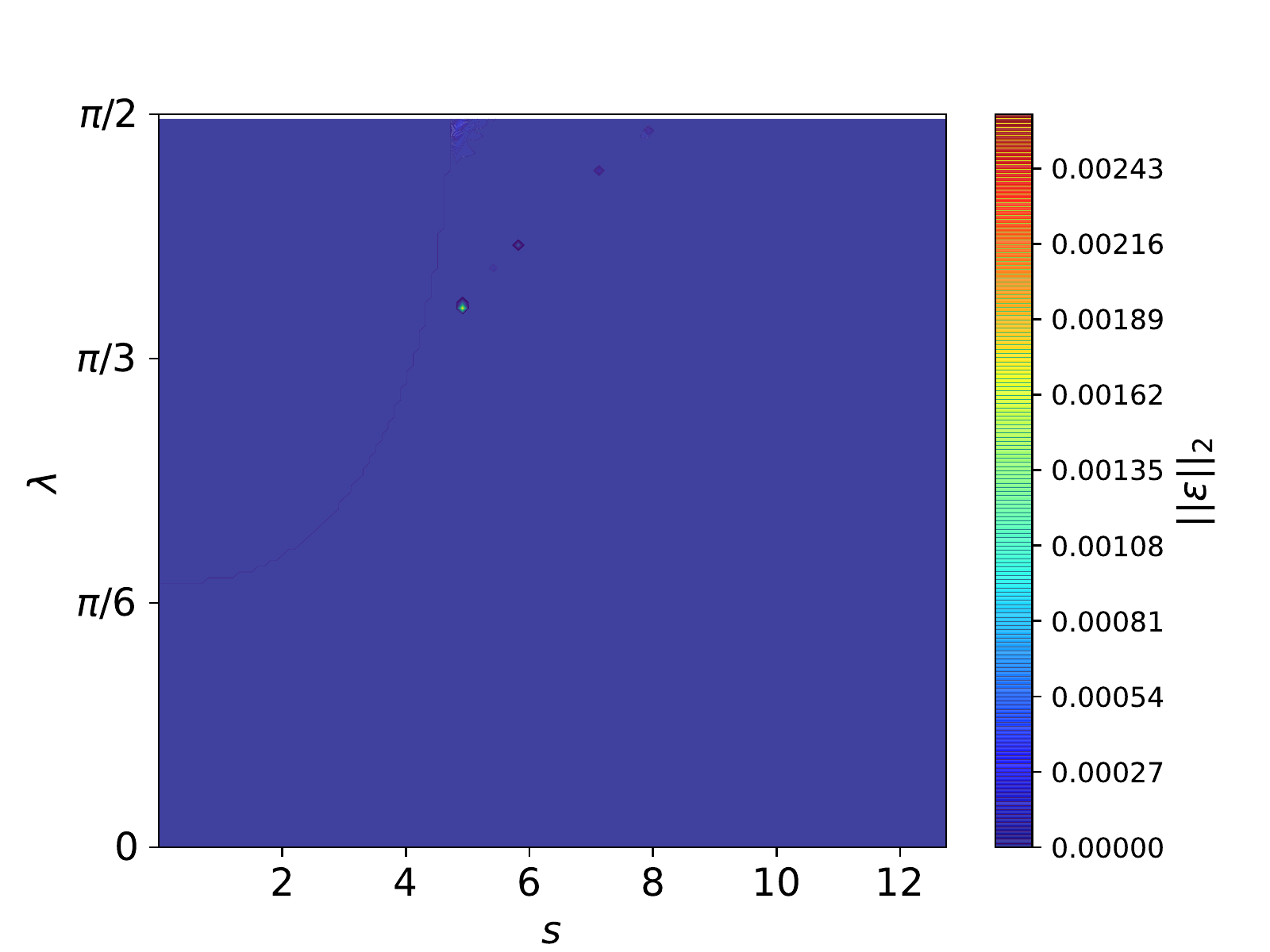}
   \end{minipage}
    \caption{Colormap of $||\boldsymbol{\epsilon}^{n_s,n_\lambda}||$ for $a=0.99$ (right) and $a=0.999$ (left).}
    \label{fig-error-a-sup-0.86}
\end{figure}

The main characteristic of the solutions are shown in Tab.\ref{tab-GCM+FrameResult}. The error $||\boldsymbol{\epsilon}^{n_s,n_\lambda}||_{\infty}$ on GCM evolves with the spin. In particular above $a_{\rm lim}$, the error increases but remains lower than $5.50\times 10^{-3}$. The right part of the equations Eq.(\ref{Eq-uvGCM-SystemQuasiLinear-General2}) concerns the propagation (with velocities and directions related to the eigenvector and eigenvalues of the $A$ matrix), and the left part has a source term. The introduction of the sub-domain $\mathcal{U}_{s_{\rm max}}^{a,f}$ induces a stair edge slicing and then the boundary condition computation leads to the propagate errors as shown on Fig.(\ref{fig-error-a-sup-0.86}). Nevertheless, except for a few "small" areas of the domain, the error remains small in most of the domain. Errors appear when $\lambda$ reaches the edge of the domain, and these errors propagate along $s$ direction.
 
The reduced area of errors on the GCM equations explains why this jump on $||\boldsymbol{\epsilon}^{n_s,n_\lambda}||_{\infty}$ does not impact much the isometric default $||\boldsymbol{\mu}-f^\star<\cdot,\cdot>_{\R^3}||$, which always remains below $10^{-3}$. The most important contribution to the isometric default comes from the term $f^\star<\boldsymbol{e}_s,\boldsymbol{e}_s>_{\R^3}=<\boldsymbol{\partial}_s,\boldsymbol{\partial}_s>$.\\

\subsection{Second form analysis}
 
 \begin{figure}[h!]
    \centering
    \hspace{0cm}\begin{minipage}[c]{.33\linewidth}
    \includegraphics[scale=0.3, trim=0cm 0cm 0cm 0cm]{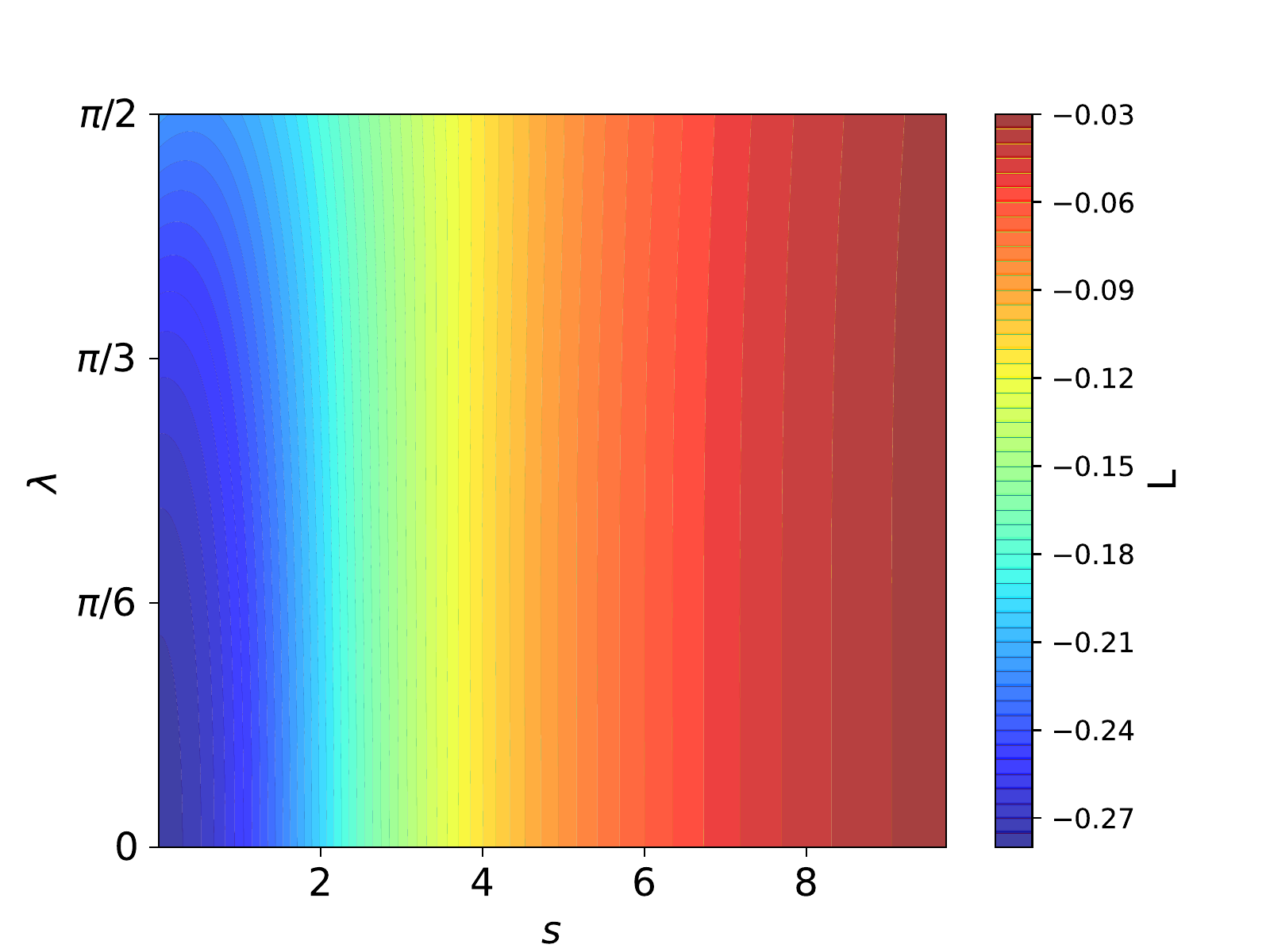}\\
    \includegraphics[scale=0.3, trim=0cm 0cm 0cm 0cm]{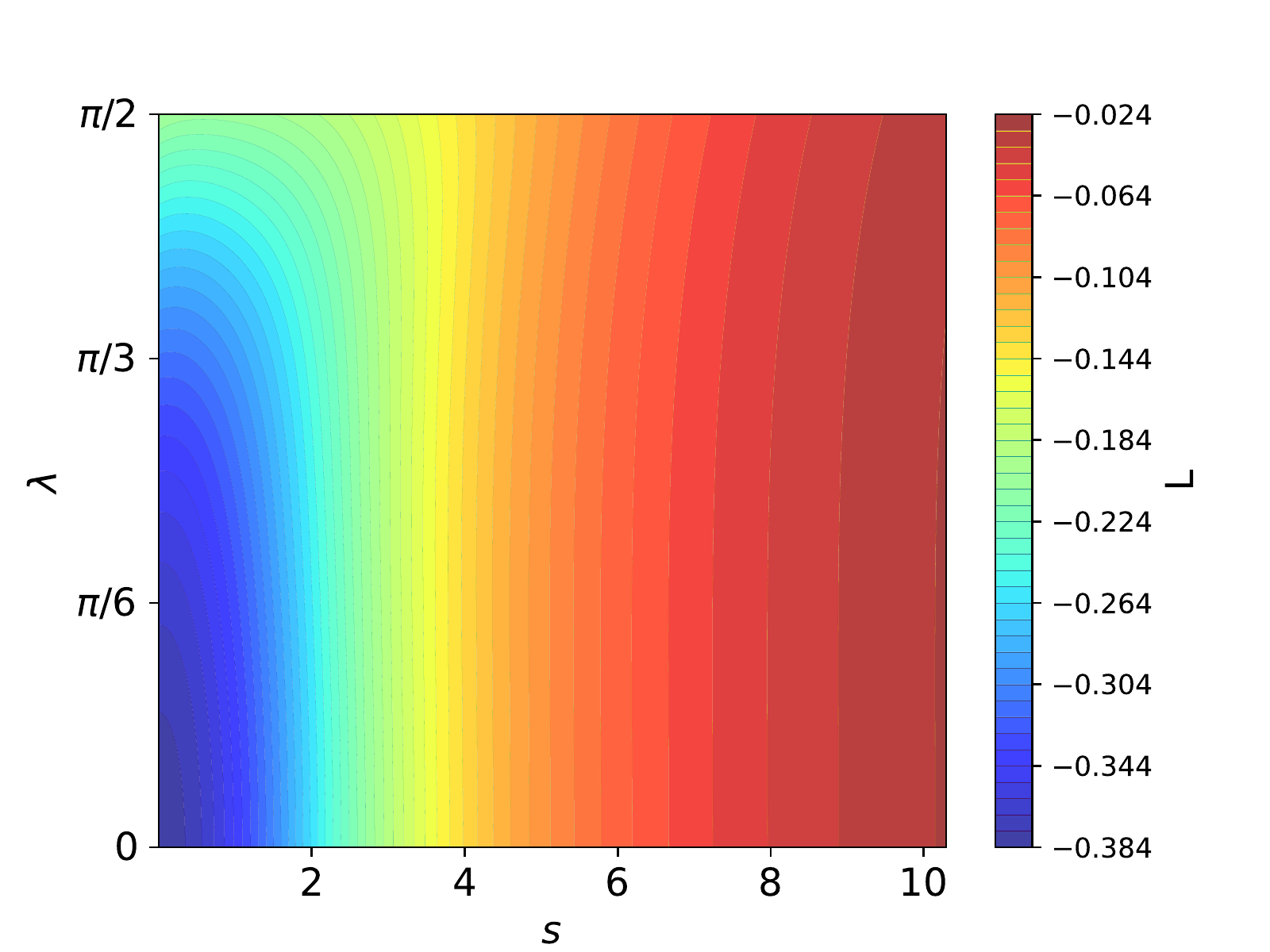}
   \end{minipage}\begin{minipage}[c]{.33\linewidth}
    \includegraphics[scale=0.3, trim=0cm 0cm 0cm 0cm]{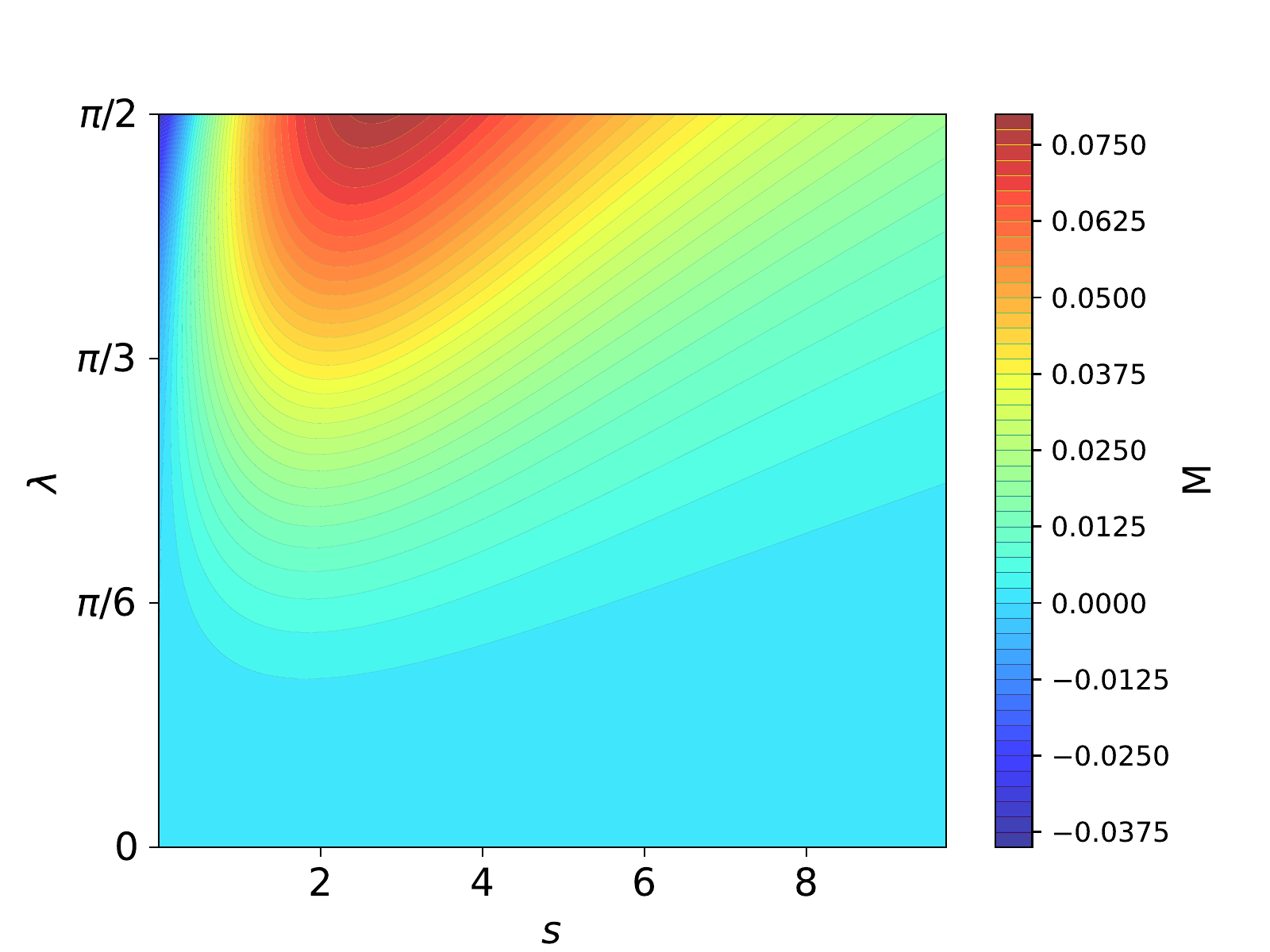}\\
    \includegraphics[scale=0.3, trim=0cm 0cm 0cm 0cm]{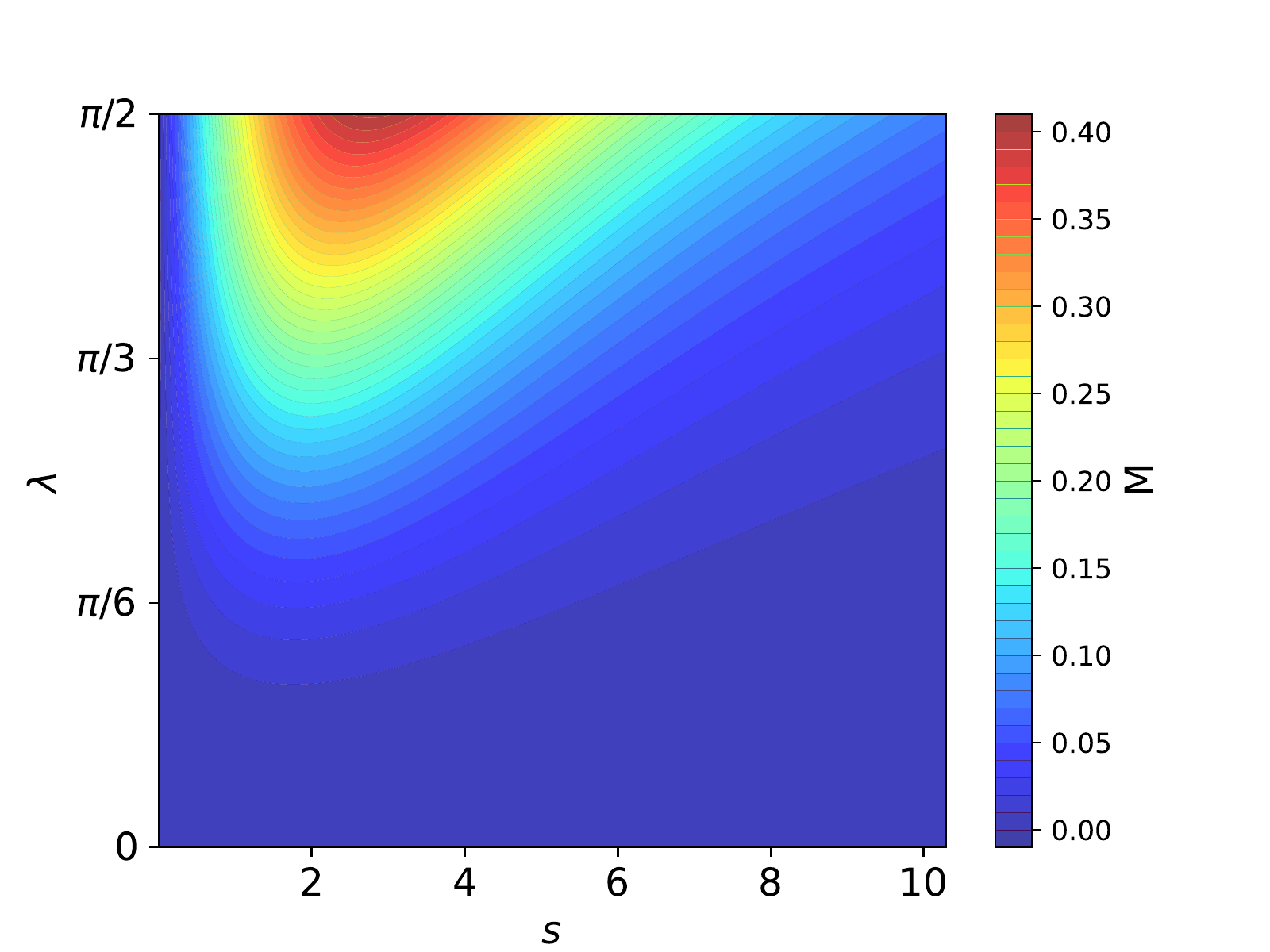}
   \end{minipage}\begin{minipage}[c]{.33\linewidth}
    \includegraphics[scale=0.3, trim=0cm 0cm 0cm 0cm]{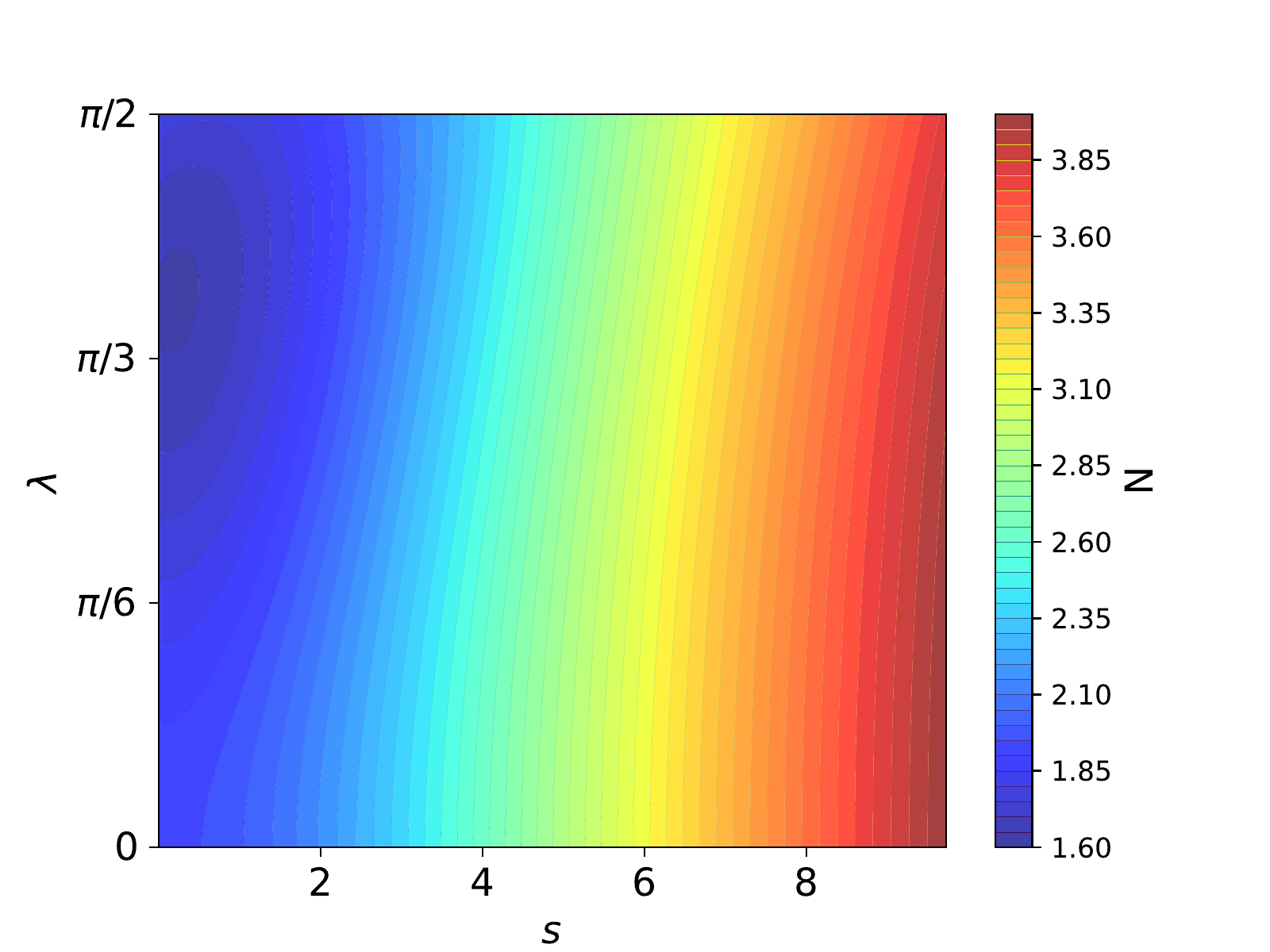}\\
    \includegraphics[scale=0.3, trim=0cm 0cm 0cm 0cm]{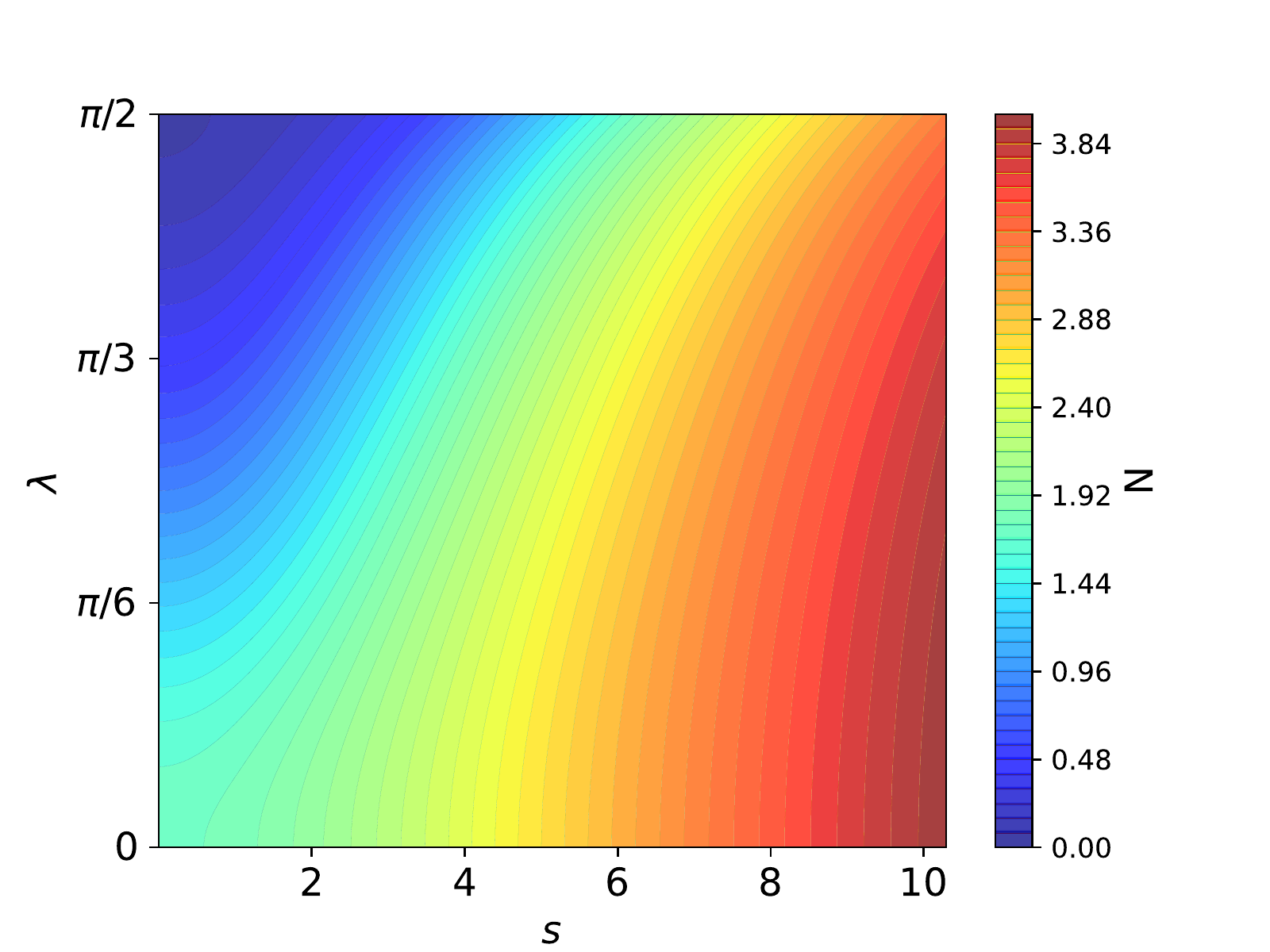}
   \end{minipage}
    \caption{Colormap of $L$ (left), $M$ (medium) and $N$ (right) functions of second fondamental form Eq.(\ref{Eq-defsecondform-1},\ref{Eq-defsecondform-2}) for the value of black hole spin $a=0.5$ (up) and $0.86$ (down). }
    \label{fig-secondform}
\end{figure}
As expected in sub-section 2, the value $a_{\rm lim}=\sqrt{3}/2$ is a limiting value. In Eqs.(\ref{Eq-uvGCM-SystemQuasiLinear-General2}, \ref{eq-matrix-GCM-reduced} and \ref{eq-source-GCM-reduced}) singularities can appear for $K+u^2=0$ or $v=0$, equivalent to $L=0$ or $N=0$. We present in Fig.(\ref{fig-secondform}) the color map of the functions $L,M$ and $N$ for two values of black hole spin. The quantity $M(0,\lambda)$ remains close to $0$, especially when $a$ is close just bellow $a_{\rm lim}$. Gauss equation could be written $L(0,\lambda)N(0,\lambda)\approx\tilde{K}(0,\lambda)$. For $a\geq a_{\rm lim}$, $\tilde{K}(0,\lambda)$ reaches zero for $\lambda_0\in[0,\pi/2]$, then $L(0,\lambda)$ or $N(0,\lambda)$ would reach zero for some $\lambda$ value near $\lambda_0$. 

The numerical integration of GCM equations systematically diverges to infinity near the region where the Gaussian curvature gets null. It explains why we need to slice our domain to avoid infinite values. 

\begin{figure}[ht]
    \centering
    \hspace{-1cm}
      \includegraphics[scale=.5, trim=0cm 0cm 0cm 0cm]{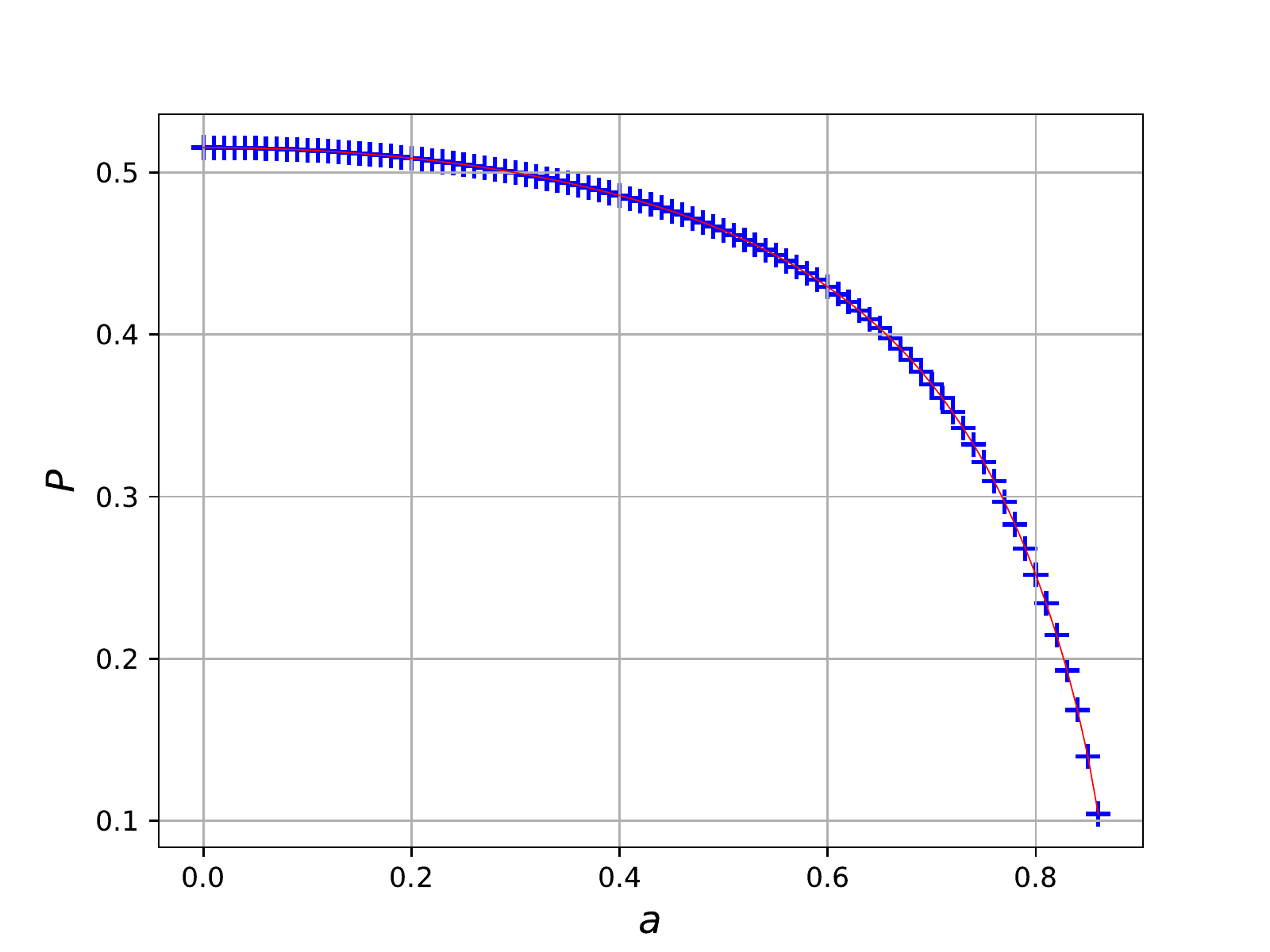}
    \caption{Evolution of $a\mapsto P(0,\pi/2)$ for $a\leq 0.86\le a_{\rm lim}$.}
    \label{fig-planar?}
\end{figure}

For $a=a_{\rm lim}$, the surface point corresponding to $s=0$ and $\lambda=\pi/2$ is a quasi-planar point. For $a=0.86$, which is the last two-digit precision computed value where we could get the solution without slicing our domain, the values $L(0,\pi/2)$, $N(0,\pi/2)$ and $M(0,\pi/2)$ are close to $0$. We can confirm the property of the point at $s=0$ and $\lambda=\pi/2$ by considering the function,
\begin{equation*}
    P : (s,\lambda) \mapsto \sqrt{H^2(s,\lambda)-2K(s,\lambda)} \, ,
\end{equation*}
where $H$ is the usual mean curvature. On a planar point, the function $P$ is equal to $0$. On the Fig.(\ref{fig-planar?}), the function $a\mapsto P(0,\pi/2)$ decreases strongly when $a\rightarrow a_{\rm lim}$. We could expect that it reaches zero for $a=a_{\rm lim}$. Even if $P$ does not reach $0$ for $a=a_{\rm lim}$, its value remains small and the surface point $(0,\pi/2)$ quasi-planar.

\subsection{Isometric embedding analysis for black hole spin below \texorpdfstring{$a_{\rm lim}$}{alim} }

\begin{figure}[ht]
    \centering
    \hspace{-1cm}\begin{minipage}[c]{.33\linewidth}
      \includegraphics[scale=0.43, trim=0cm 0cm 0cm 0cm]{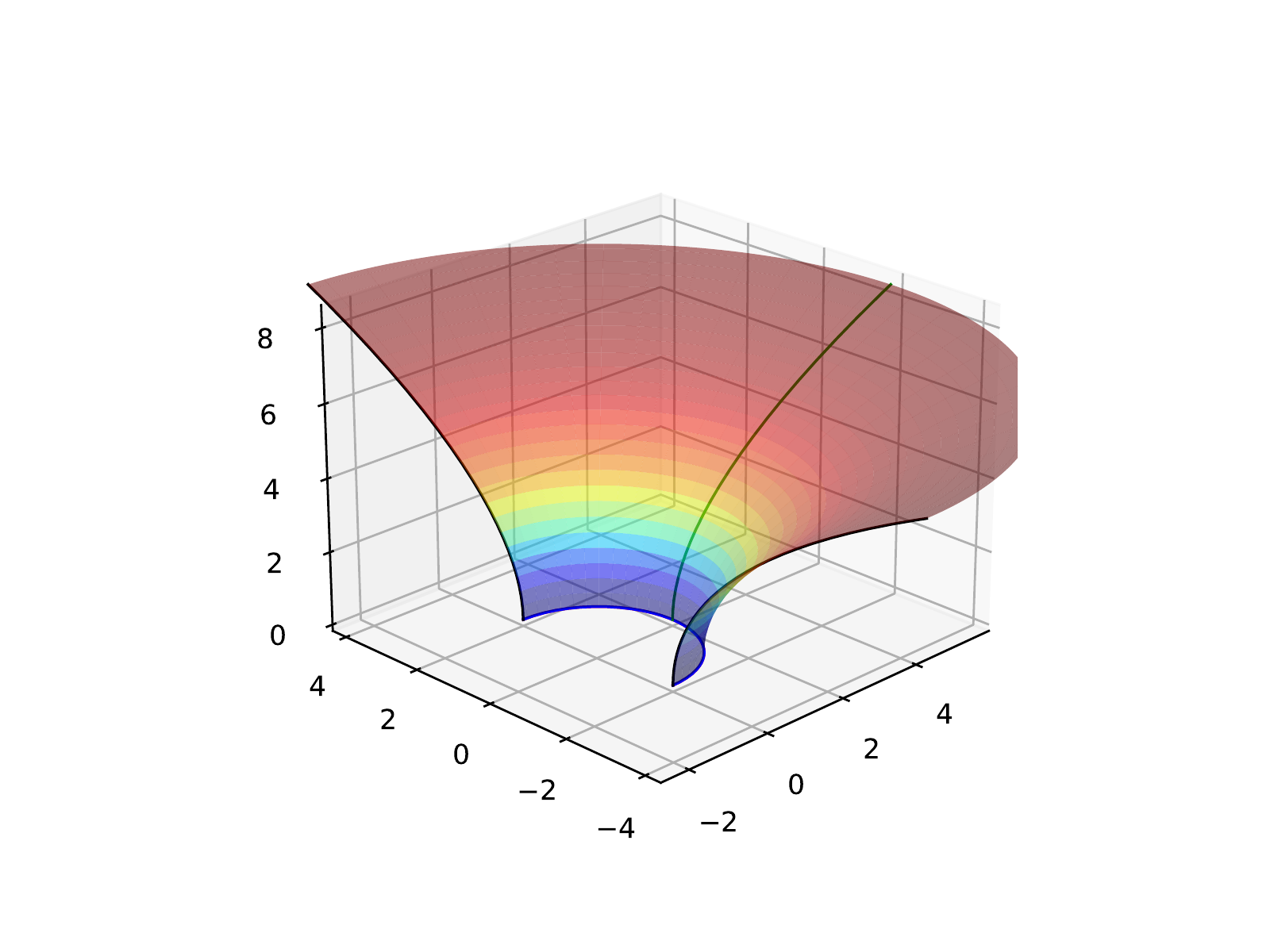}
   \end{minipage}\hfill\begin{minipage}[c]{.33\linewidth}
      \includegraphics[scale=0.43, trim=0cm 0cm 0cm 0cm]{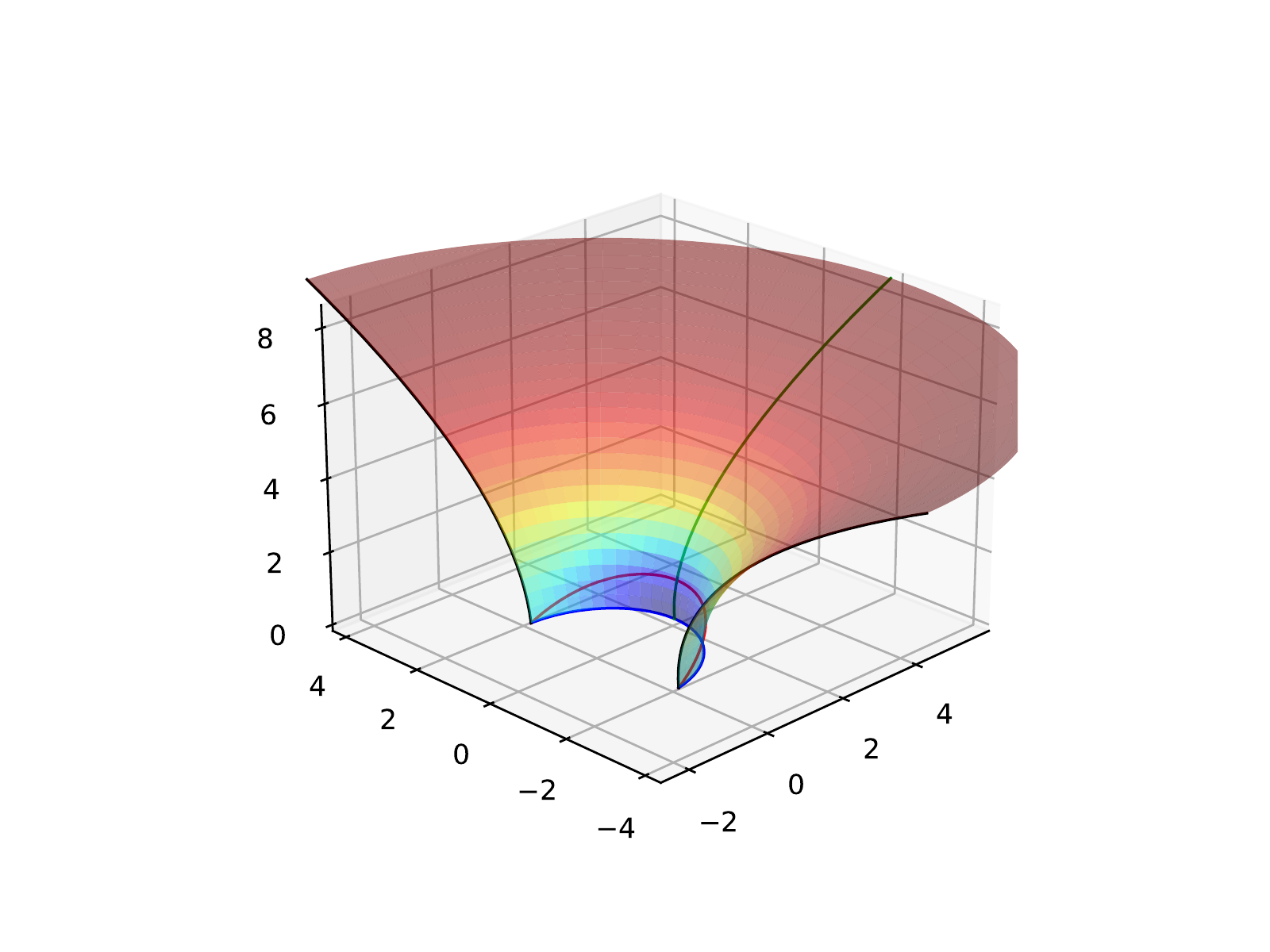}
   \end{minipage}\hfill\begin{minipage}[c]{.33\linewidth}
      \includegraphics[scale=0.43, trim=0cm 0cm 0cm 0cm]{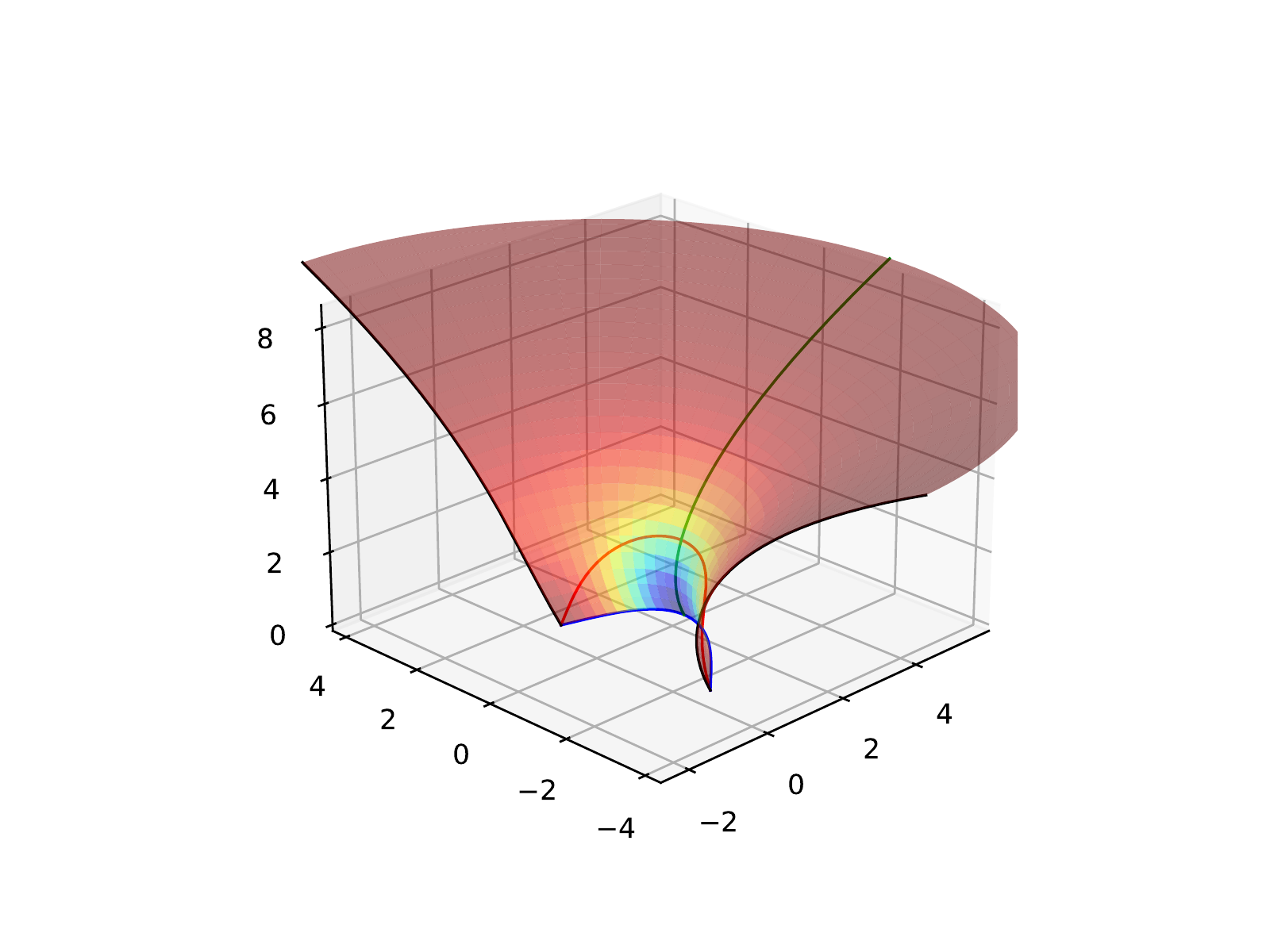}
   \end{minipage}
    \caption{From the left to the right, the isometric embedding in the 3D Euclidean space for $a=0$, $0.5$ and $0.86$. In green the equatorial axis ($\lambda=0$), in black the polar axis $(\lambda=\pm\pi/2)$, in red the ergosphere and in blue the horizon.}
    \label{fig-evolutionplongamen-a-0-0.86}
\end{figure}

In Fig.(\ref{fig-evolutionplongamen-a-0-0.86}), for $a=0.86$ the point $(s=0,\lambda=\pi/2)$ is a quasi-planar point.



\begin{figure}[ht]
    \centering
    \hspace{0cm}\begin{minipage}[c]{.46\linewidth}
        \includegraphics[scale=0.4, trim=0cm 0cm 0cm 0cm]{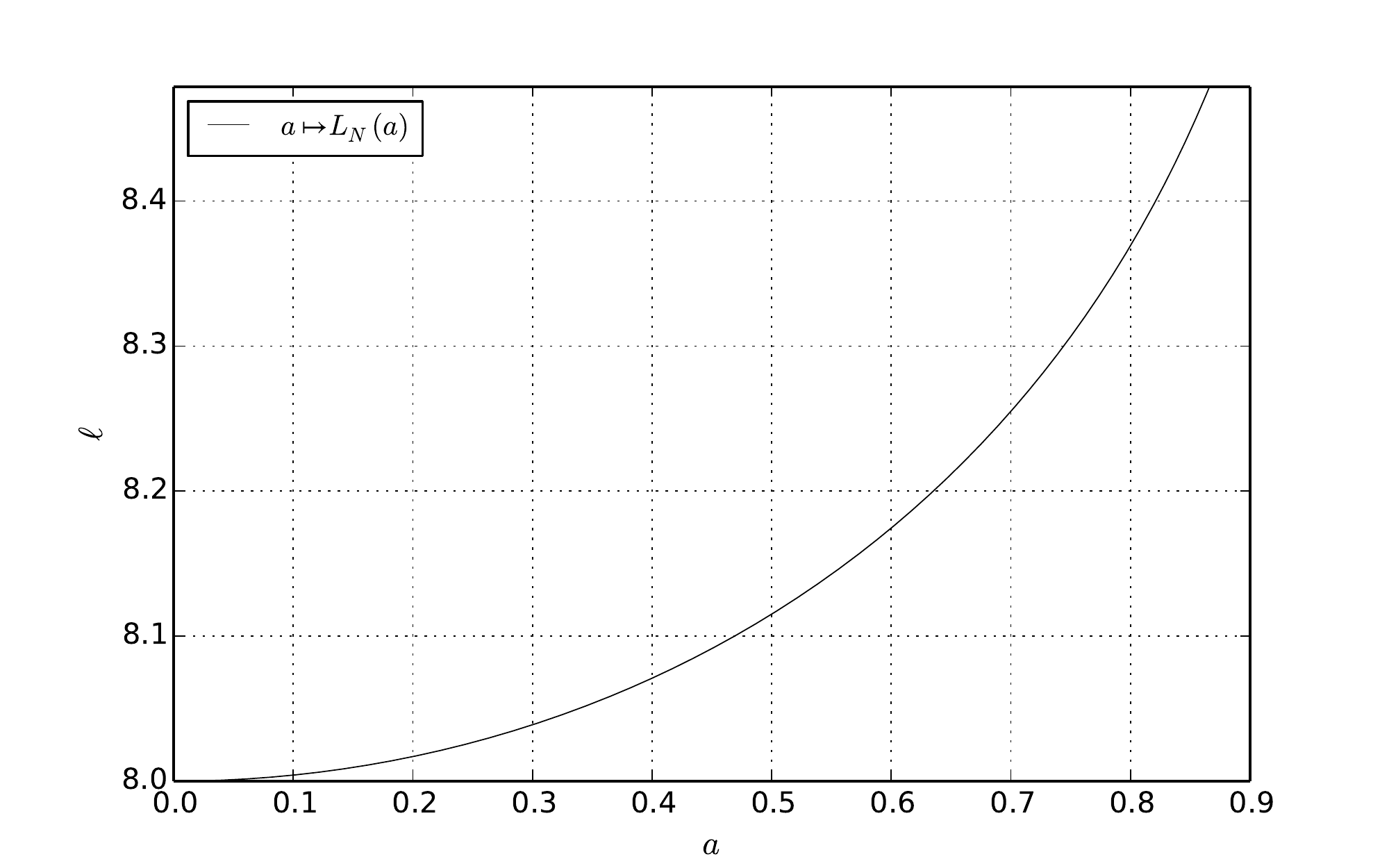}
   \end{minipage}\hfill\begin{minipage}[c]{.46\linewidth}
        \includegraphics[scale=0.4, trim=0cm 0cm 0cm 0cm]{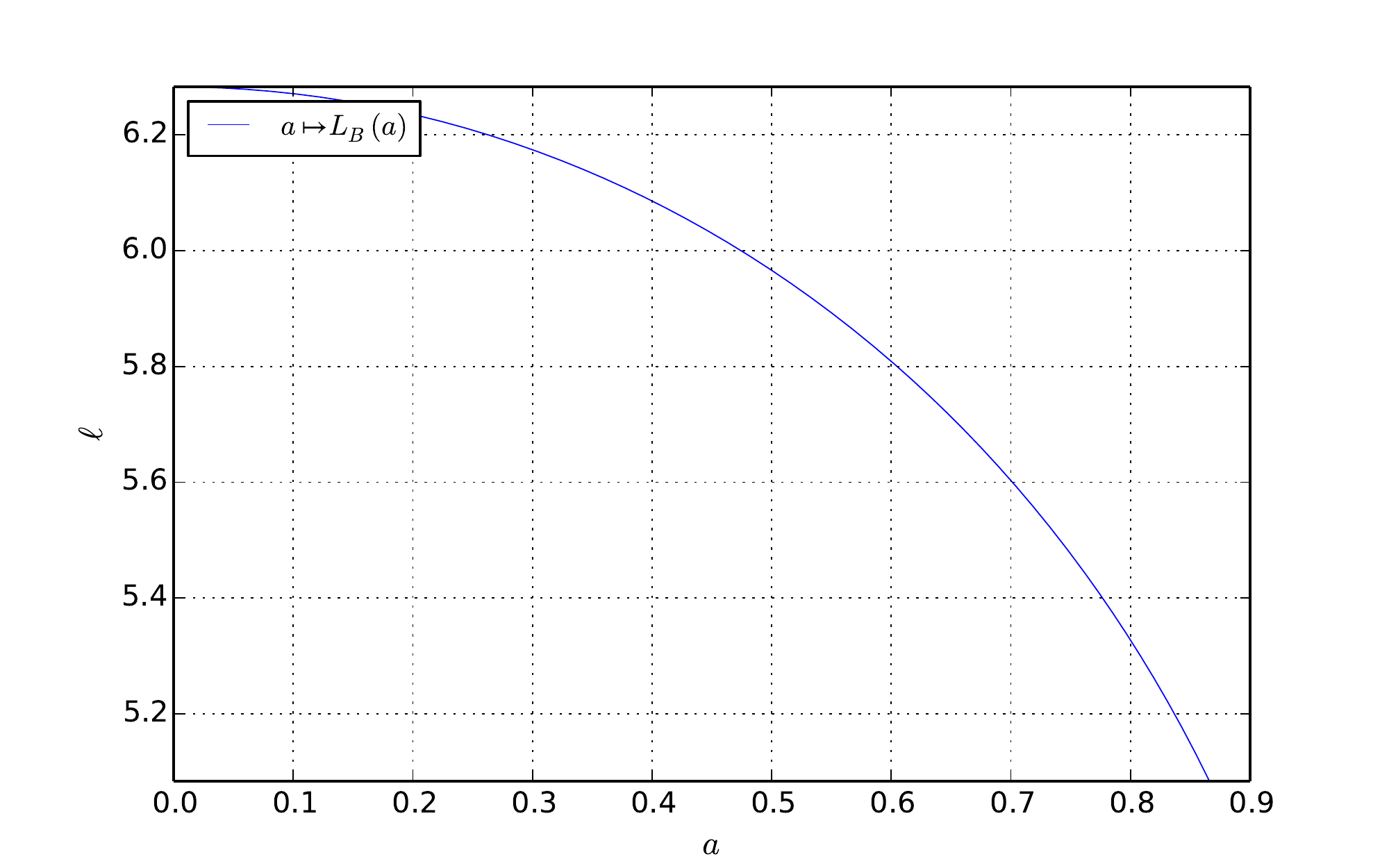}
   \end{minipage}
        \caption{On the left, evolution of the polar axis length for $s\in[0,s_{\rm max}]$ and on the right, evolution of the horizon length in function of $a\in[0,\sqrt{3}/2]$.}
        \label{fig-evolutionlenght}
\end{figure}


To understand how the isometric embedding evolves {with respect to} $a\in[0;\sqrt{3}/2]$ (cf Fig.(\ref{fig-evolutionplongamen-a-0-0.86})), we consider the evolution of polar axis, horizon and equatorial line lengths for $s\in[0,s_{\rm max}]$ (see Fig.(\ref{fig-evolutionlenght})). The equatorial line length is constant and equal to $s_{\rm max}$. The length of the line $\lambda\mapsto {\boldsymbol{f}}(\lambda,s=s_{\rm max})$ is nearly constant with $a$ for $s_{\rm max}$ sufficiently large. The length of the polar axis increases while the length of the horizon decreases in the respect of $a$. It is coherent with the surface twist when $a$ increases as seen in Fig.(\ref{fig-evolutionplongamen-a-0-0.86}).

\subsection{Torsion analysis}
\begin{figure}[h!]
    \centering
    \hspace{0cm}\begin{minipage}[c]{.46\linewidth}
      \includegraphics[scale=0.5, trim=0cm 0cm 0cm 0cm]{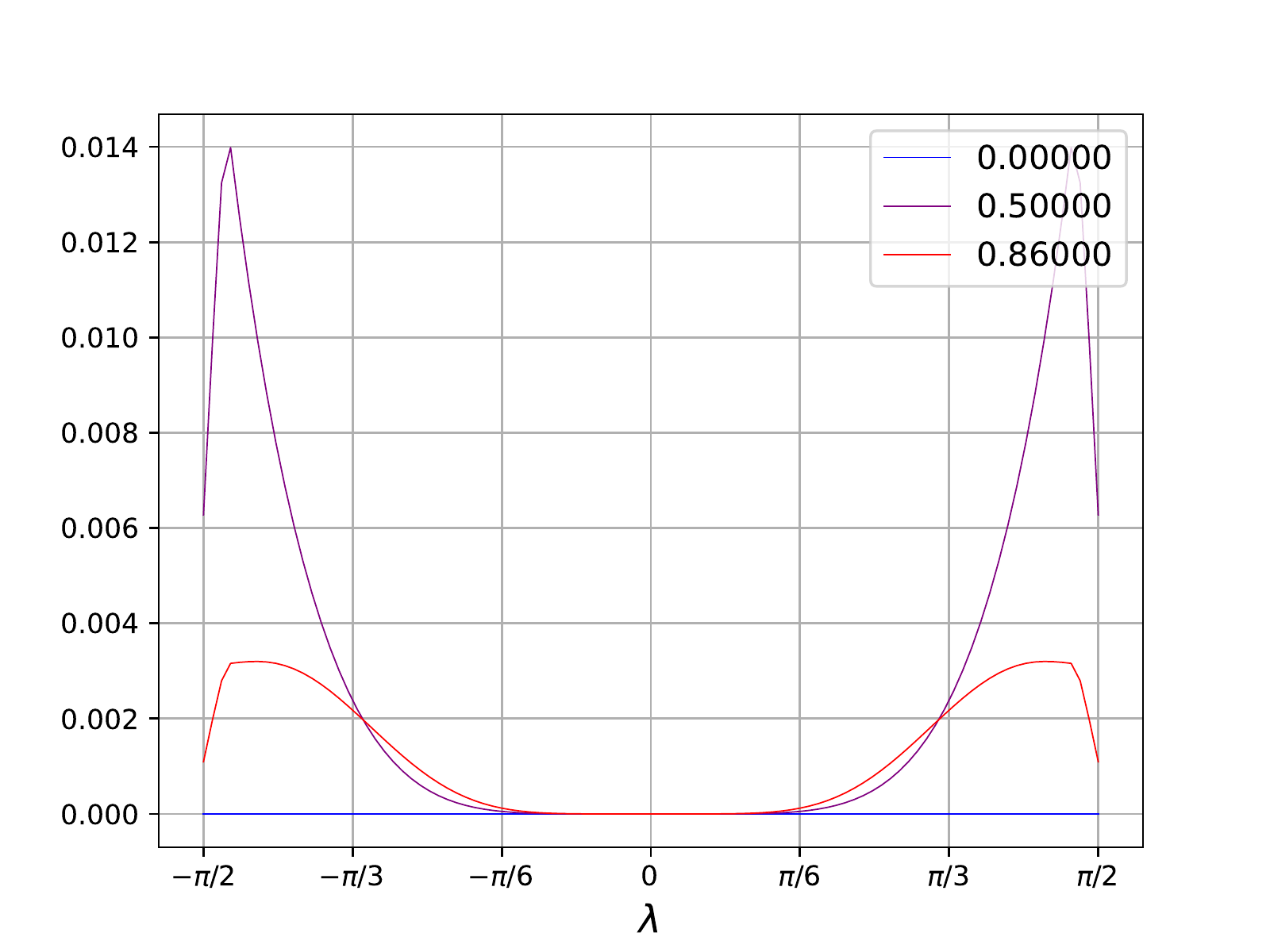}
   \end{minipage}\hfill\begin{minipage}[c]{.46\linewidth}
      \includegraphics[scale=0.5, trim=0cm 0cm 0cm 0cm]{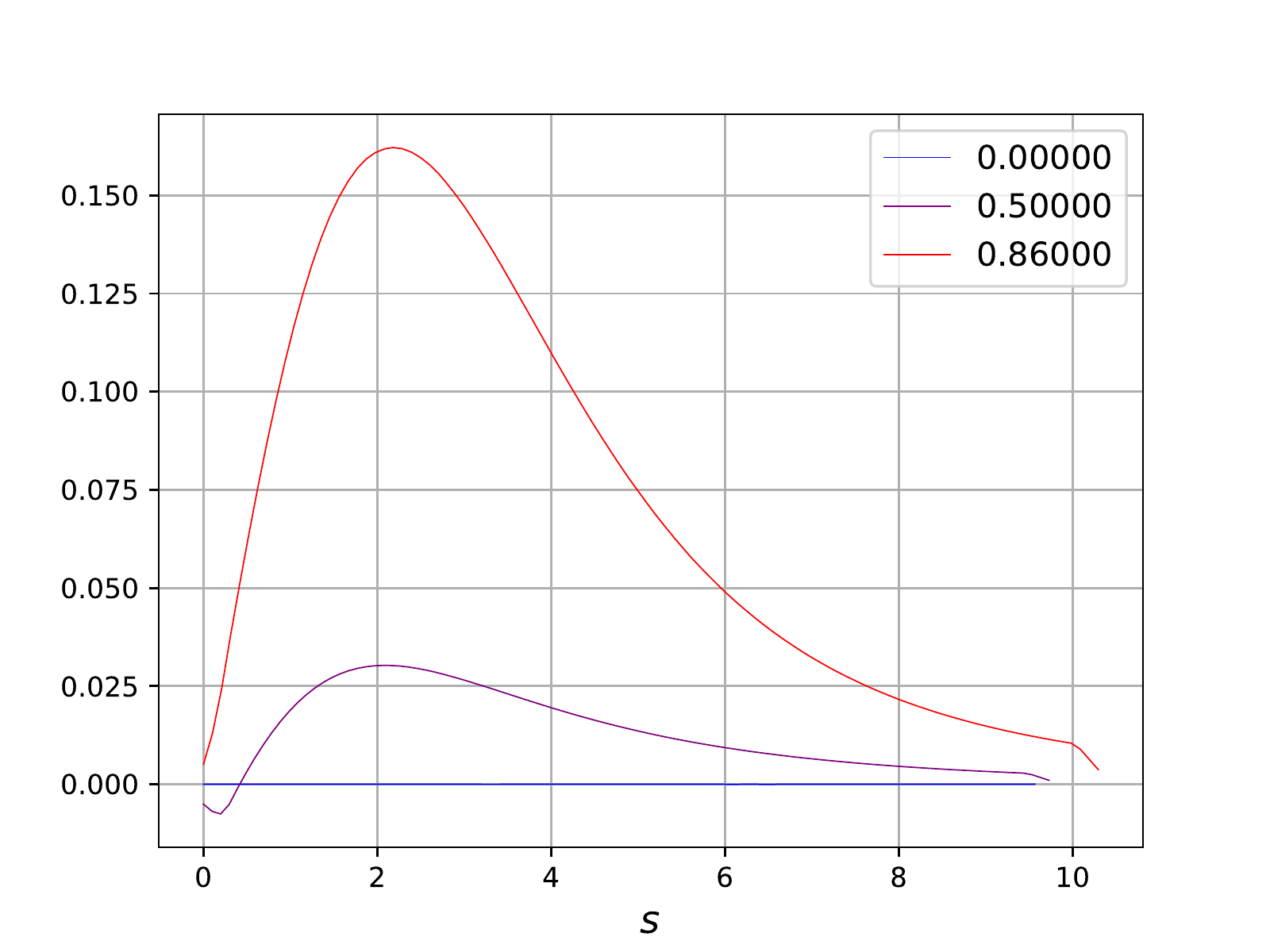}
   \end{minipage}
    \caption{On the left, evolution of torsion for the horizon line (black line in Fig.\ref{fig-evolutionplongamen-a-0-0.86}) in function of $s$. On the right, evolution of torsion for polar axis (blue line in Fig.\ref{fig-evolutionplongamen-a-0-0.86}) in function of $\lambda$.}
    \label{fig-torsion}
\end{figure}

In Fig.(\ref{fig-torsion}), we compare the torsion of the polar axis and horizon lines for different values of $a$. 
The horizon torsion gets its maximum value $\tau_{\rm H, max}(a)$ around $\lambda\approx\pm\pi/2$. Moreover, $\tau_{\rm H, max}(0.86)$ is lower than $\tau_{\rm H, max}(0.5)$. It implies that for $a=0.86$ the horizon gets closer to a planar curve than the horizon for $a=0.5$. Polar axis torsion reaches its maximum value around $s \approx 2$ that increases with $a$. For $a=0.86$, the horizon torsion is one order of magnitude lower than the one of the polar axis. For $a=0.86$ and $s$ approximately between the values $1$ and $4$, the polar axis is strongly non-planar curve.


\subsection{Isometric embedding for black hole spin greater than \texorpdfstring{$a_{\rm lim}$}{alim2}}

\begin{figure}[h!]
    \centering
    \hspace{-1cm}\begin{minipage}[c]{.48\linewidth}
      \includegraphics[scale=0.50, trim=0cm 0cm 0cm 0cm]{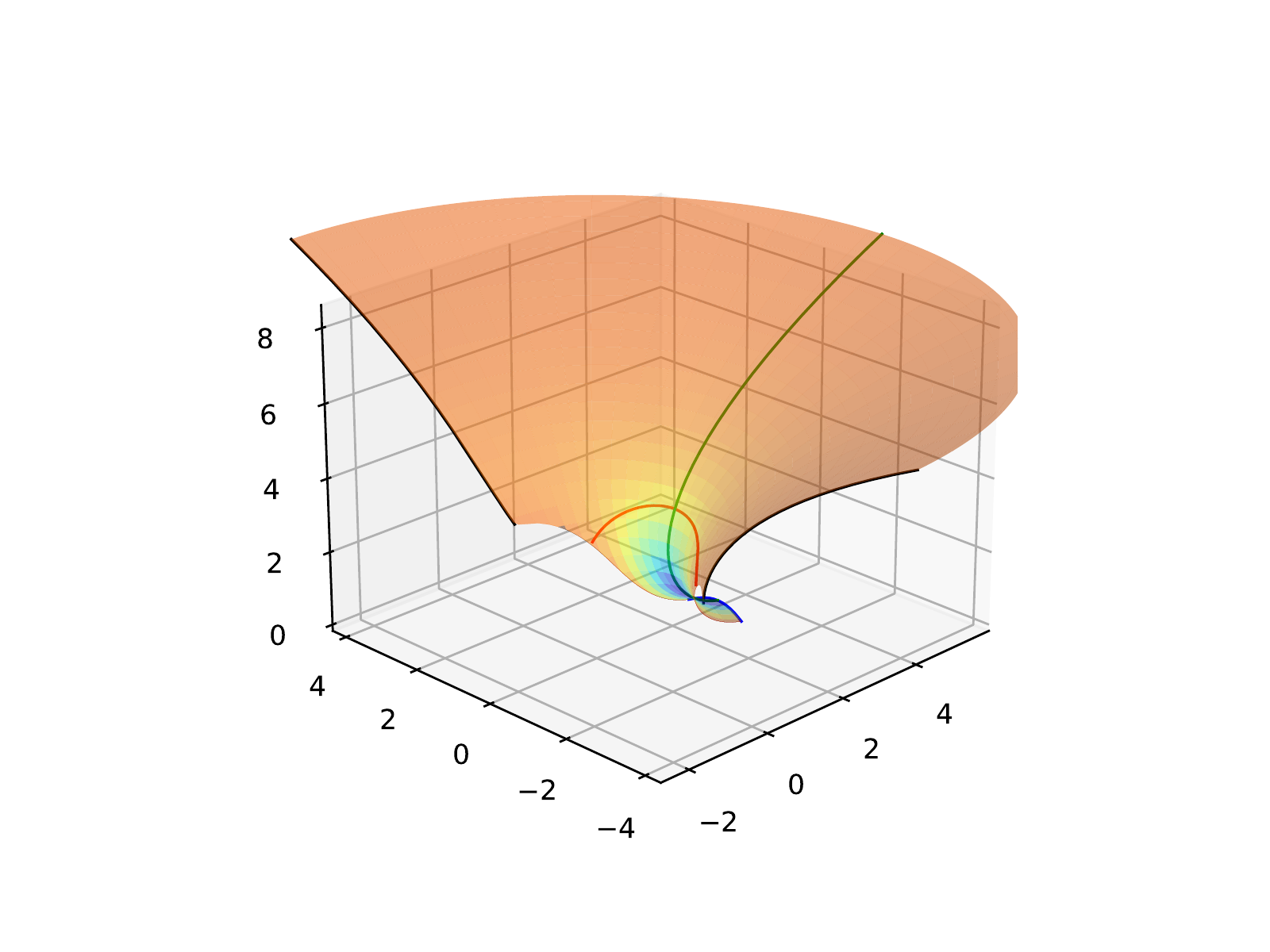}
      \includegraphics[scale=0.50, trim=0cm 0cm 0cm 0cm]{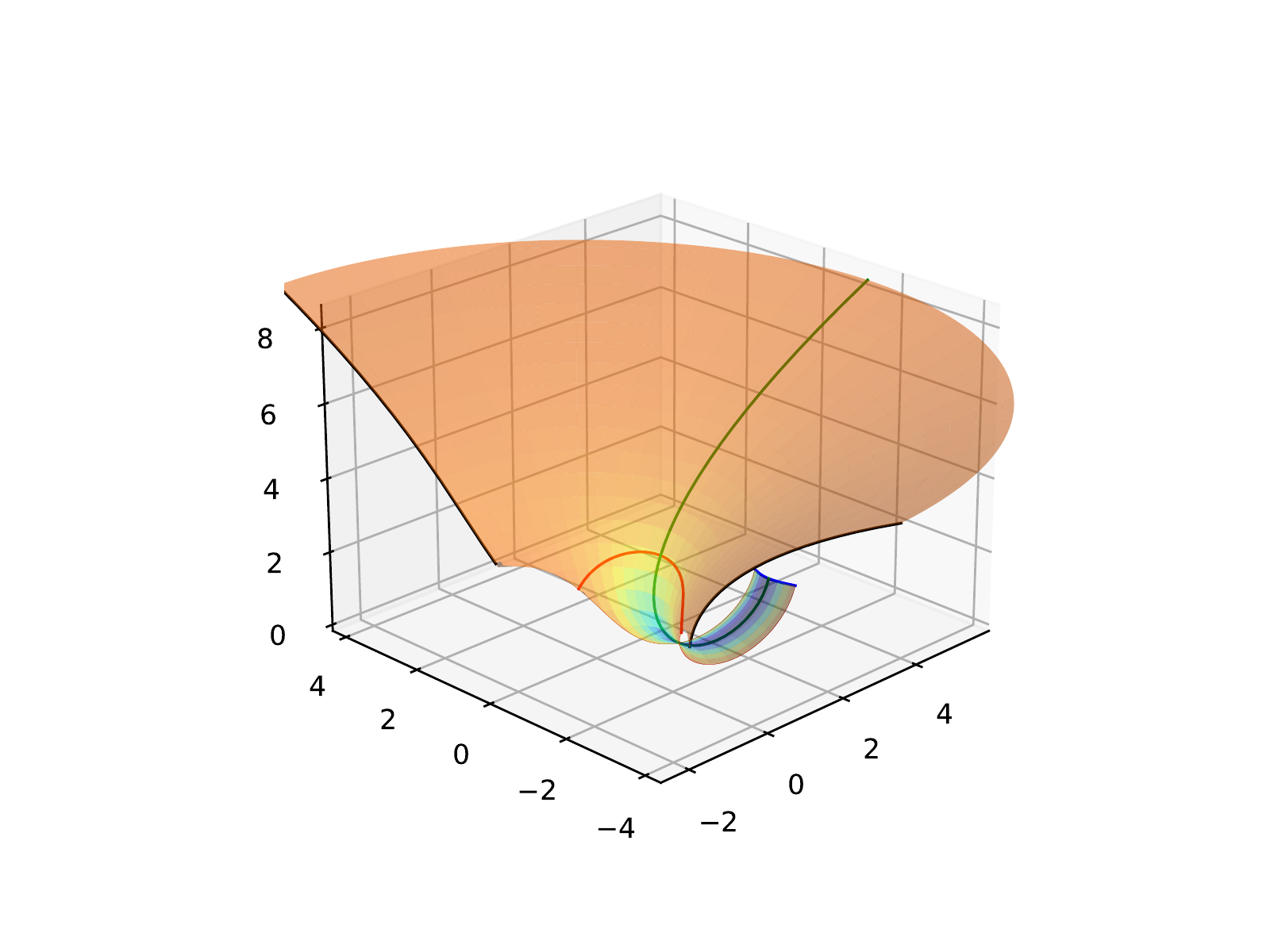}
   \end{minipage}\begin{minipage}[c]{.48\linewidth}
      \includegraphics[scale=0.50, trim=0cm 0cm 0cm 0cm]{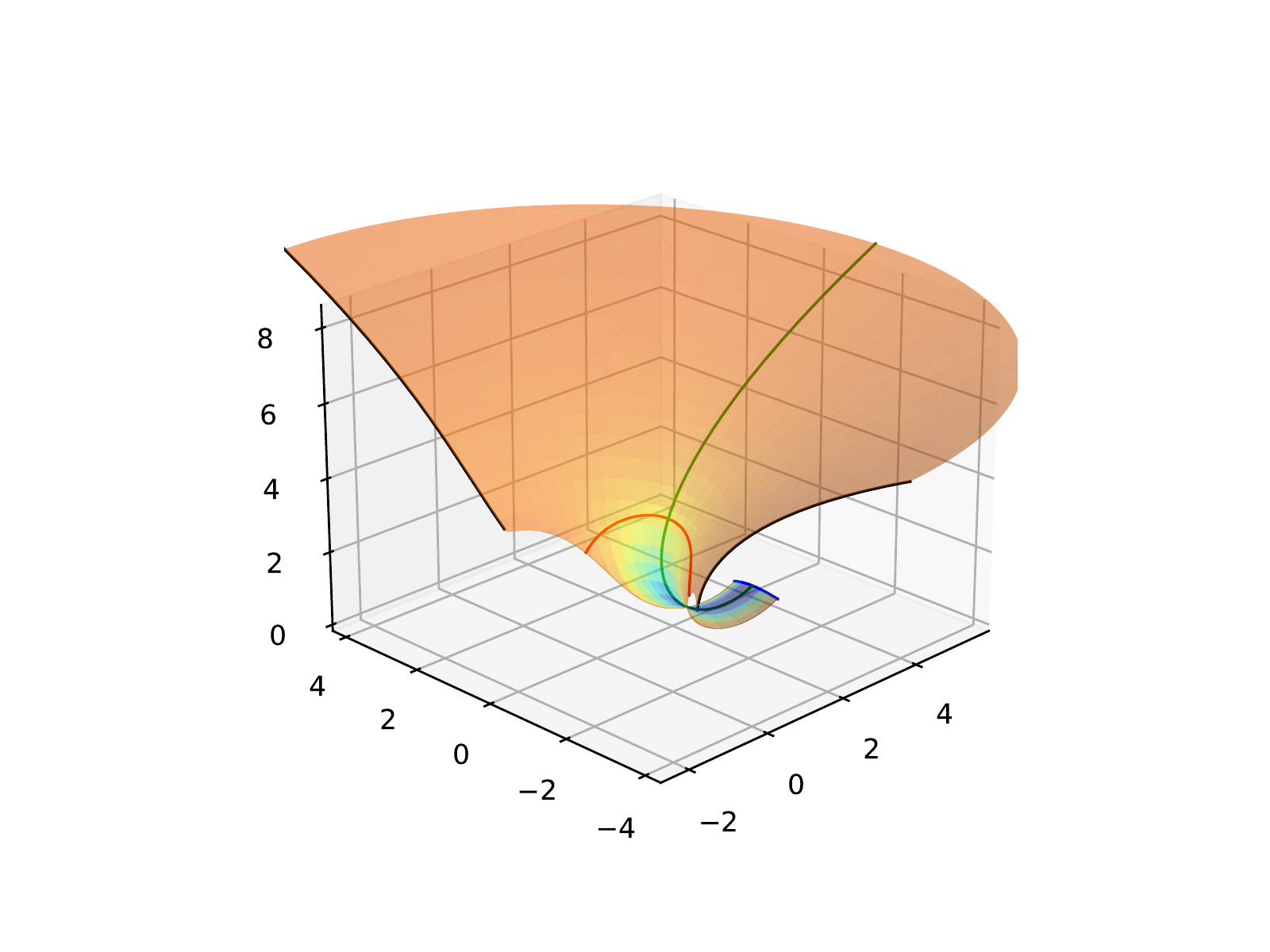}
      \includegraphics[scale=0.50, trim=0cm 0cm 0cm 0cm]{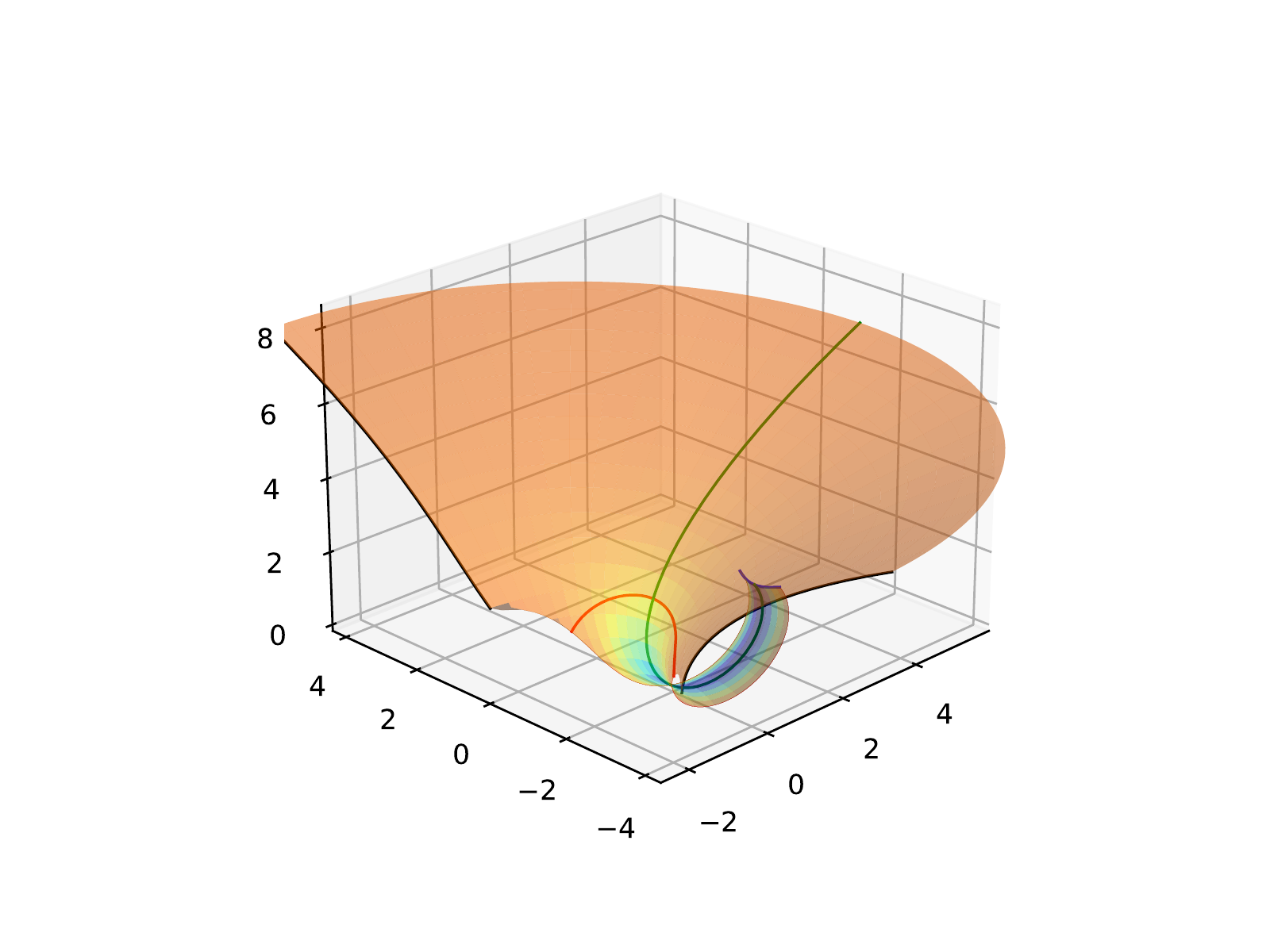}
   \end{minipage}
    \caption{Evolution of isometric embedding in the sub domain $\mathcal{U}_{s_{\rm max}}^{a,f}$ for $a=0.99$ (top-left), $a=0.999$ (top-right), $a=0.9999$ (bottom-left), $a=0.99999$ (bottom-right).}
    \label{fig-evolutionplongamen-a-0.99-0.99999}
\end{figure}

In order to obtain the isometric embedding, we have worked on the sub-domain $\mathcal{U}_{s_{\rm max}}^{a,f}=\left\{(s,\lambda)\in\mathcal{U}_{s_{\rm max}}\,|\, r^2(s)-3fa^2\sin^2\lambda>0\right\}$. For $f=1.27$ we are able to obtain the solutions for $a\in [0.86,0.99999]$. 

On Fig.(\ref{fig-evolutionplongamen-a-0.99-0.99999}), we can observe the coiling of the ergoregion as $a$ increases toward $1$. This coiling is visible only for values of $a$ extremely close to $1$. 

\begin{figure}[h!]
    \centering
    \hspace{0cm}\begin{minipage}[c]{.46\linewidth}
      \hspace{-2cm}\includegraphics[scale=0.55, trim=0cm 0cm 2cm 0cm]{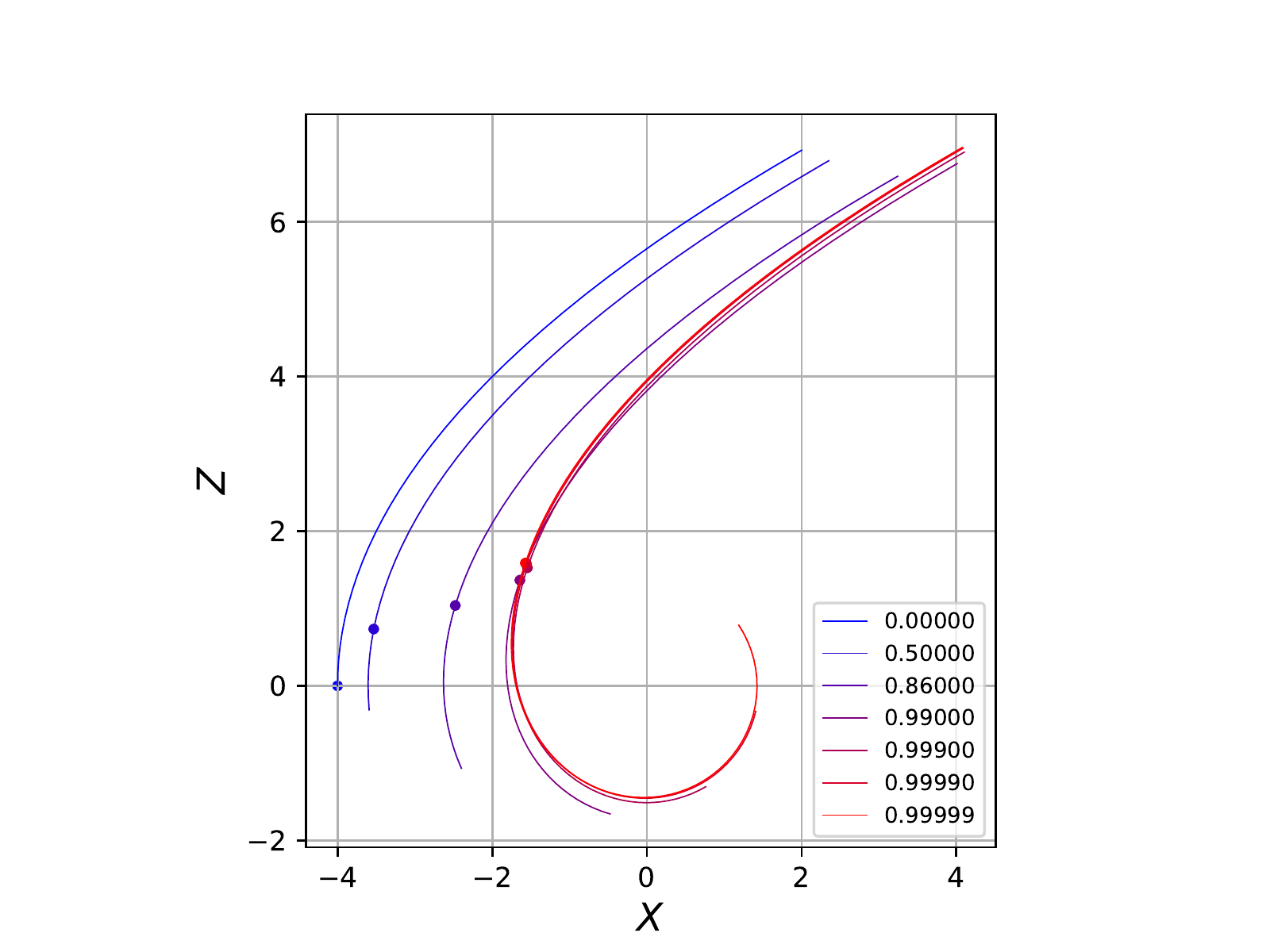}
   \end{minipage}\begin{minipage}[c]{.46\linewidth}
      \includegraphics[scale=0.50, trim=2cm 0cm 0cm 0cm]{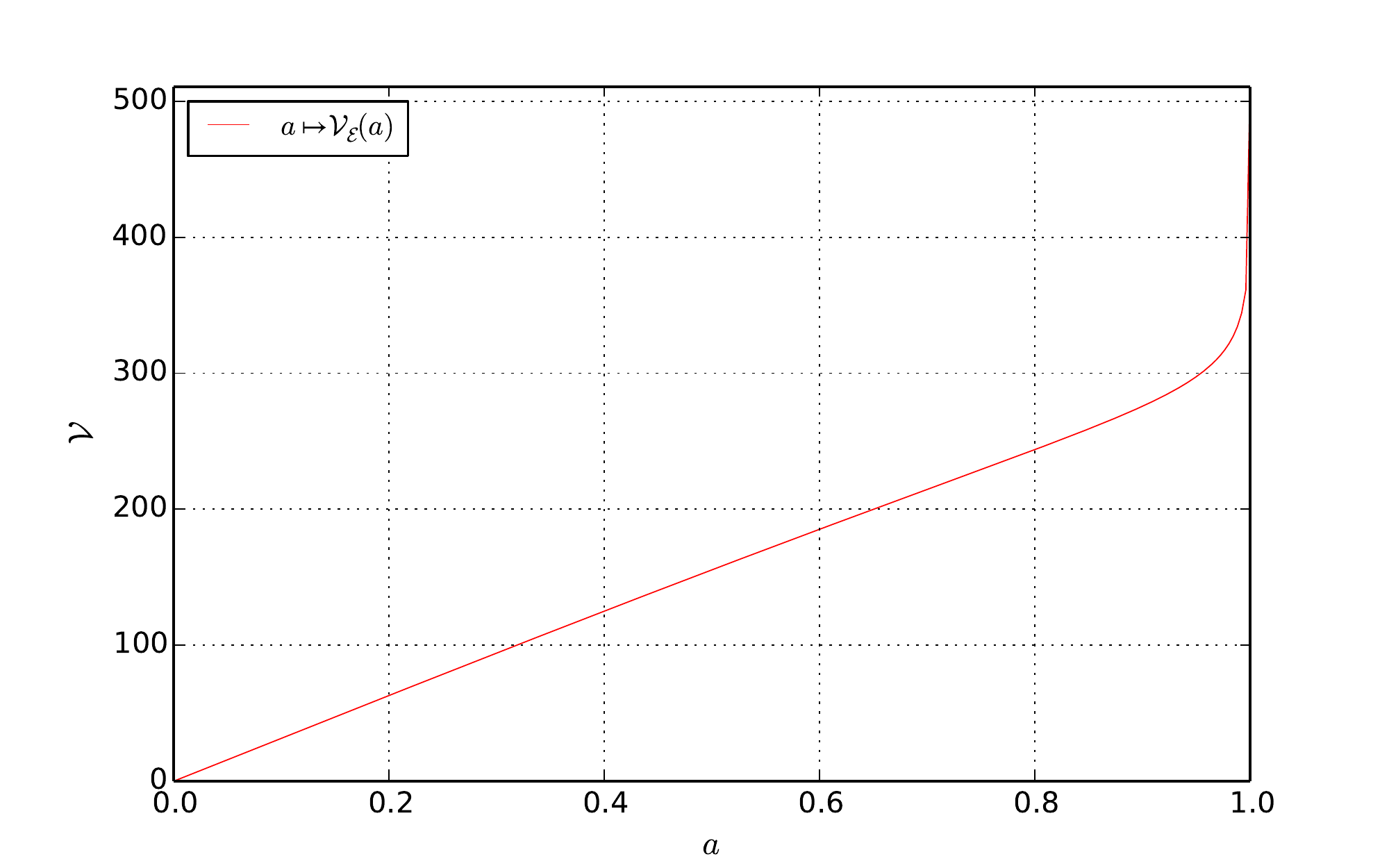}
   \end{minipage}
    \caption{On the left, the evolution of equatorial axis of isometric embeddings
    for different values of $a$ in $0_{XZ}$ plane. The dots correspond to the location of the ergosphere. On the right, the evolution of ergoregion volume (in $r_g^3$ units) measured by ZAMO frame.}
    \label{fig-evolveequatorialaxis}
\end{figure}

The coiling of the equatorial axis on the left of  Fig.(\ref{fig-evolveequatorialaxis}) is entirely determined by the choice of the initial conditions in Eq.(\ref{Eq-BoundaryCondition-3}). This coiling gets a nearly constant principal curvature along $s$ on the equatorial line when $s$ is close to $0$. When $a\rightarrow 1^{-}$, the equatorial line is coiling around a circle of radius ${\sqrt{2}}$. We already noticed (see Fig.(\ref{fig-evolutionangleprimitif})), that the angle $\psi(0){\rightarrow} -\infty$, when ${a\rightarrow 1^{-} }$, and it explains this coiling. 

Since the length of the equatorial axis remains the same for $s\in[0;s_{\rm max}]$, the whole surface is ``sliding'' by the coiling. The size of the ergoregion increases accordingly to this sliding. On the equatorial line, the length of the line included in the ergoregion is increasing with $a$ (see Fig. \ref{fig-evolveequatorialaxis}).

The  significant increase of this length - see Fig.(\ref{fig-evolveequatorialaxis}) - or the corresponding area in the ergoregion - see Fig.(\ref{fig-evolutionplongamen-a-0.99-0.99999}) - are in good agreement with the evolution of the ergoregion volume measured by the ZAMO observers,


\begin{equation}
    \mathcal{V}_{\mathcal{E}}(a)=2\pi \int_0^\pi \sin\theta \d \theta \int_{r_H}^{r_\mathcal{E}(\theta)} \sqrt{\frac{r(r^2+a^2)\Omega^2+2a^2 r\sin\theta}{r\Delta}}r^2  \d r \,.
\end{equation}

We plot in the right part of Fig.(\ref{fig-evolveequatorialaxis}) the evolution of ergoregion volume measured by ZAMO observers in function of black hole spin. This volume is increasing with $a$, and it diverges for $a=1$.


\section{Conclusion}

We exploited two different methods in order to obtain an approximate isometric embeddings of Kerr poloidal sub-manifold in the ordinary $3$-dimensional Euclidean space.

On one hand, the strictly short primitive embedding corrugation has the advantage of giving the explicit formula of the embeddings. The corrugated embeddings are furthermore defined over the entire starting domain $\mathcal{U}$. Nevertheless, the numerical computation of the explicit formula can be excessively long when the corrugation frequency increases. Moreover, these embeddings are difficult to exploit. The gaussian curvature of poloidal sub-manifold cannot be retrieved. These embeddings could not help for visualization because the fast oscillations of the tangent plane make the perspective difficult to exploit. 



On the other hand, the integration of GCM and frame equations gives good results for metric calculation particularly in the case of $a\leq a_{\rm lim} = \sqrt{3}/2$. Beyond this limiting value $a_{\rm lim}$, from computational reasons, we had to shrink down the initial set but we succeed to get the embedding near the equatorial axis for spin $a<0.99999$. It is very likely that the $\mathcal{C}^3$ isometric embedding of Kerr poloidal sub-manifold in the $3$-dimensionnal Euclidean ordinary space (like for Kerr black hole horizon \cite{PhysRevD.7.289}) could not exist in a global way beyond the limiting value $a_{\rm lim}$. Besides, we showed that the ergoregion volume increase in function of $a$ is mainly due to the length increase of the equatorial axis inside ergosphere and the ergoregion surface on the embedding. 

Finally, a possible option for getting a global embedding of the poloidal sub-manifold is, as in \cite{2009PhRvD..80d4014G} for the horizon, to consider an isometric embedding in 3-dimensional hyperbolic space. This method could probably be applied to represent different sub-manifolds of space-time (especially those prescribed in numerical relativity), on condition of being able to obtain boundary conditions.

\ack{L. Chantry thanks the Paris Observatory for the ATER position which enabled him to complete this work. }

\appendix

\section{Curvature}

We introduce the reduced Gaussian curvature,
\begin{equation}\label{Eq-GaussianCurvature-2}
\tilde{K}(s,\lambda)\hat{=}K(s,\lambda)\Omega^4(s,\lambda)=-\frac{r^2(s)-3a^2\sin^2\lambda}{r(s)(r^2(s)+a^2\sin^2\lambda)}
,\end{equation}
and its derivative that appears in Eq.(\ref{eq-source-GCM-reduced}),
\begin{equation}\label{Eq-GaussianCurvature-3}
\partial_\lambda \tilde{K}(s,\lambda)=\frac{8a^2 r^2 \cos\lambda\sin\lambda}{r\left(r^2+a^2\sin^2\lambda\right)^2}
.\end{equation}

\section{Compact form of GCM}\label{ap-derivGCMform}

The ordinary way to solve Eqs.\ref{Eq-GCM-1} is to introduce a change for the functions $L,M,N \leftrightarrow u,v$ which solves the Gauss equation,
\begin{equation}\label{Eq-utildevtile}
    \left\{\begin{array}{ccc}
        L(s,\lambda)&=&f_L(s,\lambda,u(s,\lambda),v(s,\lambda)) \\
        M(s,\lambda)&=&f_M(s,\lambda,u(s,\lambda),v(s,\lambda)) \\
        N(s,\lambda)&=&f_N(s,\lambda,u(s,\lambda),v(s,\lambda)) \\        
    \end{array}\right.
,\end{equation}
with,
\begin{equation}\label{Eq-CNS-Change}
    \forall u,v \,\,\,\,\,\, f_L(s,\lambda,u,v)f_N(s,\lambda,u,v)-f_M^2(s,\lambda,u,v)=\tilde{K}(s,\lambda)
,\end{equation}
where $\tilde{K}$ is defined in Eq.(\ref{Eq-GaussianCurvature-2}). 

Finally we obtain a semi-linear partial derivative form and the system can be written as:
\begin{equation}\label{Eq-uvGCM-SystemQuasiLinear-General1}
    \boldsymbol{A}_\lambda\left(s,\lambda,\boldsymbol{U}\right)\partial_\lambda\boldsymbol{U}+\boldsymbol{A}_s\left(s,\lambda,\boldsymbol{U}\right)\partial_s \boldsymbol{U}=\boldsymbol{S}^0\left(s,\lambda,\boldsymbol{U}\right)
,\end{equation}
with,
\begin{equation}\fl{
    \boldsymbol{U}=\left(\matrix{u \cr v \cr}\right)\mbox{,}\quad \boldsymbol{S}^0\left(s,\lambda,\boldsymbol{U}\right)=\left(\matrix{\partial_s f_M-\partial_\lambda f_L + \Gamma_{s\lambda}^s L+\left(\Gamma_{s\lambda}^\lambda-\Gamma_{ss}^s\right)M-\Gamma_{ss}^\lambda N \cr \partial_s f_N-\partial_\lambda f_M+ \Gamma_{\lambda\lambda}^s L+\left(\Gamma_{\lambda\lambda}^\lambda-\Gamma_{\lambda s}^s\right)M-\Gamma_{\lambda s}^\lambda N \cr}\right)}
,\end{equation}
and
\begin{equation}\fl{
     \boldsymbol{A}_\lambda\left(s,\lambda,\boldsymbol{U}\right)=\left(\matrix{\partial_u f_L & \partial_v f_L \cr \partial_u f_M & \partial_v f_M\cr}\right)\mbox{,}\quad \boldsymbol{A}_s\left(s,\lambda,\boldsymbol{U}\right)=-\left(\matrix{\partial_u f_M & \partial_v f_M \cr \partial_u f_N & \partial_v f_N\cr}\right)}
.\end{equation}
In most of the cases, the matrix $\boldsymbol{A}_\lambda$ could be inverted, and the system takes the form,
\begin{equation}
    \partial_\lambda\boldsymbol{U}+\boldsymbol{A}\left(s,\lambda,\boldsymbol{U}\right)\partial_s \boldsymbol{U}=\boldsymbol{S}\left(s,\lambda,\boldsymbol{U}\right)
,\end{equation}
with the matrix $\boldsymbol{A}\left(s,\lambda,\boldsymbol{U}\right)$ and the source term $\boldsymbol{S}\left(s,\lambda,\boldsymbol{U}\right)$ given by,
\begin{eqnarray}
    \boldsymbol{A}\left(s,\lambda,\boldsymbol{U}\right)=\boldsymbol{A}^{-1}_\lambda\left(s,\lambda,\boldsymbol{U}\right)\boldsymbol{A}_s\left(s,\lambda,\boldsymbol{U}\right)\\
    \boldsymbol{S}\left(s,\lambda,\boldsymbol{U}\right)=\boldsymbol{A}^{-1}_\lambda\left(s,\lambda,\boldsymbol{U}\right)\boldsymbol{S}^0\left(s,\lambda,\boldsymbol{U}\right)
\end{eqnarray}
We choose the Eq.(\ref{Eq-uvGCM-BF3}) to make explicit the Eq.(\ref{Eq-utildevtile}) in our specific case. The condition given in  Eq.(\ref{Eq-CNS-Change}) for the Eq.(\ref{Eq-uvGCM-BF3}) is verified. The matrix and source term are given by,
\begin{equation}\label{eq-matrix-GCM-reduced}
\boldsymbol{A}\left(s,\lambda,\boldsymbol{U}\right)=\Matrixd{0}{-\sqrt{2r}}{\frac{\sqrt{2r}{v}^2}{\tilde{K}+{u}^2}}{-\frac{2\sqrt{2r}{u}{v}}{\tilde{K}+{u}^2}}
,\end{equation}
and,
\begin{equation}\label{eq-source-GCM-reduced}
\boldsymbol{S}=\left( \matrix{ S^u \cr 
S^v \cr } \right)  
,\end{equation} 
with,
\begin{eqnarray*}
S^u&=& {v}\sqrt{\frac{\Delta}{2r^3}}+
\Gamma_{\lambda\lambda}^s \left(\frac{\tilde{K}+{u}^2}{\sqrt{2r}{v}}\right)+\left(\Gamma_{\lambda\lambda}^\lambda-\Gamma_{\lambda s}^s\right){u}-\sqrt{2r}\Gamma_{\lambda s}^\lambda {v}\\
S^v&=&\frac{{v}}{\tilde{K}+u^2}\left[uv\sqrt{2\frac{\Delta}{r^3}}+\partial_\lambda\tilde{K}+2r\Gamma_{ss}^\lambda v^2+2u^2\left(\Gamma_{\lambda\lambda}^\lambda-\Gamma_{s\lambda}^s\right)\right.\\
&&+\left.2\Gamma_{\lambda\lambda}^s\left(\frac{\tilde{K}+u^2}{\sqrt{2r}v}\right)u+\sqrt{2r}uv\left(\Gamma_{ss}^s-3\Gamma_{\lambda s}^\lambda\right)-\Gamma_{s\lambda}^s\left(\tilde{K}+u^2\right)\right]
\end{eqnarray*}

\newcommand{\newblock}{}
\bibliographystyle{plainnat}
\bibliography{Bibli.bib} 

\end{document}